\documentclass[12pt,american]{article}
\usepackage[T1]{fontenc}
\UseRawInputEncoding
\usepackage{color}
\usepackage{babel}
\usepackage{array}
\usepackage{float}
\usepackage{booktabs}
\usepackage{mathtools}
\usepackage{enumitem}
\usepackage{multirow}
\usepackage{varwidth}
\usepackage{amsmath}
\usepackage{amsthm}
\usepackage{amssymb}
\usepackage{graphicx}
\usepackage{geometry}
\usepackage{setspace}
\usepackage[authoryear,comma]{natbib}
\usepackage[pdfusetitle,
 bookmarks=true,bookmarksnumbered=false,bookmarksopen=false,
 breaklinks=false,pdfborder={0 0 1},backref=false,colorlinks=false]
 {hyperref}
\hypersetup{
 colorlinks,linkcolor={blue!50!black},citecolor={blue!50!black},urlcolor={blue}}

\makeatletter

\providecommand{\tabularnewline}{\\}
\newenvironment{cellvarwidth}[1][t]
    {\begin{varwidth}[#1]{\linewidth}}
    {\@finalstrut\@arstrutbox\end{varwidth}}

\theoremstyle{definition}
 \newtheorem{example}{\protect\examplename}
\theoremstyle{plain}
\newtheorem{assumption}{\protect\assumptionname}
\theoremstyle{remark}
\newtheorem{rem}{\protect\remarkname}
\theoremstyle{plain}
\newtheorem{thm}{\protect\theoremname}
\theoremstyle{plain}
\newtheorem{lem}{\protect\lemmaname}
\theoremstyle{plain}
\newtheorem{prop}{\protect\propositionname}

\usepackage{babel}
\usepackage{lmodern}
\usepackage{mathrsfs}
\usepackage{breakurl}
\usepackage{bbm}
\usepackage{float}
\usepackage{xcolor}
\usepackage[toc,page,titletoc]{appendix}
\usepackage{booktabs}
\usepackage{chngcntr}
\usepackage{mathtools}
\usepackage{threeparttable}

\numberwithin{equation}{section}

\usepackage{titling}
\allowdisplaybreaks
\makeatother

\providecommand{\assumptionname}{Assumption}
\providecommand{\examplename}{Example}
\providecommand{\lemmaname}{Lemma}
\providecommand{\propositionname}{Proposition}
\providecommand{\remarkname}{Remark}
\providecommand{\theoremname}{Theorem}

\begin{document}
\global\long\def\a{\alpha}%
 
\global\long\def\b{\beta}%
 
\global\long\def\g{\gamma}%
 
\global\long\def\d{\delta}%
 
\global\long\def\e{\epsilon}%
 
\global\long\def\l{\lambda}%
 
\global\long\def\t{\theta}%
 
\global\long\def\o{\omega}%
 
\global\long\def\s{\sigma}%
\global\long\def\ve{\varepsilon}%

\global\long\def\G{\Gamma}%
 
\global\long\def\D{\Delta}%
 
\global\long\def\L{\Lambda}%
 
\global\long\def\T{\Theta}%
 
\global\long\def\O{\Omega}%
 
\global\long\def\R{\mathbb{R}}%
 
\global\long\def\N{\mathbb{N}}%
 
\global\long\def\Q{\mathbb{Q}}%
 
\global\long\def\I{\mathbb{I}}%
 
\global\long\def\P{\mathbb{P}}%
 
\global\long\def\E{\mathbb{E}}%
\global\long\def\B{\mathbb{\mathbb{B}}}%
\global\long\def\S{\mathbb{\mathbb{S}}}%
\global\long\def\V{\mathbb{\mathbb{V}}}%
\global\long\def\C{\mathbb{\mathbb{C}}}%

\global\long\def\X{{\bf X}}%
\global\long\def\cX{\mathscr{X}}%
 
\global\long\def\cY{\mathscr{Y}}%
 
\global\long\def\cA{\mathscr{A}}%
 
\global\long\def\cB{\mathscr{B}}%
 
\global\long\def\cM{\mathscr{M}}%
\global\long\def\cN{\mathcal{N}}%
\global\long\def\cG{\mathcal{G}}%
\global\long\def\cC{\mathcal{C}}%
\global\long\def\sp{\,}%

\global\long\def\es{\emptyset}%
 
\global\long\def\mc#1{\mathscr{#1}}%
 
\global\long\def\ind{\mathbf{\mathbbm1}}%
\global\long\def\indep{\perp}%

\global\long\def\any{\forall}%
 
\global\long\def\ex{\exists}%
 
\global\long\def\p{\partial}%
 
\global\long\def\cd{\cdot}%
 
\global\long\def\Dif{\nabla}%
 
\global\long\def\imp{\Rightarrow}%
 
\global\long\def\iff{\Leftrightarrow}%

\global\long\def\up{\uparrow}%
 
\global\long\def\down{\downarrow}%
 
\global\long\def\arrow{\rightarrow}%
 
\global\long\def\rlarrow{\leftrightarrow}%
 
\global\long\def\lrarrow{\leftrightarrow}%

\global\long\def\abs#1{\left|#1\right|}%
 
\global\long\def\norm#1{\left\Vert #1\right\Vert }%
 
\global\long\def\rest#1{\left.#1\right|}%

\global\long\def\bracket#1#2{\left\langle #1\middle\vert#2\right\rangle }%
 
\global\long\def\sandvich#1#2#3{\left\langle #1\middle\vert#2\middle\vert#3\right\rangle }%
 
\global\long\def\turd#1{\frac{#1}{3}}%
 
\global\long\def\ellipsis{\textellipsis}%
 
\global\long\def\sand#1{\left\lceil #1\right\vert }%
 
\global\long\def\wich#1{\left\vert #1\right\rfloor }%
 
\global\long\def\sandwich#1#2#3{\left\lceil #1\middle\vert#2\middle\vert#3\right\rfloor }%

\global\long\def\abs#1{\left|#1\right|}%
 
\global\long\def\norm#1{\left\Vert #1\right\Vert }%
 
\global\long\def\rest#1{\left.#1\right|}%
 
\global\long\def\inprod#1{\left\langle #1\right\rangle }%
 
\global\long\def\ol#1{\overline{#1}}%
 
\global\long\def\ul#1{\underline{#1}}%
 
\global\long\def\td#1{\tilde{#1}}%
\global\long\def\bs#1{\boldsymbol{#1}}%

\global\long\def\upto{\nearrow}%
 
\global\long\def\downto{\searrow}%
 
\global\long\def\pto{\overset{p}{\longrightarrow}}%
 
\global\long\def\dto{\overset{d}{\longrightarrow}}%
 
\global\long\def\asto{\overset{a.s.}{\longrightarrow}}%
\global\long\def\gto{\rightarrow}%

\title{Identification and Estimation in a Time-Varying Endogenous Random
Coefficient Panel Data Model\thanks{I am deeply grateful to Donald W. K. Andrews, Yuichi Kitamura, and
Steven T. Berry for their invaluable guidance and support. I also
thank St\'ephane Bonhomme, Xiaohong Chen, \'Aureo de Paula, Wayne
Gao, Guido Imbens, Whitney Newey, Alexandre Poirier, Liangjun Su,
Frank Vella, and participants of conferences and seminars at the Duke
Class of 2020 and 2021 Microeconometrics Conference, Georgetown, Harvard
and MIT joint seminar, ITAM, NTU, NUS, PKU, PSU, THU, UCL, UGA, UQ,
Yale, and various Econometric Society Meetings for their helpful comments.
I am also grateful to the anonymous referees for their suggestions,
which greatly improved the paper. All remaining errors are my own.}}
\author{Ming Li\thanks{Department of Economics and Risk Management Institute, National University
of Singapore, 1 Arts Link, Singapore 117570, \protect\href{mailto:mli@nus.edu.sg}{mli@nus.edu.sg}.}}
\date{\today}
\maketitle
\begin{abstract}
This paper proposes a correlated random coefficient linear panel data
model, where regressors can be correlated with time-varying and individual-specific
random coefficients through both a fixed effect and a time-varying
random shock. I develop a new panel data-based method to identify
the average partial effect and the local average response function.
The identification strategy employs a sufficient statistic to control
for the fixed effect and a  control variable for the random shock.
Conditional on these two controls, the residual variation in the regressors
is driven solely by the exogenous instrumental variables, and thus
can be exploited to identify the parameters of interest. The constructive
identification analysis leads to three-step series estimators, for
which I establish rates of convergence and asymptotic normality. To
illustrate the method, I estimate a heterogeneous Cobb-Douglas production
function for manufacturing firms in China, finding substantial variations
in output elasticities across firms that can be related to various
firm characteristics.

\medskip{}

\noindent\textbf{Keywords:} Correlated random coefficients, panel
data, time-varying endogeneity, semiparametric estimation.
\end{abstract}
\newpage{}

\section{Introduction\label{sec:Introduction}}

Correlated random coefficient (CRC) linear panel models have proven
useful due to their ability to accommodate complex forms of unobserved
heterogeneity that are empirically relevant (\citet{wooldridge2010econometric,hsiao2022analysis}).
A crucial consideration in these models is the correlation between
random coefficients and regressors. Classical methods typically address
this issue by allowing a time-invariant, individual-specific fixed
effect---either in the form of an additive intercept or an individual-specific
coefficient---to be correlated with the regressors (e.g., \citet{hausman1981panel,hsiao2008random}).
While convenient, this approach may not fully capture agents' optimization
behavior. For example, even high-capability firms may cut inputs in
some periods when input prices are flat because input choices also
respond to transitory, time-varying shocks to output efficiency or
productivity. Throughout the sample, I assume capability is time-invariant.
Thus, both persistent capability and transitory shocks can affect
realized output efficiency and may be reflected in input choices,
thereby generating endogeneity that is not fully absorbed by a fixed
effect.

In this paper, I aim to address this gap by proposing a new time-varying
endogenous random coefficient (TERC) linear panel model, where regressors
can be correlated with time-varying and individual-specific random
coefficients not only through a fixed effect but, more importantly,
through a time-varying random shock. Specifically, the baseline TERC
model consists of two equations:
\begin{align}
Y_{it} & =X_{it}'\b\left(A_{i},\ve_{it}\right),\label{eq:output eq}\\
X_{it} & =g\left(Z_{it},A_{i},\eta_{it}\right).\label{eq:input eq}
\end{align}
The vector of random coefficients $\b_{it}=\b\left(A_{i},\ve_{it}\right)$
is modeled as a vector of unknown functions $\b$ of a fixed effect
$A_{i}$ and a time-varying random shock $\ve_{it}$. $Y_{it}$ is
then determined by the inner product between $X_{it}$ and $\b_{it}$.
In \eqref{eq:input eq}, I model the vector of regressors $X_{it}$
as a vector-valued, time-varying function $g$ of the vector of exogenous
instrumental variables (IV) $Z_{it}$, fixed effect $A_{i}$, and
per-period information $\eta_{it}$ about $\b_{it}$.

The motivation for \eqref{eq:input eq} is that the agent $i$ in
period $t$ optimally chooses values of her regressors $X_{it}$ by
solving an optimization problem (e.g., firm profit maximization),
with $Z_{it},\ A_{i},\text{ and }\eta_{it}$ in her information set,
leading to \eqref{eq:input eq}. All of $A_{i}$, $\ve_{it}$, and
$\eta_{it}$ can be scalar- or vector-valued, and any pair of the
three variables can be correlated. As an analyst, I observe $\left\{ X_{it},Y_{it},Z_{it}\right\} $
for $i=1,\ldots,n$ and $t=1,\ldots,T$ and aim to identify the pooled
average partial effect $T^{-1}\sum_{t=1}^{T}\E\b_{it}$ (APE, see
\citet{graham2012identification}) and the local average response
function $\E\left[\rest{\b_{it}}X_{it}\right]$ (LAR, see \citet{altonji2005cross}).\footnote{Note that the joint distribution of $(\ve_{it},\eta_{it})$ may vary
over $t$. Consequently, any functionals of their distribution (e.g.,
$\E\b_{it}$ and $\E\left[\rest{\b_{it}}X_{it}\right]$) are in principle
time-varying. I suppress the subscript $t$ whenever it is clear from
context.}

A key feature of the TERC model is that $X_{it}$ can be correlated
with $\b_{it}$ through both $A_{i}$ and $\eta_{it}$, potentially
in a complicated manner. I refer to this feature as ``time-varying
endogeneity through the random coefficients.'' Allowing such correlation
is important to applications when agent $i$ in period $t$ has information
about $\b_{it}$ through both $A_{i}$ and $\eta_{it}$ when optimally
deciding $X_{it}$. In Section \ref{sec:Model}, I present three empirical
examples---heterogeneous production function estimation, a labor
supply model, and demand estimation---that demonstrate the validity
of this correlation (\citet*{deaton1986demand,blundell2007labor,li2024identification,aureo2024production}).
The associated technical challenge is to control for the time-varying
endogeneity through the random coefficients when both $A_{i}$ and
$\eta_{it}$ enter $g$ in a nonlinear and nonseparable manner in
\eqref{eq:input eq}. Moreover, I do not impose parametric assumptions
on the distributions of $A_{i}$, $\ve_{it}$, or $\eta_{it}$, nor
do I specify the functional form of $\b$ or $g$. In this sense,
the TERC model is a fixed-effects panel data model, and addressing
these challenges requires a new method.

I propose a new panel data-based method for identifying the APE and
LAR in the TERC model. The idea is to control for $A_{i}$ and $\eta_{it}$
via a \textit{sufficient statistic} and\textit{ a generalized residual},
respectively, such that, conditional on these two controls, the residual
variation in $X_{it}$ is driven solely by the exogenous IV $Z_{it}$.
In other words, the residual variation in $X_{it}$ is causal and
thus can be exploited to identify the parameters of interest. Specifically,
I first impose an index exclusion assumption that supplies a sufficient
statistic $W_{i}$ for $A_{i}$. I present justifications for this
assumption from the literature. Next, I construct a feasible conditional
cumulative distribution function (CDF) $V_{it}=F_{\rest{X_{it}}Z_{it},W_{i}}\left(\rest{X_{it}}Z_{it},W_{i}\right)$
to control for $\eta_{it}$. Given $(A_{i},W_{i})$, I establish a
one-to-one mapping between $V_{it}$ and $\eta_{it}$. Finally, conditioning
on these two controls---which effectively fixes both $A_{i}$ and
$\eta_{it}$---\eqref{eq:input eq} implies that the residual variation
in $X_{it}$ is driven solely by $Z_{it}$. This residual, instrument-induced
variation can therefore be exploited to identify the APE and LAR via
perturbation arguments and iterated expectations. At the end of Section
\ref{sec:Identification}, I present three extensions: (i) allowing
$g$ to depend on multiple components of $\eta_{it}$, (ii) identifying
higher-order moments of $\b_{it}$, and (iii) adding ex-post shocks
to $\b_{it}$ and $Y_{it}$, and including exogenous covariates into
$X_{it}$.

I estimate the parameters of interest using a three-step series procedure.
In the first step, I estimate the conditional CDF $V_{it}$, which
serves as a control for $\eta_{it}$. In the second step, I exploit
the linear structure of (\ref{eq:output eq}) to estimate $\E[\rest{Y_{it}}X_{it},V_{it},W_{i}]$.
In the third step, I use these estimates to recover the APE and LAR.
When working with large datasets, the main computational challenge
comes from the first-step estimation of $V_{it}$ and the second-step
estimation of $\E[\rest{Y_{it}}X_{it},V_{it},W_{i}]$. I address these
issues in Remark \ref{rem:comp-issue}. 

Inference is less standard because these estimators are multi-step:
they rely on intermediate nonparametric estimates, include derivatives
of an earlier-step estimate, and then apply an additional unknown
but estimable mapping to a conditional expectation. As a result, sampling
error from the first and second steps can affect the limiting distribution
of the final estimators, so valid standard errors must account for
how uncertainty propagates across steps. I therefore derive convergence
rates and asymptotic normality by explicitly tracking the contribution
of each step, building on existing results for multi-step series estimators
(\citet{don1991asymnormality,newey1997convergence,imbens2009identification}).
A discussion of how the errors from each step aggregate into the final
standard error is provided at the end of Subsection \ref{subsec:est and inf -rates and normality}.

To illustrate the method, I estimate a Cobb--Douglas production function
with heterogeneous output elasticities using data on Chinese manufacturing
firms. The instruments are inspired by \citet*{blp1995io} and constructed
as competitors\textquoteright{} leave-one-out weighted-average input
prices within the same city and industry. The resulting APE estimates
are broadly consistent with those obtained from classical constant-coefficient
approaches applied to the same data (\citet*{olleypakes1996teltfp,levinsohnpetrin2003res,ackerberg2015identification}).
I also estimate the LAR functions evaluated at each firm\textquoteright s
realized input choices, and I find substantial cross-firm variation.
I then relate heterogeneity in output elasticities to firm characteristics
and obtain economically intuitive patterns. Finally, I conduct extensive
robustness checks in Appendix \ref{sec:Additional-Empirical-Results}
and find that the results are stable.

To complement the empirical analysis, Appendix \ref{sec:Simulation}
presents Monte Carlo simulations designed to mirror the empirical
environment; the results show that the proposed estimator performs
well and reinforce the empirical findings. In particular, in the baseline
design ($n=1,000$, $T=2$, and 1,000 replications, using second-degree
spline bases), the APE estimator exhibits small normalized bias (about
2.8--3.2\% relative to the length of the true parameter) and low
normalized rMSE (about 2.9--3.7\%) across coefficients. Estimating
the function-valued LAR is naturally less precise, but still yields
normalized rMSEs on the order of 3.4--5.6\% for all the coefficients.
Performance improves with larger $n$ and additional time periods
$T$. Robustness checks also indicate that second-degree splines outperform
first- or third-degree bases, and that adding ex-post shocks increases
rMSEs only slightly.

Before turning to the related literature, I note two limitations of
the paper. First, the baseline model assumes each coordinate of $g$
is monotone in a single coordinate of $\eta_{it}$, which (as in \citet{imbens2009identification})
rules out certain nonseparable supply-and-demand systems. Subsection
\ref{subsec:Extension-2:Vector-Valued-U} discusses remedies under
additional, empirically motivated assumptions. Second, the approach
requires IVs and a tractable way to summarize the fixed effect $A_{i}$
via a sufficient statistic $W_{i}$; while I provide examples for
the applications considered, finding credible instruments (and justifying
the required structure) may be challenging in some settings.

\subsection{Related Literature}

This paper contributes to the literature on CRC linear panel models.
Related work includes \citet{chamberlain1992efficiency} on regular
identification of first moments when $T>d_{X}$; \citet{graham2012identification}
on irregular identification when $T=d_{X}$ using movers/stayers;
and \citet*{arellano2012identifying} on identifying first (and, under
additional assumptions, higher) moments and distributions via within--between
variation and restrictions on residual dependence. \citet{wooldridge2005fixed,wooldridge2009estimating}
develops estimation strategies for APEs---respectively under mean-independence
conditions and with sample selection in unbalanced panels---while
\citet{murtazashvili2008fixed} study consistency of fixed-effects
IV estimators in time-invariant CRC models. Additional contributions
include identification of quantiles (\citet*{graham2018quantile}),
random-coefficient simultaneous equations (\citet{masten2018random}),
and a two-step identification approach with additive fixed effects
and time-invariant random coefficients (\citet{laage2024correlated}).

My contribution to the CRC literature is to study a model, (\ref{eq:output eq})
and (\ref{eq:input eq}), with time-varying random coefficients that
can be correlated with regressors through both fixed effects and time-varying
shocks. The closest paper to my work is \citet{graham2012identification},
who identify moments by exploiting movers and stayers under a time-stationarity
restriction. This time-stationarity restriction conditions on the
regressors for all periods and thus effectively rules out the time-varying
endogeneity central here. I instead introduce controls for $A_{i}$
and importantly $\eta_{it}$ to identify the APE and LAR. The two
approaches rely on non-nested assumptions; I view this paper as a
step toward addressing time-varying endogeneity through random coefficients.

In addition to the aforementioned papers, my work is also related
to the literature on: (i) time-varying parameter panel data models
(e.g., \citet*{li2011non,li2016panel}), (ii) latent structure panel
data models (e.g., \citet*{su2016identifying,su2019sieve,wang2024panel}),
and (iii) linear panel data model with interactive fixed effects,
which can be treated as imposing a specific structure on the random
coefficient associated with the constant term (e.g., \citet{pesaran2006estimation,bai2009panel,moon2015linear}).
The modeling technique and identification method of my paper are different
from these seminal papers.

The second line of research concerns the techniques used in this paper.
For nonseparable panel models with generalized fixed effects, restrictions
on the fixed effect are generally needed to identify global objects
such as the APE and LAR (\citet[p.2]{hoderlein2012nonparametric};
see also \citet*{GaoLiXu2023LogicalDifferencing} and \citet{GaoLi2025SemiparametricPanelMultinomial}
on dealing with the issue of fixed effects in network formation and
panel multinomial choice models, respectively). My method requires
a sufficient statistic $W_{i}$ for $A_{i}$ and a control variable
$V_{it}$ for $\eta_{it}$. To justify the sufficiency of $W_{i}$
for $A_{i}$, I adapt several techniques from the literature (e.g.,
\citet*{mundlak1978pooling,altonji2005cross,bester2009identification,WOOLDRIDGE2019JoE,arkhangelsky2022role}).
See \citet*{liu2024identification} for a discussion on the index
sufficiency condition in nonlinear semiparametric panel data models.
To justify $V_{it}$ as a valid control for $\eta_{it}$, I generalize
the technique of \citet{imbens2009identification} in two nontrivial
ways and discuss them in relation to (\ref{eq:feas-V}). More recently,
\citet{Nagasawa2024NoisyConditioning} studies treatment effect estimation
with noisy conditioning variables using control variables from the
joint distribution of noisy proxy measures.

The asymptotic analysis builds on foundational results for series
and control-function estimators. \citet{don1991asymnormality} and
\citet{newey1997convergence} provide conditions for convergence rates
and asymptotic normality of series estimators, which I use to handle
vector-valued functionals of regression functions. \citet*{newey1999nonparametric}
establish asymptotic normality for two-step nonparametric estimators
in triangular models with a separable first stage, while \citet{imbens2009identification}
extend the triangular framework to nonseparable first stages and derive
rates and asymptotic normality for known functionals of conditional
expectations. I build on \citet{imbens2009identification} to obtain
asymptotic normality for unknown but estimable functionals of conditional
expectation functions.

\textbf{Organization}. The rest of the paper is organized as follows.
Section \ref{sec:Model} formally introduces the TERC model and provides
several empirical examples that fit within its framework. Section
\ref{sec:Identification} outlines the identification idea and presents
the assumptions, key identification theorem, and three extensions.
Section \ref{sec:Est and Inf} presents the series estimators for
the APE and LAR and establishes their asymptotic properties. Section
\ref{sec:Empirical-Illustration} provides an empirical illustration.
Finally, Section \ref{sec:Conclusion} concludes. Proofs, additional
empirical results, and a simulation study are included in Appendix
\ref{sec:Proofs-addl-theory}, \ref{sec:Additional-Empirical-Results},
and \ref{sec:Simulation}, respectively.

\textbf{Notation}. Let $i\in\left\{ 1,\ldots,n\right\} $ index agents
and $t\in\left\{ 1,\ldots,T\right\} $ index periods, with finite
$T\geq2$. I use boldface upper case letters for random matrices,
regular upper case letters for random variables and random vectors,
and lower case letters for their values. Let ${\bf X}_{i}=\left(X_{i1},...,X_{iT}\right)'$,
$Y_{i}=\left(Y_{i1},...,Y_{iT}\right)'$, ${\bf Z}_{i}=\left(Z_{i1},...,Z_{iT}\right)'$,
$\ve_{i}=\left(\ve_{i1},...,\ve_{iT}\right)'$, and $\eta_{i}=\left(\eta_{i1},...,\eta_{iT}\right)'$.
I assume $\left({\bf X}_{i},{\bf Z}_{i},A_{i},\ve_{i},\eta_{i}\right)$
are i.i.d. across $i$. Let $d_{X}$ be the dimension of $X_{it}$
and $X_{it,l}$ denote the $l^{\text{th}}$ coordinate of vector $X_{it}$.
Similar notation is used for all other variables. I use $\coloneqq$
to define a new random variable, $\sim_{d}$ to indicate that two
random variables are identically distributed, $F$ for CDF, $f$ for
probability density function (PDF), $\E$ for expectation, $\V$ for
variance, $I_{d}$ for $d\times d$ identity matrix, $\pto$ for convergence
in probability, and $\dto$ for convergence in distribution. All convergence
results are stated under $n\to\infty$.

\section{Model \label{sec:Model}}

Consider the following baseline TERC model
\begin{align}
Y_{it} & =X_{it}'\b\left(A_{i},\ve_{it}\right),\label{eq:Yeq}\\
X_{it} & =g\left(Z_{it},A_{i},\eta_{it}\right),\label{eq:main eqn 2 X}
\end{align}
where $\b_{it}=\b\left(A_{i},\ve_{it}\right)\in\R^{d_{X}}$, the central
object of interest, is a vector of random coefficients modeled as
an unknown vector-valued function $\b$ of $(A_{i},\ve_{it})$. Here,
$A_{i}$ is a fixed effect and $\ve_{it}$ governs the time-varying
behavior of $\b_{it}$; both may have arbitrary dimension. The mapping
$\b$ itself can be time-varying, because $\ve_{it}$ captures time
variation in the functional form. $\eta_{it}\in\R^{d_{\eta}}$ is
a continuously distributed, time-varying random vector that may be
correlated with $A_{i}$ and $\ve_{it}$; its coordinates may also
be mutually correlated. This correlation can be important in applications;
see the examples below. Let $X_{it}\in\R^{d_{X}}$ denote choice variables
(e.g., capital and labor), $Y_{it}\in\R$ the outcome, and $Z_{it}\in\R^{d_{Z}}$
instruments. Finally, $g$ is an unknown vector-valued function of
$(Z_{it},A_{i},\eta_{it})$ that determines each coordinate of $X_{it}$. 

The TERC model is widely used in empirical studies. It extends the
classic constant-coefficient linear panel model while preserving directly
interpretable targets---such as the APE and LAR---that carry clear
policy relevance. It also accommodates natural extensions, including
higher-order moments of the APE and LAR. In what follows, I describe
three applications of the TERC model.

\begin{example}[Production Function Estimation]
 Suppose firm $i$ in year $t$ has a heterogeneous Cobb-Douglas
production function in the form of (\ref{eq:Yeq}). The capital and
labor elasticities $\b_{it}=\b\left(A_{i},\ve_{it}\right)\in\R^{2}$
are modeled as a two-dimensional function of the firm fixed effect
$A_{i}$ (e.g., manager ability that is assumed to be stable over
time) and a time-varying random shock $\ve_{it}$ (e.g., realized
per-period technology shock). When deciding its $X_{it}\in\R^{2}$
of capital and labor, assume firm $i$ knows $A_{i}$ and $\eta_{it}$
(e.g., expected per-period technology shock). Clearly, $\eta_{it}$
is correlated with $A_{i}$ and $\ve_{it}$. Assume firm $i$ also
observes $Z_{it}$, which captures local input market dynamics that
affect $X_{it}$---for example, competitors' debt-weighted interest
rates and employment-weighted wages. Let the cost function be given
by $c\left(x,z\right)$. Then, firm $i$ selects $X_{it}$ by maximizing
its expected profit with the knowledge of $\left(Z_{it},A_{i},\eta_{it}\right)$,
i.e., 
\[
X_{it}=\arg\max_{x\in\R^{2}}\left[\E\left[\rest{x'\b\left(A_{i},\ve_{it}\right)}Z_{it},A_{i},\eta_{it}\right]-c\left(x,Z_{it}\right)\right],
\]
leading to (\ref{eq:main eqn 2 X}). Subsequently, firm $i$ obtains
its output $Y_{it}$ via (\ref{eq:Yeq}).
\end{example}
\begin{example}[Labor Supply]
\label{exa:labor} Suppose worker $i$ has a linear labor supply
function of the form (\ref{eq:Yeq}), where $Y_{it}$ is the number
of hours worked and $X_{it}$ includes the endogenous hourly wage
($X_{it,1}$) along with other exogenous demographic variables. The
coordinate of $\b_{it}$ corresponding to the wage is the key object
of interest because it quantifies the elasticity of labor supply with
respect to the wage rate. Given exogenous factors $Z_{it}$ (e.g.,
county minimum-wage changes or earned income tax credit expansions),
individual ability $A_{i}$, and shock $\eta_{it}$ (e.g., her expectation
of a near-term surge in labor demand in her sector, which increases
the expected marginal product of labor and wage offers), she chooses
a job with the wage that maximizes her expected utility, resulting
in (\ref{eq:main eqn 2 X}). Finally, worker $i$ commits $Y_{it}$
hours of working via (\ref{eq:Yeq}).
\end{example}
\begin{example}[Almost Ideal Demand System (AIDS)]
 Suppose household $i$'s gasoline budget share $Y_{it}$ at time
$t$ depends on the gasoline price $X_{it,1}$ and total expenditure
$X_{it,2}$ as in (\ref{eq:Yeq}). Let $\b_{it}\in\R^{2}$ denote
the possibly heterogeneous elasticities with respect to $X_{it}$,
modeled as unknown functions of a household fixed effect $A_{i}$
(e.g., a persistent propensity to drive versus use transit) and a
time-varying wealth shock $\ve_{it}$. Given instruments $Z_{it}$
(e.g., head of household's earned income and gas tax change), $A_{i}$,
and a signal $\eta_{it}$ about $\b_{it}$ (e.g., expected near-term
income change such as upcoming bonus), the household searches for
gasoline prices and sets its total expenditure budget by maximizing
expected utility, which yields (\ref{eq:main eqn 2 X}).
\end{example}
The time-varying correlation between $X_{it}$ and $\b_{it}$ illustrates
a distinct source of endogeneity: optimization-induced selection into
regressors through random coefficients. Classic work (\citet*{manski1987semiparametric,altonji2005cross,graham2012identification,chernozhukov2013average})
allows $X_{it}$ to correlate with $A_{i}$ (e.g., better-managed
firms choose higher inputs). But empirically, $X_{it}$ can also co-move
with $\b_{it}$ beyond $A_{i}$, because time variation in choices
is rarely fully spanned by instrument variation $Z_{it}$. For example,
even with little movement in input prices, a well-managed firm may
temporarily scale inputs up or down in response to private information
about returns. This motivates a time-varying shock $\eta_{it}$ that
affects $X_{it}$ and is informative about $\b_{it}$.

A second departure from standard triangular models (\citet*{newey1999nonparametric,imbens2009identification})
is that $A_{i}$ enters both the outcome equation and the first-step
choice equation: $X_{it}=g\left(Z_{it},A_{i},\eta_{it}\right)$, not
$X_{it}=g\left(Z_{it},\eta_{it}\right)$. Consequently, residual-
or CDF-based control-function approaches that rely on a one-to-one
mapping between the control and $\eta_{it}$ no longer apply. Moreover,
$g$ is nonseparable and can depend on $A_{i}$ nonlinearly, as implied
by optimization, so demeaning or first-differencing cannot eliminate
$A_{i}$. I instead use an index exclusion condition to handle the
fixed effects.

\section{Identification\label{sec:Identification}}

\subsection{Identification Idea}

\subsubsection*{Why Standard Linear Identification Fails}

The main obstacle to identifying the APE and the LAR function in the
TERC model using standard linear identification is that the regressor
is chosen by the agent using information that is also informative
about the random coefficients. To see where the standard argument
breaks down, take conditional expectations in (\ref{eq:Yeq}) given
$X_{it}$:
\begin{equation}
\E[\rest{Y_{it}}X_{it}]=X_{it}'\E[\rest{\beta_{it}}X_{it}].\label{eq:usual id lr}
\end{equation}
In a constant-coefficient model, the term $\E[\rest{\beta_{it}}X_{it}]$
would be a constant vector and could be recovered from moment conditions.
Here, however, $\E[\rest{\beta_{it}}X_{it}]$ is generally a function
of $X_{it}$. As a result, the familiar argument based on multiplying
both sides of (\ref{eq:usual id lr}) by $X_{it}$, taking expectations,
and inverting $\E[X_{it}X_{it}']$ does not identify $\E[\rest{\beta_{it}}X_{it}]$:
\begin{equation}
\E[X_{it}Y_{it}]=\E\left[X_{it}X_{it}'\E[\rest{\beta_{it}}X_{it}]\right],\label{eq:id fail 1}
\end{equation}
because $X_{it}X_{it}'$ and $\E[\beta_{it}\mid X_{it}]$ move together
inside the expectation. Likewise, the ``derivative/perturbation''
argument for linear models does not isolate $\E[\rest{\beta_{it}}X_{it}]$,
since differentiating (\ref{eq:usual id lr}) yields
\begin{equation}
\frac{\partial\E[\rest{Y_{it}}X_{it}]}{\partial X_{it}}=\E[\rest{\beta_{it}}X_{it}]+\left(\frac{\partial\E[\rest{\beta_{it}}X_{it}]}{\partial X_{it}}\right)'X_{it},\label{eq:id fail 2}
\end{equation}
and the second term captures how selection into different $X_{it}$
changes the conditional distribution of $\beta_{it}$.

\subsubsection*{Motivation and Interpretation of the Two Control Variables}

The identification strategy in this paper is to recreate, using observables,
the thought experiment ``vary $X_{it}$ while holding fixed everything
the agent knows that is related to $\beta_{it}$.'' Doing so requires
two controls---one for the time-invariant heterogeneity $A_{i}$
and one for the time-varying information $\eta_{it}$.

The first control is a statistic $W_{i}=W({\bf X}_{i},{\bf Z}_{i})$
constructed from the individual\textquoteright s panel history. The
maintained index exclusion condition implies that once I condition
on $W_{i}$, the remaining variation in $(X_{it},Z_{it})$ for a fixed
$t$ contains no additional information about $A_{i}$:
\begin{equation}
\rest{A_{i}}(X_{it},Z_{it},W_{i})\sim_{d}\rest{A_{i}}W_{i}.\label{eq:index-res}
\end{equation}
Intuitively, $W_{i}$ is chosen to summarize the persistent, individual-specific
component of behavior that is linked to $A_{i}$. For example, in
many panel applications a time average of $X_{it}$ (possibly together
with time averages of instruments) is a natural candidate: firms with
higher managerial ability or individuals with higher baseline productivity
tend to exhibit persistently different choices, and $W_{i}$ is designed
to capture that persistent component.

Even after accounting for $A_{i}$, endogeneity remains because agents
may respond to time-varying information $\eta_{it}$ when choosing
$X_{it}$. The second control is constructed as the conditional CDF
(a ``rank'' or \textquotedblleft generalized residual\textquotedblright ):
\begin{equation}
V_{it}:=F_{\rest{X_{it}}Z_{it},W_{i}}\left(\rest{X_{it}}Z_{it},W_{i}\right).\label{eq:def-V}
\end{equation}
The interpretation is that $V_{it}$ records where the realized $X_{it}$
sits in the distribution of $X_{it}$ among agents with the same $(Z_{it},W_{i})$.
Since $A_{i}$ and $Z_{it}$ appear in the first-step equation (\ref{eq:input eq})
and $W_{i}$ controls for $A_{i}$ by (\ref{eq:index-res}), the residual
variation of $X_{it}$ given $(Z_{it},W_{i})$ is only driven by $\eta_{it}$
which is assumed to enter (\ref{eq:input eq}) strictly monotonically.
For example, consider two firms with the same observed average inputs
over all periods ($W_{i}$) and input prices in period $t$ ($Z_{it}$).
If firm 1's choice of capital ($X_{it}$) in this period is higher
than that of firm 2, it can be inferred that firm 1 receives a larger
productivity shock ($\eta_{it}$) to its capital choice function in
period $t$ than firm 2 since (\ref{eq:index-res}) ensures that the
residual variation in $A_{i}$ given $W_{i}$ is not informative about
the choice of $X_{it}$.

As a result, $V_{it}$ records where the realized $\eta_{it}$ sits
in the distribution of $\eta_{it}$ among agents with the same $(Z_{it},A_{i},W_{i})$.
By an exogeneity assumption between $Z_{it}$ and $\eta_{it}$, it
drops out from the conditioning set and hence $V_{it}$ controls for
$\eta_{it}$ given $(A_{i},W_{i})$. Thus, conditioning on $(V_{it},W_{i})$
is intended to hold fixed both the individual\textquoteright s persistent
type (through $W_{i}$) and the individual\textquoteright s period-$t$
information that drives endogenous adjustment (through $V_{it}$).

This two-control structure matches the economics of the motivating
examples in Section \ref{sec:Model}. In the production function estimation
example, $W_{i}$ captures persistent firm heterogeneity (e.g., management
skill) that affects both average input choices and average elasticities,
while $V_{it}$ captures time-$t$ private information or expectations
about technology that shift input choices and are correlated with
realized productivity shocks. In the labor supply application, $W_{i}$
summarizes persistent ability/tastes and $V_{it}$ captures time-varying
job-match information that affects wage choice. In the AIDS application,
$W_{i}$ captures persistent household heterogeneity and $V_{it}$
captures time-varying wealth or liquidity information that affects
expenditure plans.

\subsubsection*{Steps to Identify APE and LAR}

Given the two controls $V_{it}$ and $W_{i}$, the central step to
identification of the APE and LAR is to show that once I condition
on the two controls, $X_{it}$ no longer carries additional information
about $\beta_{it}$. The argument proceeds in two layers.

The first step tackles the \textit{time-varying} part of the endogeneity
by assuming $A_{i}$ is known and showing 
\begin{equation}
\E[\rest{\beta_{it}}X_{it},A_{i},V_{it},W_{i}]=\E[\rest{\beta_{it}}A_{i},V_{it},W_{i}].\label{eq:tv-endo}
\end{equation}
Specifically, since $V_{it}$ records where the realized $\eta_{it}$
sits in the distribution of $\eta_{it}$ among agents with the same
$(A_{i},W_{i})$, holding $(A_{i},V_{it},W_{i})$ fixed amounts to
holding $(A_{i},\eta_{it})$ fixed. In this step, Assumption \ref{assu:mono}
(Componentwise Monotonicity) is used to invert $\eta_{it}$ out from
function $g$, Assumption \ref{assu:index exclusion} (Index Exclusion)
allows one to add $A_{i}$ into the conditioning set of the CDF $V_{it}$,
and Assumption \ref{assu:control for eta}\ref{enu:assu 3 Z indep eta ve}
(Exogeneity of IV) enables one to exclude $Z_{it}$ from the conditioning
set of the CDF $V_{it}$. Assumption \ref{assu:control for eta}\ref{enu:assu 3 control eta CDF mono}
(Monotonicity of the Conditional CDF) then implies that $V_{it}$
is a one-to-one transformation of $\eta_{it}$ given $(A_{i},W_{i})$,
proving the claim that $(A_{i},V_{it},W_{i})$ uniquely pins down
$(A_{i},\eta_{it},W_{i})$. Hence, equation (\ref{eq:input eq}) implies
that the remaining variation in $X_{it}$ is generated only by $Z_{it}$.
Under the maintained IV exogeneity condition, this residual variation
is independent of $\beta_{it}$. This is the point at which the time-varying
part of the endogeneity problem is resolved: within cells defined
by $(A_{i},V_{it},W_{i})$, movements in $X_{it}$ are \textquotedblleft as
good as randomized\textquotedblright{} by $Z_{it}$. In the production
function estimation example, if two firms have the same management
ability ($A_{i}$) and expected per-period technology shock ($\eta_{it}$)
but selects different capital or labor, such difference can only be
explained by the variation in exogenous $Z_{it}$, which is causal
and thus can be exploited to identify the APE and LAR. 

The second step tackles the \textit{time-invariant} part of the endogeneity.
Since $A_{i}$ is not observed, it is not feasible to condition on
it in (\ref{eq:tv-endo}). I deal with unknown $A_{i}$ by integrating
it out. This involves iterating expectations and the index exclusion
restriction of $W_{i}$. Specifically, I show 
\begin{equation}
\E[\rest{\E[\rest{\beta_{it}}A_{i},V_{it},W_{i}]}X_{it},V_{it},W_{i}]=\E[\rest{\beta_{it}}V_{it},W_{i}].\label{eq:invariant-endo}
\end{equation}
Notice that on the left-hand side of (\ref{eq:invariant-endo}), $X_{it}$
affects the conditional expectation only through the conditional density
$f_{\rest{A_{i}}X_{it},V_{it},W_{i}}$. The index exclusion restriction
(Assumption \ref{assu:index exclusion}) implies that, conditional
on $W_{i}$, $A_{i}$ is independent of $(X_{it},Z_{it})$ and thus
of $V_{it}$. For instance, in the production function estimation
example, $W_{i}$ captures the firm's time-invariant management ability
$A_{i}$, so once $W_{i}$ is held fixed, period-$t$ inputs $X_{it}$
and control for productivity shock $V_{it}$ only reflect transitory
variation and add no extra information about $A_{i}$. Therefore,
I have 
\begin{equation}
f_{\rest{A_{i}}X_{it},V_{it},W_{i}}(\rest{A_{i}}X_{it},V_{it},W_{i})=f_{\rest{A_{i}}W_{i}}(\rest{A_{i}}W_{i}),\label{eq:excl-X-V}
\end{equation}
which ensures that (\ref{eq:invariant-endo}) holds.

Combining (\ref{eq:tv-endo}) and (\ref{eq:invariant-endo}) gives
\begin{equation}
\E[\rest{\beta_{it}}X_{it},V_{it},W_{i}]=\E\left[\rest{\E[\rest{\beta_{it}}X_{it},A_{i},V_{it},W_{i}]}X_{it},V_{it},W_{i}\right]=\E[\rest{\beta_{it}}V_{it},W_{i}].\label{eq:E beta X V W}
\end{equation}
Equation (\ref{eq:E beta X V W}) is the key identification result:
after conditioning on $(V_{it},W_{i})$, the dependence of $\beta_{it}$
on $X_{it}$ is removed.

Given (\ref{eq:E beta X V W}), equation (\ref{eq:Yeq}) implies a
conditional linear representation
\begin{equation}
\E[\rest{Y_{it}}X_{it},V_{it},W_{i}]=X_{it}'\E[\rest{\beta_{it}}V_{it},W_{i}].\label{eq:id excl}
\end{equation}
With sufficient residual variation in $X_{it}$ conditional on $(V_{it},W_{i})$,
$\E[\rest{\beta_{it}}V_{it},W_{i}]$ can be identified either from
conditional second moments
\begin{equation}
\E[\rest{\beta_{it}}V_{it},W_{i}]=\E[\rest{X_{it}X_{it}'}V_{it},W_{i}]^{-1}\E[\rest{X_{it}Y_{it}}V_{it},W_{i}],\label{eq:id suc 1}
\end{equation}
or from the derivative of the conditional mean
\begin{equation}
\E[\rest{\beta_{it}}V_{it},W_{i}]=\frac{\partial\E[\rest{Y_{it}}X_{it},V_{it},W_{i}]}{\partial X_{it}}.\label{eq:id suc 2}
\end{equation}
Finally, identification of the APE and LAR follows by the law of iterated
expectations
\begin{equation}
\ol b=\dfrac{1}{T}\sum_{t=1}^{T}\E\left[\E[\rest{\beta_{it}}V_{it},W_{i}]\right],\quad\text{and}\quad b_{t}(X_{it})=\E\left[\rest{\E[\rest{\beta_{it}}V_{it},W_{i}]}X_{it}\right].\label{eq:id idea lie}
\end{equation}
In the production function estimation example, $\ol b$ summarizes
output elasticities by averaging across firms and time, so it plays
the same role as the constant coefficient in standard linear panel
regressions. $b_{t}(x)$ denotes the period-$t$ average elasticity
conditional on operating at input level $x$, a policy-relevant local
parameter when elasticities are systematically related to firms\textquoteright{}
input choices. Analogous interpretations apply to labor supply (average
and local wage elasticities) and demand estimation (average and local
expenditure/price elasticities). The purpose of $(W_{i},V_{it})$
is precisely to make these interpretations credible by ensuring that
the remaining variation used to identify (\ref{eq:id idea lie}) is
driven by the exogenous instruments rather than by unobserved, coefficient-relevant
information.

\subsection{Assumptions}

Next, I provide a list of assumptions on the model primitives required
for the identification analysis and discuss them in relation to model
(\ref{eq:Yeq})--(\ref{eq:main eqn 2 X}).
\begin{assumption}[\textbf{Componentwise Monotonicity}]
\label{assu:mono} For every $(Z_{it},A_{i})$ and each $j\in\{1,\ldots,d_{X}\}$,
$g_{j}\left(Z_{it},A_{i},\eta_{it}\right)$ depends on $\eta_{it}$
only through $\eta_{it,j}$ and is strictly monotone in $\eta_{it,j}$.
\end{assumption}
The role of Assumption \ref{assu:mono} is to deliver the invertibility/control-function
step used in the identification argument. This assumption is mild
when $\eta_{it}$ is scalar: it is automatically satisfied in additive
triangular specifications (e.g., \citet*{newey1999nonparametric}),
and in the nonseparable scalar case it is essentially the same as
the monotonicity condition in \citet[eq. (2)]{imbens2009identification},
except that I also allow for time-invariant heterogeneity $A_{i}$.
When $d_{X}>1$, as in production-function estimation (\citet*{olleypakes1996teltfp,levinsohnpetrin2003res,ackerberg2015identification}),
one can further weaken the requirement to monotonicity in only one
endogenous input, mirroring the identifying restriction in those classic
approaches.

Assumption \ref{assu:mono} requires justification when $\eta_{it}$
is vector-valued and each coordinate of $X_{it}$ admits one distinct
coordinate of $\eta_{it}$. In the production function estimation
example, this is plausible if firms make hiring and investment decisions
in different units (e.g., HR and finance), each responding primarily
to its own market-specific signal---labor-market conditions for hiring
and financial conditions for investment---satisfying the \textquotedblleft single-coordinate\textquotedblright{}
dependence. Moreover, positive shocks in each market naturally raise
the corresponding choice (more hiring when labor conditions improve,
more investment when financing conditions improve), implying that
$g_{j}$ is strictly monotone in $\eta_{it,j}$ for each coordinate
$j$ and thus satisfying componentwise monotonicity.

Importantly, Assumption \ref{assu:mono} places no restrictions on
the dependence structure across coordinates of $\eta_{it}$, nor on
the dependence between $A_{i}$ and $\eta_{it}$. Accordingly, without
loss of generality, one may normalize $\eta_{it}$ so that $d_{\eta}=d_{X}$
and each coordinate of $g$ depends on the corresponding coordinate
of $\eta_{it}$.
\begin{rem}
Assumption \ref{assu:mono} excludes settings where multiple components
of $\eta_{it}$ enter a single coordinate of $X_{it}$, as in simultaneous
supply--demand systems. Subsection \ref{subsec:Extension-2:Vector-Valued-U}
relaxes this restriction via timing assumptions, global univalence
conditions, or single-index structure.
\end{rem}
\begin{assumption}[\textbf{Index Exclusion}]
\label{assu:index exclusion}  Let $W_{i}\coloneqq W\left({\bf X}_{i},{\bf Z}_{i}\right)$,
where $W:\R^{T\times\left(d_{X}+d_{Z}\right)}\mapsto\R^{d_{W}}$ is
known. Suppose $\rest{A_{i}}\left(X_{it},Z_{it},W_{i}\right)\sim_{d}\rest{A_{i}}W_{i}$
for each $t$.
\end{assumption}
Assumption \ref{assu:index exclusion} serves two roles. First, it
makes $W_{i}$ a sufficient statistic for $A_{i}$, which is used
to establish (\ref{eq:E beta X V W}). Second, it is used to construct
$V_{it}$ and establish its connection with $\eta_{it}$ by holding
$A_{i}$ fixed. Under this assumption, the channel of endogeneity
operating through $A_{i}$ is handled entirely via $W_{i}$, while
the remaining ``time-varying endogeneity through the random coefficients''
is driven by the relationship between $\eta_{it}$ and $\ve_{it}$
and is addressed via $V_{it}$. Assumption \ref{assu:index exclusion}
is similar to Assumption 2.1 of \citet{altonji2005cross} and Assumption
A2 of \citet*{liu2024identification}. Considering the heterogeneity
and endogeneity afforded in the TERC model, some restriction that
isolates $A_{i}$ is typically needed.

To motivate $W_{i}$ in empirical settings, consider steady-state
input policies with convex adjustment costs in the production application.
In standard investment and hiring models with convex (e.g., quadratic)
adjustment costs (\citet{cooper2006nature}), optimality implies convergence
to a firm-specific long-run target $X_{i}^{\star}(A_{i})$ that is
increasing in ability $A_{i}$. Observed inputs fluctuate around this
target due to transitory shocks and gradual adjustment. Averaging
$X_{it}$ over time filters out these high-frequency deviations and
provides a consistent estimator of $X_{i}^{\star}(A_{i})$. Because
$X_{i}^{\star}(\cdot)$ is monotone, conditioning on $\ol X_{i}=T^{-1}\sum_{t=1}^{T}X_{it}$
induces a control for $A_{i}$, satisfying Assumption \ref{assu:index exclusion}.

Next, I provide three sufficient conditions for Assumption \ref{assu:index exclusion}. 

First, Assumption \ref{assu:index exclusion} can be justified by
nonparametrically generalizing equation (2.4) of \citet{mundlak1978pooling}
to be $A_{i}=h\left(\ol X_{i},\nu_{i}\right)$, where $\nu_{i}\perp\left(X_{it},Z_{it}\right)$
for all $t$. Here, the functional form of $h$ and distribution of
$\nu_{i}$ can be unknown. Then, Assumption \ref{assu:index exclusion}
is satisfied by taking $W_{i}$ to be $\ol X_{i}$. This is true because
conditioning on $W_{i}$, any $t$-specific $X_{it}$ and $Z_{it}$
does not affect the distribution of $A_{i}$ as the residual randomness
in $A_{i}$ is only driven by $\nu_{i}$. 

Second, insights from treatment assignment models in panel data (e.g.,
\citet{arkhangelsky2022role}) can be exploited to find candidate
$W_{i}$. For example, if $(X_{it},Z_{it})$ conditional on $A_{i}$
is a two-dimensional\textit{ }normal random vector i.i.d. through
time with known covariance and ${\bf Z}_{i}\perp A_{i}$, then the
density of $(X_{it},Z_{it})$ given $A_{i}$ depends on $A_{i}$ only
through $\ol X_{i}$. Hence, by the sufficient statistic for bi-variate
normal random vectors with known covariance matrix (\citet*[ch. 7]{hogg2019introduction}),
$W_{i}=(\ol X_{i},\ol Z_{i})$ suffices for Assumption \ref{assu:index exclusion}. 

Third, the nonparametric exchangeability condition (e.g., \citet{altonji2005cross})
can be adapted to justify Assumption \ref{assu:index exclusion}.
I present its details in Proposition \ref{prop:sufficient condition for assu 2}
of Appendix \ref{subsec:Prop-1}. The key idea is time exchangeability:
conditional on $A_{i}$, the joint behavior of $(X_{it},Z_{it})$
is unchanged if I relabel periods, so the time order contains no information
about $A_{i}$. Hence only order-invariant summaries of the panel
matter for learning about $A_{i}$, and symmetric functions of the
observed pairs $(X_{it},Z_{it})$ can be used as $W_{i}$ to satisfy
Assumption \ref{assu:index exclusion}.

$\text{\ }$

Next, I control for the time-varying $\eta_{it}$. When $A_{i}$ is
not present in (\ref{eq:main eqn 2 X}) and under Assumption \ref{assu:mono},
\citet{imbens2009identification} propose
\begin{equation}
V_{it}^{\text{IN09}}\coloneqq F_{\rest{X_{it}}Z_{it}}\left(\rest{X_{it}}Z_{it}\right)\overset{\text{by IN09}}{=}F_{\eta_{it}}\left(\eta_{it}\right)\label{eq:IN09-no-A}
\end{equation}
as a control for $\eta_{it}$. The intuition is that, holding input
prices $Z_{it}$ fixed, a firm choosing a higher capital $X_{it}$
must be experiencing a higher productivity shock $\eta_{it}$; otherwise
the higher input choice would not be optimal. However, this argument
breaks down in the TERC model because the unobserved, time-invariant
component $A_{i}$ also shifts input choices: firm 1\textquoteright s
higher $X_{it}$ could reflect greater managerial ability rather than
a larger transitory shock $\eta_{it}$. Therefore $A_{i}$ must also
be conditioned on, but this makes $F_{\rest{X_{it}}A_{i},Z_{it}}\left(\rest{X_{it}}A_{i},Z_{it}\right)$
infeasible since $A_{i}$ is unobserved. 

I proceed in two steps to generalize the argument of \citet{imbens2009identification},
adapting it to the TERC model. First, to make the control for $\eta_{it}$
feasible, I replace $A_{i}$ with the observable proxy $W_{i}$ and
define 
\begin{equation}
V_{it}=F_{\rest{X_{it}}Z_{it},W_{i}}\left(\rest{X_{it}}Z_{it},W_{i}\right).\label{eq:feas-V}
\end{equation}
Then, I show in (\ref{eq:cond_U}) that, under the next Assumption,
$V_{it}$ controls for $\eta_{it}$ conditional on $A_{i}$ and $W_{i}$.
Second, I further extend their analysis to allow coordinates of $g$
to depend on multiple elements of $\eta_{it}$ in Subsection \ref{subsec:Extension-2:Vector-Valued-U}.
\begin{assumption}[\textbf{Control for $\boldsymbol{\eta_{it}}$}]
\label{assu:control for eta}  Suppose the following conditions hold: 
\begin{enumerate}[label=\textup{(\alph*)}]
\item \label{enu:assu 3 Z indep eta ve}$\rest{Z_{it}\perp\left(\eta_{it},\ve_{it}\right)}\left(A_{i},W_{i}\right)$.
\item \label{enu:assu 3 control eta CDF mono}For every $(A_{i},W_{i})$
and each $l=1,...,d_{X}$, $F_{\rest{\eta_{it,l}}A_{i},W_{i}}(\rest{\eta_{it,l}}A_{i},W_{i})$
is strictly increasing in $\eta_{it,l}$.
\end{enumerate}
\end{assumption}
Assumption \ref{assu:control for eta} enables construction of a feasible
control variable $V_{it}$ for $\eta_{it}$ given $A_{i}$ and $W_{i}$.
To present it clearly, suppose $d_{X}=1$ and recall that $V_{it}=F_{\rest{X_{it}}Z_{it},W_{i}}\left(\rest{X_{it}}Z_{it},W_{i}\right)$.
Then, 
\begin{align}
V_{it} & =F_{\rest{X_{it}}Z_{it},A_{i},W_{i}}\left(\rest{X_{it}}Z_{it},A_{i},W_{i}\right)=F_{\rest{\eta_{it}}Z_{it},A_{i},W_{i}}\left(\rest{\eta_{it}}Z_{it},A_{i},W_{i}\right)\nonumber \\
 & =F_{\rest{\eta_{it}}A_{i},W_{i}}\left(\rest{\eta_{it}}A_{i},W_{i}\right),\label{eq:V map eta}
\end{align}
where the first equality holds by Assumption \ref{assu:index exclusion},
the second equality holds by Assumption \ref{assu:mono} and a change
of variable argument, and the last equality holds by Assumption \ref{assu:control for eta}\ref{enu:assu 3 Z indep eta ve}.
Thus, $V_{it}$ uniquely determines $\eta_{it}$ given $A_{i}$ and
$W_{i}$ by Assumption \ref{assu:control for eta}\ref{enu:assu 3 control eta CDF mono}.
Note that (\ref{eq:V map eta}) relies on Assumption \ref{assu:mono};
when multiple elements of $\eta_{it}$ appear in one coordinate of
$X_{it}$, the one-to-one relationship does not hold in general, and
additional structure or assumption is needed to recover (\ref{eq:V map eta}).
I discuss them in Subsection \ref{subsec:Extension-2:Vector-Valued-U}.

Assumption \ref{assu:control for eta}\ref{enu:assu 3 Z indep eta ve}
is similar to condition (i) of Theorem 1 in \citet{imbens2009identification},
except that I further condition on $(A_{i},W_{i})$. Since $W_{i}$
can be viewed as summarizing all the time-invariant information about
$A_{i}$ in the data, Assumption \ref{assu:control for eta}\ref{enu:assu 3 Z indep eta ve},
loosely speaking, requires $\rest{Z_{it}\perp\left(\ve_{it},\eta_{it}\right)}A_{i}$,
which is already implied by the standard exogeneity condition $Z_{it}\perp\left(A_{i},\ve_{it},\eta_{it}\right)$.
Note that $Z_{it}\perp\left(A_{i},\ve_{it},\eta_{it}\right)$ is stronger
than Assumption \ref{assu:control for eta}\ref{enu:assu 3 Z indep eta ve}
as the former rules out the possibility that $Z_{it}$ and $A_{i}$
are correlated.\footnote{For instance, Bartik instruments use national industry shocks, weighted
by pre-existing local industry shares. The intuition is the national
shocks are plausibly \textquotedblleft as-good-as-random\textquotedblright{}
conditional on the fixed effects and controls, so the conditional
exogeneity is plausible. But the baseline shares can reflect deep
local traits---so unconditional exogeneity is harder to defend.} Nonetheless, the unconditional exogeneity condition remains a useful
benchmark since it is widely used in applied work.

Assumption \ref{assu:control for eta}\ref{enu:assu 3 Z indep eta ve}
can be justified in many applications with the usual choice of IVs
from the literature. For example, in labor-supply applications, $Z_{it}$
can be county minimum-wage changes or EITC expansions---policy-driven
shocks that, conditional on time-invariant ability $A_{i}$ and its
sufficient statistic $W_{i}$, vary independently of expected sectoral
labor-demand shocks $\eta_{it}$ and labor-supply-elasticity shocks
$\varepsilon_{it}$. In gasoline-demand estimation, $Z_{it}$ can
be the head of household\textquoteright s earned income or gas-tax
changes, assumed (conditional on a household fixed effect $A_{i}$
and $W_{i}$) to be independent of anticipated near-term income changes
$\eta_{it}$ (e.g., an upcoming bonus) and transitory wealth shocks
$\varepsilon_{it}$.

For the empirical application of Section \ref{sec:Empirical-Illustration},
I construct the instruments toward BLP-style (\citet*{blp1995io})
cost shifters: leave-one-out, competitor-weighted input-price (wages
and interest rates) averages at the industry--city--year level.
These instruments are relevant because firms compete locally for labor
and capital, so competitors' input prices shift firm $i$'s cost environment
and input choices. They are plausibly exogenous to firm $i$\textquoteright s
idiosyncratic productivity shocks conditional on the fixed effect
and its control variable because firm $i$\textquoteright s own input
prices are excluded from the construction, so identification comes
from competitors' input-price variation rather than firm $i$'s unobserved
shocks. Robustness checks in Appendix \ref{sec:Additional-Empirical-Results}
provide evidence that these IVs are plausibly valid and satisfy Assumption
\ref{assu:control for eta}\ref{enu:assu 3 Z indep eta ve}.

Assumption \ref{assu:control for eta}\ref{enu:assu 3 control eta CDF mono}
is analogous to condition (ii) of Theorem 1 in \citet{imbens2009identification}.
It is mild because it concerns a conditional CDF, which is typically
strictly increasing for continuous random variables. 

The last assumption concerns the residual variation in $X_{it}$ given
$V_{it}$ and $W_{i}$, which is used to identify the APE and LAR
by (\ref{eq:id suc 1}).
\begin{assumption}[\textbf{Residual Variation in $\boldsymbol{X_{it}}$}]
\label{assu:residual variation in X} There are at least $d_{X}$
linearly independent points in the support of $X_{it}$ conditional
on $(V_{it},W_{i})$ almost surely.
\end{assumption}
Assumption \ref{assu:residual variation in X} ensures that $\E\left[\rest{X_{it}X_{it}'}V_{it},W_{i}\right]$
in (\ref{eq:id suc 1}) is invertible. Thus, the APE and LAR can be
identified by (\ref{eq:id suc 1}) and the law of iterated expectations. 

Assumption \ref{assu:residual variation in X} is best interpreted
as a conditional support (non-degeneracy) requirement: for each $t$,
$X_{it}$ must retain sufficient variation conditional on $(V_{it},W_{i})$
to identify the target objects. Requiring residual variation in $X_{it}$
conditional on $V_{it}$ is generally not restrictive (\citet{imbens2009identification}).
Essentially, it requires $Z_{it}$ to induce enough variation in $X_{it}$
so that, for each $v\in\text{Supp}(V_{it})$, the set $\left\{ X_{it}:V_{it}=v\right\} $
is not a singleton. Therefore, it corresponds to the classical IV
relevance condition of $\p g(z,a,\eta)/\p z\neq0$.

The restriction on the support of $X_{it}$ in Assumption \ref{assu:residual variation in X}
mainly comes from conditioning on $W_{i}$, which also concerns Assumption
\ref{assu:index exclusion}. There is an inherent trade-off between
Assumptions \ref{assu:index exclusion} and \ref{assu:residual variation in X}.
To see this, suppose $W_{i}=(\mathbf{X}_{i},\mathbf{Z}_{i})$ which
satisfies Assumption \ref{assu:index exclusion} trivially. However,
because $X_{it}$ is included in $W_{i}$, it makes the support of
$X_{it}$ given $W_{i}$ a singleton. Hence, Assumption \ref{assu:residual variation in X}
does not hold unless $X_{it}$ is a scalar and the point in the support
of $X_{it}$ given $W_{i}$ is nonzero. More generally, including
more elements in $W_{i}$ tends to make Assumption \ref{assu:index exclusion}
easier to satisfy while making Assumption \ref{assu:residual variation in X}
harder to maintain. Because of this trade-off, I describe next how
the conditions used to justify Assumption \ref{assu:index exclusion}
affect the plausibility of Assumption \ref{assu:residual variation in X},
and give corresponding sufficient conditions.

When justifications for Assumption \ref{assu:index exclusion} are
used so that $W_{i}$ only includes averages of $X_{it}$ and/or $Z_{it}$
through time (e.g., \citet*{mundlak1978pooling,arkhangelsky2022role,liu2024identification}),
Assumption \ref{assu:residual variation in X} is not restrictive.
For instance, with $d_{X}=2$, $T=2$, and $W_{i}=\ol X_{i}$, Assumption
\ref{assu:residual variation in X} holds whenever $X_{i1}$ and $X_{i2}$
are not perfectly linearly dependent, so that $\mathrm{Supp}(X_{it}\mid\ol X_{i})$
remains non-degenerate. By (\ref{eq:input eq}) and the relevance
condition above, it in turn requires that the serial dependence in
$Z_{it}$ cannot be so high that the value of $Z_{it}$ (hence $X_{it}$)
for any $t$ is uniquely determined once $W_{i}$ is fixed.

By contrast, when a high-dimensional $W_{i}$ is used to justify Assumption
\ref{assu:index exclusion} (e.g., via the exchangeability condition
of \citet{altonji2005cross}), Assumption \ref{assu:residual variation in X}
becomes more restrictive. Intuitively, a high-dimensional $W_{i}$
imposes many constraints on the path $(X_{i1},\ldots,X_{iT})$, so
one typically needs longer panel to leave enough feasible configurations
and satisfy Assumption \ref{assu:residual variation in X}. 

In this case, one possibility is to explore the symmetry through time
in the solution to $\left\{ {\bf X}_{i}:V_{it}=v,W_{i}=w\right\} $.
In particular, when $W_{i}$ includes averages through time of the
polynomials of $\left(X_{it},Z_{it}\right)$ up to the $T^{\text{th}}$
order, Assumption \ref{assu:index exclusion} is satisfied under a
nonparametric exchangeability condition on $f_{\rest{\eta_{i}}A_{i}}$
by adapting the proof of \citet{altonji2005cross} in Proposition
\ref{prop:sufficient condition for assu 2}. Then, since $W_{i}$
is symmetric in $\left(X_{it},Z_{it}\right)$ through time, one may
permute the order of $\left(X_{it},Z_{it}\right)$ in time without
changing the value of $W_{i}$. This creates enough linearly independent
points in the support of $X_{it}$ given $W_{i}$ when $T\geq d_{X}$.
When such permutation also does not change the value of $V_{it}$
which is usually satisfied under the relevance condition, Assumption
\ref{assu:residual variation in X} is satisfied. I provide an example
in Appendix \ref{subsec:As245-Discuss}.

Assumption \ref{assu:residual variation in X} pertains to the inversion-based
argument in (\ref{eq:id suc 1}). To facilitate the derivative-based
argument in (\ref{eq:id suc 2}), I introduce the next assumption,
which requires more residual variation of $X_{it}$ conditional on
$V_{it}$ and $W_{i}$ than Assumption \ref{assu:residual variation in X}.

\addtocounter{assumption}{-1}
\begingroup
\let\oldtheassumption\theassumption
\renewcommand{\theassumption}{\oldtheassumption\ensuremath{'}}
\let\oldtheHassumption\theHassumption
\renewcommand{\theHassumption}{\oldtheHassumption-p}
\begin{assumption}[\textbf{More Variation in $\boldsymbol{X_{it}}$}]
\label{assu:more_var_X}The support of $X_{it}$ conditional on $(V_{it},W_{i})$
contains an open ball of positive radius almost surely.
\end{assumption}
\endgroup

Assumption \ref{assu:more_var_X} allows one to perturb $X_{it}$
in $\E[\rest{Y_{it}}X_{it},V_{it},W_{i}]$ when $(V_{it},W_{i})$
is fixed so that conditional moments of $\b_{it}$ can be recovered
by (\ref{eq:id suc 2}). It is similar to Assumption 2.2 of \citet{altonji2005cross}. 

The same trade-off between Assumptions \ref{assu:index exclusion}
and \ref{assu:residual variation in X} also applies to Assumption
\ref{assu:more_var_X}. When Assumption \ref{assu:index exclusion}
is justified such that $W_{i}$ is low-dimensional (e.g., $\ol X_{i}$
as in \citet{mundlak1978pooling}), Assumption \ref{assu:more_var_X}
is generally not restrictive. In essence, $Z_{it}$ must vary enough
to generate sufficient residual variation in $X_{it}$ conditional
on $(V_{it},W_{i})$.

However, when Assumption \ref{assu:index exclusion} is justified
using nonparametric arguments (e.g., the exchangeability condition
of \citet{altonji2005cross}), $d_{W}$ is large and Assumption \ref{assu:more_var_X}
can be restrictive. To address this issue, I present two sets of solutions
in Appendix \ref{subsec:As245-Discuss}. First, the \textit{unconditional}
variation of exogenous regressors outside of $W_{i}$ can be leveraged.
Second, following the suggestions of \citet{altonji2005cross}, one
may restrict how $W_{i}$ enters $f_{\rest{A_{i}}W_{i}}$ so that
the relevant conditional expectations only concern a subset of $W_{i}$
or a linear index of its components. 

Assumption \ref{assu:more_var_X} allows the straightforward perturbation-based
method (\ref{eq:id suc 2}) to identify the APE and LAR without requiring
the computation of the inverse of $\E\left[\rest{X_{it}X_{it}'}V_{it},W_{i}\right]$.
When residual variation is not a concern---such as when $W_{i}$
contains only the mean of $X_{it}$ through time or when exogenous
regressors are included in (\ref{eq:Yeq})---Assumption \ref{assu:more_var_X}
is preferred for simpler analysis. I compare these two approaches
in Remark \ref{rem:inversion-vs-derivative} below.

$\text{\ }$

The next theorem summarizes the main identification result of this
paper.
\begin{thm}[\textbf{Identification}]
\label{thm:main id}If Assumptions \ref{assu:mono}--\ref{assu:control for eta}
and either Assumption \ref{assu:residual variation in X} or \ref{assu:more_var_X}
are satisfied, then the APE $\ol b=T^{-1}\sum_{t=1}^{T}\E\b_{it}$
and the LAR function $b_{t}(x)=\E\left[\rest{\b_{it}}X_{it}=x\right]$
are both identified.
\end{thm}

\subsection{Extensions}

\subsubsection*{Vector-Valued $\boldsymbol{\eta_{it}}$\label{subsec:Extension-2:Vector-Valued-U}}

I present four methods for incorporating vector-valued $\eta_{it}$
into $X_{it}$ under additional assumptions. To illustrate the idea
clearly, I assume $d_{X}=d_{\eta}=2$.

First, timing assumptions on the choice of regressors can be exploited.
For instance, in production function estimation, choices of later
inputs such as capital typically do not depend on random shocks to
earlier chosen inputs such as labor after conditioning on the level
of the earlier chosen inputs; see \citet*[(Appendix 1 of 2006 Working Paper)]{ackerberg2015identification}
for a similar idea. Under this timing assumption, I modify (\ref{eq:Yeq})--(\ref{eq:main eqn 2 X})
to be:
\begin{align}
Y_{it} & =X_{it}'\b\left(A_{i},\ve_{it}\right),\nonumber \\
X_{it,1} & =g_{1}\left(Z_{it},A_{i},\eta_{it,1}\right),\text{ and }X_{it,2}=g_{2}\left({\color{red}X_{it,1}},Z_{it},A_{i},\eta_{it,2}\right).
\end{align}
Note that now $X_{it,2}$ depends on the whole vector of $\eta_{it}$
if $X_{it,1}=g_{1}\left(Z_{it},A_{i},\eta_{it,1}\right)$ is substituted
into $g_{2}$. I first use $V_{it,2}\coloneqq F_{\rest{X_{it,2}}X_{it,1},Z_{it},W_{i}}\left(\rest{X_{it,2}}X_{it,1},Z_{it},W_{i}\right)$
to control for $\eta_{it,2}$ first. Then, I use $F_{\rest{X_{it,1}}Z_{it},W_{i}}\left(\rest{X_{it,1}}Z_{it},W_{i}\right)$
to control for $\eta_{it,1}$.

Second, interaction across different coordinates of $X_{it}$ can
provide useful information about $\eta_{it}$. One possibility is
to assume that, although the full vector $\eta_{it}$ enters each
coordinate of $X_{it}$, the ratio $X_{it,1}/X_{it,2}$ depends only
on a single component of $\eta_{it}$. For example, in production
function estimation, $\eta_{it,1}$ represents the Hicks neutral productivity
and $\eta_{it,2}$ denotes the labor-augmenting technology. Although
both capital and labor choices depend on vector $\eta_{it}$, it is
plausible that \textit{capital per worker }depends only on the labor-augmenting
technology $\eta_{it,2}$. See Proposition 2.1 of \citet{demirer2020production}
for more discussions about this assumption. Given this assumption,
I can proceed to control for $\eta_{it,2}$ first by 
\begin{equation}
V_{it,2}\coloneqq F_{\rest{X_{it,1}/X_{it,2}}Z_{it},W_{i}}\left(\rest{X_{it,1}/X_{it,2}}Z_{it},W_{i}\right),
\end{equation}
and then for $\eta_{it,1}$ by 
\begin{equation}
V_{it,1}\coloneqq F_{\rest{X_{it,1}}Z_{it},V_{it,2},W_{i}}\left(\rest{X_{it,1}}Z_{it},V_{it,2},W_{i}\right).
\end{equation}

Third, general ``global univalence'' results---such as \citet{gale1965jacobian}
or \citet{palais1959natural,hadamard1906b,hadamard1906a}---suffice
to establish the invertibility required here. In production function
estimation, the inverse isotonicity approach in \citet*{berry2013connected}
appears promising to obtain global injectivity of $g$ in $\eta_{it}$.
In particular, suppose each input choice increases in its own component
of $\eta_{it}$, while cross-partials are weakly negative due to substitutability.
This happens when, for example, inputs are substitutes in the firm\textquoteright s
optimal input demand; a favorable shock to one input in the form of
higher efficiency or lower effective cost leads the firm to substitute
toward that input and away from others. Then, the Jacobian matrix
of $g$ with respect to $\eta_{it}$, denoted by $J_{g}(\eta_{it})$,
has positive diagonal entries and nonpositive off-diagonal entries,
i.e., the $Z$-matrix sign pattern required for an $M$-matrix. If,
in addition, the own effects are strong enough that the matrix is
strictly diagonally dominant, then $J_{g}(\eta_{it})$ is an $M$-matrix.
This, in turn, guarantees a unique monotone inverse mapping from observables
back to the vector $\eta_{it}$.

Finally, if $\eta_{it}$ enters the structural function $g$ through
a scalar index $\eta_{it}^{\prime}\tau$ and $g$ is monotone in that
index (\citet{ichimura1991semiparametric}), then the effective unobservable
is scalar, satisfying Assumption \ref{assu:mono}. This aligns with
common practice in asset pricing, where a scalar market factor emerges
as an index of aggregate shocks and affects stock returns monotonically.
A similar structure arises in production function estimation, where
scalar productivity summarizes richer underlying states and affects
input choices strictly increasingly.

\subsubsection*{Higher-Order Moments of $\boldsymbol{\protect\b_{it}}$\label{subsec:Extension-1:-Higher-order}}

I identify the $P^{\text{th}}$-order moments of $\b_{it}$. For clarity
of illustration, suppose the regressor is $X_{it}=(K_{it},L_{it})'$
and $\b_{it}=(\b_{it,K},\b_{it,L})'$. The extension to $d_{X}>2$
is straightforward. Since the residual variation in $X_{it}$ given
$(V_{it},W_{i})$ is driven only by exogenous $Z_{it}$, (\ref{eq:E beta X V W})
holds for any measurable function $h$ of $\b_{it}$:
\begin{equation}
\mathbb{E}\left[h(\b_{it})\mid X_{it},V_{it},W_{i}\right]=\mathbb{E}\left[h(\b_{it})\mid V_{it},W_{i}\right].\label{eq:lth moments}
\end{equation}
In particular, let $h=(h_{1},\ldots,h_{P})'$ and $h_{q}(\b_{it})=\beta_{it,K}^{q}\b_{it,L}^{P-q}$
for $q=0,\ldots,P$. By (\ref{eq:output eq}), the conditional $P^{\text{th}}$-order
moment of $Y_{it}$ given $(X_{it},V_{it},W_{i})$ is
\begin{equation}
\mathbb{E}\left[Y_{it}^{P}\mid X_{it},V_{it},W_{i}\right]=\sum_{q=0}^{P}\binom{P}{q}K_{it}^{q}L_{it}^{P-q}\mathbb{E}\left[\beta_{it,K}^{q}\b_{it,L}^{P-q}\mid V_{it},W_{i}\right].\label{eq:Kth moments}
\end{equation}
Under $P$-times differentiability of $\mathbb{E}\left[Y_{it}^{P}\mid X_{it},V_{it},W_{i}\right]$
in $X_{it}$, the coefficients on $K_{it}^{P}$ and $L_{it}^{P}$
are identified via differentiation, yielding
\begin{align}
\mathbb{E}\left[\beta_{it,K}^{q}\beta_{it,L}^{P-q}\mid V_{it},W_{i}\right] & =\dfrac{1}{P!}\frac{\partial^{P}}{\partial k^{q}\p l^{P-q}}\mathbb{E}\left[Y_{it}^{P}\mid K_{it}=k,L_{it}=l,V_{it},W_{i}\right],\text{ for }q=0,\ldots,P,
\end{align}
and hence the $P^{\text{th}}$-order moments of $\b_{it}$ follow
by the law of iterated expectations
\begin{align}
\mathbb{E}[\beta_{it,K}^{q}\beta_{it,L}^{P-q}] & =\mathbb{E}\left[\dfrac{1}{P!}\frac{\partial^{P}}{\partial k^{q}\p l^{P-q}}\mathbb{E}\left[Y_{it}^{P}\mid K_{it}=k,L_{it}=l,V_{it},W_{i}\right]\right],\text{ for }q=0,\ldots,P.
\end{align}
Repeating this argument for all $P\in\N$ identifies the entire moment
sequence of $\b_{it}$. Under standard moment-determinacy conditions
(e.g., \citet{stoyanov2000krein}), this sequence uniquely determines
the distribution of $\beta_{it}$.

\subsubsection*{Exogenous Shocks and Covariates}

The identification argument goes through when I include ex-post shocks
$\upsilon_{it}$ to $\b_{it}$ (i.e., $\b_{it}=\b\left(A_{i},\ve_{it},\upsilon_{it}\right)$)
and $\widetilde{\epsilon}_{it}$ to $Y_{it}$ (i.e., $Y_{it}=X_{it}'\b_{it}+\widetilde{\epsilon}_{it}$
with $\E\widetilde{\epsilon}_{it}=0$), where $\upsilon_{it}$ and
$\widetilde{\epsilon}_{it}$ are independent of all other variables.
Note that $\b(\cd)$ may a priori be time-varying and the joint distribution
of $(\ve_{it},\eta_{it})$ may also vary with $t$. Introducing ex-post
shocks $\upsilon_{it}$ provides an additional and economically distinct
source of time variation (e.g., unexpected technology shock). While
the identification result remains valid with the inclusion of $\upsilon_{it}$,
I present the required changes to the proof at the end of the proof
of Theorem \ref{thm:main id}. I also include $\upsilon_{it}$ into
$\b_{it}$ when estimating the APE and LAR and examine its impact
via simulations. As for $\widetilde{\epsilon}_{it}$, since it is
independent of all other variables and enters (\ref{eq:Yeq}) in an
additive way, equations (\ref{eq:E beta X V W})--(\ref{eq:id idea lie})
hold as before. Thus, the identification analysis is unaffected.

The presence of exogenous covariates in $X_{it}$ can strengthen identification.
The key point is that Assumptions \ref{assu:mono}--\ref{assu:control for eta}
pertain only to the endogenous covariates of $X_{it}$. In particular,
the control variables $V_{it}$ and $W_{i}$ depend only on the endogenous
covariates of $X_{it}$, which alleviates concerns that the controls
may be high-dimensional and makes the required residual variation
in $X_{it}$ easier to satisfy. Moreover, as discussed in Appendix
\ref{subsec:As245-Discuss}, unconditional variation in exogenous
covariates of $X_{it}$ can be directly exploited to make Assumptions
\ref{assu:residual variation in X} and \ref{assu:more_var_X} easier
to satisfy. Details on how to adapt the analysis to incorporate exogenous
covariates are provided at the end of the proof of Theorem \ref{thm:main id}.

\section{\label{sec:Est and Inf}Estimation and Large Sample Theory}

In this section, I first describe how to estimate the parameters using
three-step series estimators. I then establish the convergence rates
of the proposed estimators, and finally show that they are asymptotically
normal.

\subsection{\label{subsec:Estimation}Estimation}

The parameters of interest are 
\begin{equation}
\ol b=T^{-1}\sum_{t=1}^{T}\E\b_{it}\quad\text{and}\quad b_{t}\left(x\right)\coloneqq\E\left[\rest{\b_{it}}X_{it}=x\right].\label{eq:param of interest}
\end{equation}
$\ol b$ is the pooled APE across all firms and periods. $b_{t}\left(x\right)$
is the LAR function for a subpopulation characterized by $X_{it}=x$
in period $t$ and is also useful for answering policy-related questions.
For example, plugging realizations $x_{it}$ of $X_{it}$ into $b_{t}\left(x\right)$
provides a fine approximation to $\b_{it}$.

I propose to estimate the parameters in (\ref{eq:param of interest})
with three-step series estimators. For clarity of exposition, I set
$d_{X}=1$ and highlight the modifications required for $d_{X}>1$
when needed. To fix ideas, I follow \citet*{liu2024identification}
to let $W_{i}=T^{-1}\sum_{t=1}^{T}X_{it}$, which can be motivated
by generalizing the method of \citet{mundlak1978pooling}.

First, for each $t$, I estimate $V_{t}\left(x,z,w\right)\coloneqq F_{\rest{X_{it}}Z_{it},W_{i}}\left(\rest xz,w\right)$
by regressing $\ind\left\{ X_{it}\leq x\right\} $ on the basis functions
$q^{M_{1}}$ of $\left(Z_{it},W_{i}\right)$ with trimming function
$\tau$: 
\begin{align}
\widehat{V}_{t}\left(x,z,w\right) & =\tau\left(\widehat{F}_{\rest{X_{it}}Z_{it},W_{i}}\left(\rest xz,w\right)\right)\nonumber \\
 & =\tau\left(q^{M_{1}}\left(z,w\right)'n^{-1}\widehat{Q}_{t}^{-1}\sum_{j=1}^{n}q_{jt}\ind\left\{ x_{jt}\leq x\right\} \right)\nonumber \\
 & \eqqcolon\tau\left(q^{M_{1}}\left(z,w\right)'\widehat{\gamma}_{t}^{M_{1}}\left(x\right)\right),\label{eq:V hat}
\end{align}
where $\widehat{Q}_{t}:=n^{-1}\sum_{i=1}^{n}q_{it}q_{it}'$ and $q_{it}:=q^{M_{1}}\left(z_{it},w_{i}\right)$.
Examples of $q^{M_{1}}$ include power series and spline functions.
When $d_{X}>1$, I regress $\ind\left\{ X_{it,l}\leq x_{l}\right\} $
on the basis functions $q^{M_{1}}$ of $\left(Z_{it},W_{i}\right)$
with trimming function $\tau$ for each $l$ and obtain $\widehat{V}_{t}\left(x,z,w\right)=\left(\widehat{V}_{t,1}\left(x,z,w\right),\ldots,\widehat{V}_{t,d_{X}}\left(x,z,w\right)\right)'$.
I highlight two properties of $\widehat{V}_{t}\left(x,z,w\right)$.
First, the regression coefficient $\widehat{\gamma}_{t}^{M_{1}}\left(x\right)$
in \eqref{eq:V hat} depends on $x$ because the dependent variable
$\ind\left\{ X_{it}\leq x\right\} $ is a function of $x$. This fact
causes the convergence rate of $\widehat{V}_{t}\left(x,z,w\right)$
to be slower than the standard rates for series estimators (\citet{imbens2009identification}).
Second, a trimming function $\tau$ is needed since I am estimating
a conditional CDF. An example of $\tau$ is $\tau\left(x\right)=\ind\left\{ x>0\right\} \cd\min\left(x,1\right)$. 

Next, define $S\coloneqq\left(X,V,W\right)$ and let $\mathcal{X},\ \mathcal{Z},\ \mathcal{V},\ \mathcal{W},$
and $\mathcal{S}$ denote the supports of $X,\ Z,\ V,\ W,$ and $S$,
respectively. Let $V_{it}\coloneqq V_{t}\left(X_{it},Z_{it},W_{i}\right),$$\widehat{V}_{it}\coloneqq\widehat{V}_{t}\left(X_{it},Z_{it},W_{i}\right)$,
and $\widehat{v}_{it}\coloneqq\widehat{V}_{t}\left(x_{it},z_{it},w_{i}\right)$.
For any $s=\left(x,v,w\right)\in\mathcal{S}$, I estimate $G_{t}\left(s\right)\coloneqq\E\left[\rest{Y_{it}}S_{it}=s\right]$
by regressing $Y_{it}$ on the basis functions $p^{M_{2}}$ of $\left(X_{it},\widehat{V}_{it},W_{i}\right)$:
\begin{equation}
\widehat{G}_{t}\left(s\right)=p^{M_{2}}\left(s\right)'n^{-1}\widehat{P}_{t}^{-1}\widehat{p}_{t}'y_{t}\eqqcolon p^{M_{2}}\left(s\right)'\widehat{\alpha}_{t}^{M_{2}},
\end{equation}
where $\widehat{P}_{t}:=n^{-1}\sum_{i=1}^{n}\widehat{p}_{it}\widehat{p}_{it}'$,
$\widehat{p}_{it}:=p^{M_{2}}\left(x_{it},\widehat{v}_{it},w_{i}\right)$,
$\widehat{p}_{t}$ is an $n\times M_{2}$ matrix with $\widehat{p}_{it}'$
being its $i^{\text{th}}$ row, and $y_{t}$ is the vector of $y_{it}$s.
Following \citet*{newey1999nonparametric}, I let 
\begin{equation}
p^{M_{2}}\left(s\right)=x\otimes p^{m_{2}}\left(v,w\right)
\end{equation}
by exploiting the index structure of (\ref{eq:Yeq}). Hence, the effective
degree that matters for the convergence rate is $m_{2}$ since $M_{2}=d_{X}\times m_{2}$
and $d_{X}$ is finite.

Finally, I exploit the index structure of (\ref{eq:Yeq}) again to
estimate $b_{1t}\left(v,w\right)\coloneqq\E\left[\rest{\b_{it}}V_{it}=v,W_{i}=w\right]$.
When (\ref{eq:id suc 1}) is used, I estimate $b_{1t}\left(v,w\right)$
by
\begin{equation}
\widetilde{b}_{1t}\left(v,w\right)=\left(\widehat{\E}\left[\rest{X_{it}X_{it}'}\widehat{V}_{it}=v,W_{i}=w\right]\right)^{-1}\widehat{\E}\left[\rest{X_{it}Y_{it}}\widehat{V}_{it}=v,W_{i}=w\right],\label{eq:series_estr_sys_lr_eqn}
\end{equation}
where $\widehat{\E}\left[\rest{X_{it}Y_{it}}\widehat{V}_{it},W_{i}\right]$
is obtained by regressing each coordinate of $X_{it}Y_{it}$ on the
basis functions of $\left(\widehat{V}_{it},W_{i}\right)$ and similarly
for $\widehat{\E}\left[\rest{X_{it}X_{it}'}\widehat{V}_{it},W_{i}\right]$.
When (\ref{eq:id suc 2}) is used, I estimate $b_{1t}\left(v,w\right)$
by
\begin{equation}
\widehat{b}_{1t}\left(v,w\right)=\partial\widehat{G}_{t}\left(s\right)/\partial x=\left(I_{d_{X}}\otimes p^{m_{2}}\left(v,w\right)\right)'\widehat{\alpha}_{t}^{M_{2}}\eqqcolon\ol p^{M_{2}}\left(s\right)'\widehat{\alpha}_{t}^{M_{2}},\label{eq:derivate_estr}
\end{equation}
where the second equality holds by the chain rule. I follow (\ref{eq:derivate_estr})
in what follows, as it simplifies both implementation and the derivation
of asymptotic properties.
\begin{rem}[\textbf{Compare $\boldsymbol{\widetilde{b}_{1t}\left(v,w\right)}$
with $\boldsymbol{\widehat{b}_{1t}\left(v,w\right)}$}]
\label{rem:inversion-vs-derivative} While both approaches are valid,
this paper recommends using the derivative-based estimator $\widehat{b}_{1t}(v,w)$
rather than the inversion-based estimator $\widetilde{b}_{1t}(v,w)$
in practice when Assumption \ref{assu:more_var_X} holds. The derivative-based
route operates through the estimated one-dimensional regression function
$\widehat{G}_{t}(x,v,w)$ and obtains $\widehat{b}_{1t}(v,w)$ via
the analytic partial derivative with respect to $x$. This approach
exploits the linear random-coefficient structure directly, avoids
an additional ``matrix estimation + inversion'' step, and leads
to a cleaner and more stable asymptotic argument in the paper's framework.
By contrast, $\widetilde{b}_{1t}(v,w)$ requires estimating and inverting
the conditional second-moment matrix $\mathbb{E}[X_{it}X_{it}'\mid V_{it}=v,W_{i}=w]$,
whose sample analogue can be ill-conditioned in finite samples and
hence numerically unstable (\citet*{CarrascoFlorensRenault2007}).
This inversion also complicates the asymptotic analysis, because one
must control how estimation error in the conditional second moment
propagates through a matrix inverse.
\end{rem}
To estimate $\ol b$ and $b_{t}\left(x\right)$, by the law of iterated
expectations I regress $\widehat{b}_{1t}\left(\widehat{V}_{it},W_{i}\right)$
on the basis function $r^{M_{3}}$ of constant one and $X_{it}$,
respectively:
\begin{equation}
\widehat{\ol b}=\dfrac{1}{nT}\sum_{t=1}^{T}\sum_{i=1}^{n}\widehat{b}_{1t}\left(\widehat{v}_{it},w_{i}\right)\quad\text{and}\quad\widehat{b}_{t}\left(x\right)=r^{M_{3}}\left(x\right)'n^{-1}\widehat{R}_{t}^{-1}r_{t}'\widehat{B}_{t}\eqqcolon r^{M_{3}}\left(x\right)'\widehat{\rho}_{t}^{M_{3}},\label{eq: estimator for beta bar}
\end{equation}
where $\widehat{R}_{t}:=n^{-1}\sum_{i=1}^{n}r_{it}r_{it}'$, $r_{it}:=r^{M_{3}}(x_{it})$,
$r_{t}$ is an $n\times M_{3}$ matrix with $r_{it}'$ being its $i^{\text{th}}$
row, and $\widehat{B}_{t}$ is an $n\times d_{X}$ matrix with $\widehat{b}_{1t}(\widehat{v}_{it},w_{i})'$
being its $i^{\text{th}}$ row.
\begin{rem}[\textbf{Computational Considerations}]
\label{rem:comp-issue}While the three-step estimators are based
on standard series regressions and are typically computationally manageable,
two implementation aspects are worth noting for large datasets. First,
when estimating $V_{t}\left(x,z,w\right)$, a key computational simplification
is that the regressors---the basis functions $q^{M_{1}}(Z_{it},W_{i})$---are
identical across all thresholds, with only the dependent variable
$\ind\{X_{it}\leq x\}$ changing in $x$. Hence, I can compute the
inverse Gram matrix $\widehat{Q}_{t}^{-1}$ once and reuse it to obtain
the coefficient vector $\widehat{\gamma}_{t}^{M_{1}}\left(x\right)$
for every threshold $x_{it}$. This reuse is substantially cheaper
than running $n$ fully separate regressions. Second, when estimating
$G_{t}\left(s\right)$, I exploit the linear structure in $X_{it}$
using the separable specification $p^{M_{2}}\left(s\right)=x\otimes p^{m_{2}}\left(v,w\right)$,
which reduces the effective regressor dimensionality and alleviates
the associated computational burden.
\end{rem}

\subsection{\label{subsec:est and inf -rates and normality}Convergence Rates
and Asymptotic Normality}

Since I let $n\gto\infty$ for each fixed $t$ in the asymptotic analysis,
the $t$-subscript is suppressed when it is clear. Let $p_{i}\coloneqq p^{M_{2}}\left(X_{i},V_{i},W_{i}\right)$
and $P\coloneqq\E p_{i}p_{i}'$. In the analysis, I use results from
\citet{imbens2009identification} to establish convergence rates for
$\widehat{V}\left(x,z,w\right)$ and $\widehat{G}\left(s\right)$.
The following assumption is needed.
\begin{assumption}
\label{assu: rates-V and G} Suppose the following conditions hold:
\begin{enumerate}[label=\textup{(\alph*)}]
\item \label{enu: approx cdf} $V\left(x,z,w\right)$ is continuously differentiable
of order $d_{1}$ on the support with derivatives uniformly bounded
in $\left(x,z,w\right)$ and $\mathcal{Z}\times\mathcal{W}\subset\R^{j_{1}}$.
\item \label{enu:normalization p}$p^{m_{2}}\left(v,w\right)=p^{m_{2v}}\left(v\right)\otimes p^{m_{2w}}\left(w\right)$
and there exist constants $C,\t>0$ such that $\lambda_{\min}\left(P\right)\geq C$
and
\[
\inf_{w\in\mathcal{W}}f_{V,W}\left(v,w\right)\geq C\left[v\left(1-v\right)\right]^{\t}.
\]
\item \label{enu:approx G} $G\left(s\right)$ is continuously differentiable
of order $d_{2}>1$ on $\mathcal{S}\subset\R^{j_{2}}$.
\item \label{enu: bound var Y} $\V\left(\rest YX,Z,W\right)$ is bounded.
\end{enumerate}
\end{assumption}
Let $\zeta\left(m_{2}\right)\coloneqq m_{2v}^{\t}m_{2}$ and $\zeta_{1}\left(m_{2}\right)\coloneqq m_{2v}^{\t+2}m_{2}$.
With Assumption \ref{assu: rates-V and G} in place, the next lemma
follows directly from Theorem 12 of \citet*{imbens2009identification}.
Let $\D_{1n}^{2}:=n^{-1}M_{1}+M_{1}^{1-2d_{1}/j_{1}}$ and $\D_{2n}^{2}:=\D_{1n}^{2}+n^{-1}m_{2}+m_{2}^{-2d_{2}/j_{2}}$
in the next lemma.
\begin{lem}[\textbf{Convergence Rates of $\boldsymbol{\widehat{V}}$ and $\boldsymbol{\widehat{G}}$}]
 \label{lem: rates of V and G} If the conditions of Theorem \ref{thm:main id}
and Assumption \ref{assu: rates-V and G} are satisfied, and $m_{2}^{2}m_{2v}^{\t+2}\left(n^{-1}M_{1}+M_{1}^{1-2d_{1}/j_{1}}\right)\to0$,
then
\begin{align*}
\E\left[n^{-1}\sum_{i}\left(\widehat{V}_{i}-V_{i}\right)^{2}\right] & =O\left(\D_{1n}^{2}\right),\\
\int\left[\widehat{G}\left(s\right)-G\left(s\right)\right]^{2}dF\left(s\right) & =O_{p}\left(\D_{2n}^{2}\right),\quad\text{and}\quad\sup_{s\in\mathcal{S}}\abs{\widehat{G}\left(s\right)-G\left(s\right)}=O_{p}\left(\zeta\left(m_{2}\right)\D_{2n}\right).
\end{align*}
\end{lem}
Lemma \ref{lem: rates of V and G} states that the mean-square convergence
rate for $\widehat{G}\left(s\right)$ is the sum of the first-step
rate $\D_{1n}^{2}$, the variance term $n^{-1}m_{2}$, and the squared
bias term $m_{2}^{-2d_{2}/j_{2}}$. $d_{1}/j_{1}$ and $d_{2}/j_{2}$
are the uniform approximation rates that govern how well the unknown
functions $V\left(x,z,w\right)$ and $G\left(s\right)$ can be approximated
with $\widehat{V}\left(x,z,w\right)$ and $\widehat{G}\left(s\right)$,
respectively; see Assumptions 3 and 5 of \citet*{imbens2009identification}
for the details. 

Let $\xi_{i}\coloneqq b_{1}\left(V_{i},W_{i}\right)-b\left(X_{i}\right)$
and $\xi\coloneqq\left(\xi_{1},\ldots,\xi_{n}\right)'$. To derive
the convergence rates for the APE and LAR estimators, I impose the
following assumption.
\begin{assumption}
\label{assu: rates beta x beta bar} Suppose the following conditions
hold:
\begin{enumerate}[label=\textup{(\alph*)}]
\item \label{enu:approx beta x series}$b\left(x\right)$ is continuously
differentiable of order $d_{3}$ on $\mathcal{\mathcal{X}}\subset\R^{j_{3}}$.
\item \label{enu:sup-norm-r-bd}There is a constant $C>0$ and $\zeta\left(M_{3}\right)$,
such that for each $M_{3}$ there exists a non-singular constant matrix
$N_{r}$ such that $\widetilde{r}^{M_{3}}\left(x\right)\coloneqq N_{r}r^{M_{3}}\left(x\right)$
satisfies $\lambda_{\min}\left(\E\widetilde{r}^{M_{3}}\left(X_{i}\right)\widetilde{r}^{M_{3}}\left(X_{i}\right)'\right)\geq C$
and $\sup_{x\in\mathcal{X}}\norm{\widetilde{r}^{M_{3}}\left(x\right)}\leq C\zeta\left(M_{3}\right)$.
\item \label{enu:rates bx xi bound 2nd order}$\E\left[\rest{\xi\xi'}{\bf X}\right]$
is bounded.
\end{enumerate}
\end{assumption}
Assumption \ref{assu: rates beta x beta bar} imposes conditions on
the degree of smoothness of $b\left(x\right)$, the normalization
of basis functions $r^{M_{3}}\left(x\right)$, and the boundedness
of the second moment of $\xi_{i}$, similar to those in Assumption
\ref{assu: rates-V and G}. Since $\widehat{b}_{1}\left(v,w\right)$
and $\widehat{G}\left(s\right)$ share the same series regression
coefficient $\widehat{\a}^{M_{2}}$, the convergence rate of $\widehat{b}_{1}\left(v,w\right)$
to $b_{1}\left(v,w\right)$ is the same as that of $\widehat{G}\left(s\right)$
to $G\left(s\right)$. I use this result to prove the convergence
rates of $\widehat{\ol b}$ and $\widehat{b}\left(x\right)$, both
of which are unknown but estimable functionals of $\widehat{G}\left(s\right)$.
Let $\D_{3n}^{2}:=n^{-1}M_{3}+M_{3}^{-2d_{3}/j_{3}}+\Delta_{2n}^{2}$
in the next theorem.
\begin{thm}[\textbf{Convergence Rates of $\boldsymbol{\widehat{\ol b}}$ and $\boldsymbol{\widehat{b}\left(x\right)}$}]
\label{thm:conv rate beta x beta bar} If the conditions of Lemma
\ref{lem: rates of V and G} and Assumption \ref{assu: rates beta x beta bar}
are satisfied, and $n^{-1}M_{3}\zeta\left(M_{3}\right)^{2}\to0$,
then
\begin{align*}
\norm{\widehat{\ol b}-\ol b}^{2} & =O_{p}\left(\Delta_{2n}^{2}\right),\\
\int\norm{\widehat{b}\left(x\right)-b\left(x\right)}^{2}dF(x) & =O_{p}\left(\D_{3n}^{2}\right),\quad\text{and}\quad\sup_{x\in\mathcal{X}}\norm{\widehat{b}\left(x\right)-b\left(x\right)}=O_{p}\left(\zeta\left(M_{3}\right)\D_{3n}\right).
\end{align*}
\end{thm}
The 2002 working paper version of \citet{imbens2009identification}
(henceforth IN02) has obtained asymptotic normality for estimators
of known and scalar-valued linear functionals of $G\left(s\right)$.
I build on their results to analyze vector-valued functionals of $G\left(s\right)$
via a Cram\'er--Wold device and prove asymptotic normality for $\widehat{b}_{1}\left(v,w\right)$.
To simplify the notation, I take $\widetilde{\ol p}^{M_{2}}\left(s\right)$
and $\widetilde{r}^{M_{3}}\left(x\right)$ that satisfy Assumptions
\ref{assu: rates-V and G} and \ref{assu: rates beta x beta bar}
as $\ol p^{M_{2}}\left(s\right)$ and $r^{M_{3}}\left(x\right)$ in
what follows.
\begin{assumption}
\label{assu:bvw normality} Suppose the following conditions hold:
\begin{enumerate}[label=\textup{(\alph*)}]
\item \label{enu:bvw normality 1}There is a constant $C>0$ and $\zeta\left(M_{1}\right)$,
such that for each $M_{1}$ there exists a non-singular constant matrix
$N_{q}$ such that $\widetilde{q}^{M_{1}}\left(z,w\right)\coloneqq N_{q}q^{M_{1}}\left(z,w\right)$
satisfies $\lambda_{\min}\left(\E\widetilde{q}^{M_{1}}\left(Z_{i},W_{i}\right)\widetilde{q}^{M_{1}}\left(Z_{i},W_{i}\right)'\right)\geq C$
and $\sup_{\left(z,w\right)\in\mathcal{Z\times\mathcal{W}}}\norm{\widetilde{q}^{M_{1}}\left(z,w\right)}\leq C\zeta\left(M_{1}\right)$.
\item \label{enu:newey1997 assumption}$G\left(s\right)$ is twice continuously
differentiable with bounded first and second derivatives. For functional
$a$ of $G$ and some constant $C>0$, it is true that $\abs{a\left(G\right)}\leq C\sup_{s}\abs{G\left(s\right)}$
and either (i) there is $\d\left(s\right)$ and $\widetilde{\alpha}^{M_{2}}$
such that $\E\d\left(S_{i}\right)^{2}<\infty$, $a\left(p_{m}^{M_{2}}\right)=\E\d\left(S_{i}\right)p_{m}^{M_{2}}\left(S_{i}\right)$
for all $m=1,...,M_{2}$, $a\left(G\right)=\E\d\left(S_{i}\right)G\left(S_{i}\right)$,
and $\E\left(\d\left(S_{i}\right)-p^{M_{2}}\left(S_{i}\right)'\widetilde{\alpha}^{M_{2}}\right)^{2}\rightarrow0$;
or (ii) for some $\widetilde{\alpha}^{M_{2}}$, $\E\left[p^{M_{2}}\left(S_{i}\right)'\widetilde{\alpha}^{M_{2}}\right]^{2}\rightarrow0$
and $a\left(p^{M_{2}}{}'\widetilde{\alpha}^{M_{2}}\right)$ is bounded
away from zero as $M_{2}\rightarrow\infty$.
\item \label{enu:var y bdd below}$\E\left[\rest{\left(Y-G\left(S\right)\right)^{4}}X,Z,W\right]$
is bounded and $\V\left(\rest YX,Z,W\right)$ is bounded away from
zero.
\item $nM_{1}^{1-2d_{1}/j_{1}}$, $nM_{2}^{-2d_{2}/j_{2}}$, $nM_{3}^{-2d_{3}/j_{3}}$,
$n^{-1}M_{1}^{2}M_{2}\zeta_{1}\left(M_{2}\right)^{2}$, $n^{-1}M_{1}M_{3}\zeta_{1}\left(M_{2}\right)^{2}$,
$n^{-1}M_{1}M_{3}\zeta\left(M_{3}\right)^{2}$, $n^{-1}M_{1}^{2}\zeta\left(M_{3}\right)^{2}$,
$n^{-1}M_{2}\zeta\left(M_{3}\right)^{2}$, $n^{-1}M_{1}\zeta\left(M_{1}\right)^{4}\zeta\left(M_{2}\right)^{4}$,
$n^{-1}M_{1}\zeta\left(M_{1}\right)^{4}\zeta\left(M_{3}\right)^{4}$,
and $n^{-1}M_{1}^{4}\zeta\left(M_{2}\right)^{6}$ are all $o\left(1\right)$.
\item \label{enu:bvw normality 5} There exist $d_{4}$ and $\ol{\alpha}^{M_{2}}$
such that for each element $s_{j}$ of $s=\left(x,v,w\right)\in\mathcal{S}\subset\R^{j_{4}}$:
\[
\sup\left\{ \sup_{s\in\mathcal{S}}\abs{G\left(s\right)-p^{M_{2}}\left(s\right)'\ol{\alpha}^{M_{2}}},\sup_{s\in\mathcal{S}}\abs{\partial\left(G\left(s\right)-p^{M_{2}}\left(s\right)'\ol{\alpha}^{M_{2}}\right)/\partial s_{j}}\right\} =O\left(M_{2}^{-d_{4}/j_{4}}\right).
\]
Also, $nM_{2}^{-2d_{4}/j_{4}}$ and $M_{1}M_{2}^{-2d_{4}/j_{4}}\zeta_{1}\left(M_{2}\right)^{2}$
are $o\left(1\right)$.
\item \label{enu:bvw normality  6 jiii of andrews 1991}(Assumption J(iii)
of \citet{don1991asymnormality}) For a bounded sequence of constants
$\left\{ c_{1n}:n\geq1\right\} $ and constant pd matrix $\ol{\Omega}_{1}$,
it is true that $c_{1n}\Omega_{1}\pto\ol{\Omega}_{1}$, where $\O_{1}$
is defined in (\ref{eq:def_Omega_1}).
\end{enumerate}
\end{assumption}
Assumption \ref{assu:bvw normality}\ref{enu:bvw normality 1}--\ref{enu:bvw normality 5}
is also imposed by IN02 and is a regularity condition required for
the asymptotic normality of $\widehat{b}_{1}\left(v,w\right)$. Assumption
\ref{assu:bvw normality}\ref{enu:bvw normality  6 jiii of andrews 1991}
concerns the asymptotic covariance matrix of $\widehat{b}_{1}\left(v,w\right)$
and is used by \citet{don1991asymnormality}. It guarantees that the
normality result of IN02 applies to vector-valued functionals of $G\left(s\right)$.
Essentially, Assumption \ref{assu:bvw normality}\ref{enu:bvw normality  6 jiii of andrews 1991}
requires that all the coordinates of $\widehat{b}_{1}\left(v,w\right)$
converge at the same speed, which is mild because ex-ante I do not
distinguish any coordinate of $\b_{it}$ from the others.
\begin{lem}[\textbf{Asymptotic Normality of $\boldsymbol{\widehat{b}_{1}\left(v,w\right)}$}]
\label{lem: bvw normality}  If the conditions of Theorem \ref{thm:conv rate beta x beta bar}
and Assumption \ref{assu:bvw normality} are satisfied, then 
\[
\sqrt{n}\,\Omega_{1}^{-1/2}\left(\widehat{b}_{1}\left(v,w\right)-b_{1}\left(v,w\right)\right)\dto N\left(0,I\right),
\]
where $\Omega_{1}$ is defined in (\ref{eq:def_Omega_1}). 

Furthermore, 
\[
\widehat{\Omega}_{1}\Omega_{1}^{-1}\pto I,
\]
where $\widehat{\Omega}_{1}$ is defined in (\ref{eq:def_Omega_1_hat}).
\end{lem}
Lemma \ref{lem: bvw normality} concerns $b_{1}\left(v,w\right)$,
a \textit{known} functional of $G\left(s\right)$. Therefore, the
results of IN02 directly apply and I omit its proof in this paper.
However, the results of IN02 do not directly apply to $\widehat{\ol b}$
and $\widehat{b}(x)$ because $\ol b$ and $b(x)$ are \textit{unknown}
functionals of $G\left(s\right)$. To explain, notice that by the
law of iterated expectations
\begin{align}
\ol b=T^{-1}\sum_{t=1}^{T}\E\left[\partial G\left(S\right)/\partial X\right],\quad\text{and}\quad b\left(x\right)=\E\left[\rest{\partial G\left(S\right)/\partial X}X=x\right],\label{eq: in02 result na}
\end{align}
both of which involve integrating $b_{1}\left(V,W\right)=\partial G\left(S\right)/\partial X$
with respect to the unknown but estimable distribution of $\left(V,W\right)$.
Therefore, I need to estimate the unknown functionals in (\ref{eq: in02 result na})
and correctly account for the additional estimation bias in the asymptotic
analysis.
\begin{assumption}
\label{assu:bx and bbar normality} Suppose the following conditions
hold:
\begin{enumerate}[label=\textup{(\alph*)}]
\item \label{enu:as on Erp}$\E\ol p_{i}r_{i}'$ has full column rank.
\item \label{enu:bx normality lindfeller} $\E\left[\rest{\norm{\xi}^{4}}{\bf X}\right]$
is bounded and $\E\left[\rest{\xi\xi'}{\bf X}\right]$ is bounded
away from zero.
\item \label{enu:bx jiii da91}For a sequence of bounded constants $\left\{ c_{2n},c_{3n}:n\geq1\right\} $
and some constant pd matrix $\ol{\Omega}_{2}$ and $\ol{\Omega}_{3}$,
$c_{2n}\Omega_{2}\pto\ol{\Omega}_{2}$ and $c_{3n}\Omega_{3}\pto\ol{\Omega}_{3}$,
where $\Omega_{2}$ and $\O_{3}$ are defined in (\ref{eq:def_Omega_2})
and (\ref{eq:Omega_3}), respectively.
\item \label{enu:b^bar_cond}$\E\norm{b_{1}\left(v,w\right)-\ol b}^{4}<\infty$.
\end{enumerate}
\end{assumption}
Assumption \ref{assu:bx and bbar normality}\ref{enu:as on Erp} is
needed to show that the asymptotic covariance matrix $\Omega_{2}$
of $\sqrt{n}(\widehat{b}\left(x\right)-b\left(x\right))$ is positive
definite and similarly for $\O_{3}$. Assumption \ref{assu:bx and bbar normality}\ref{enu:bx normality lindfeller}
is a regularity condition imposed for the Lindeberg--Feller central
limit theorem (CLT). Assumption \ref{assu:bx and bbar normality}\ref{enu:bx jiii da91}
is similar to Assumption \ref{assu:bvw normality}\ref{enu:bvw normality  6 jiii of andrews 1991}
and is needed to prove that the asymptotic normality result holds
for vector-valued functionals of $G\left(s\right)$ by the Cram\'er--Wold
device.
\begin{thm}[\textbf{Asymptotic Normality of $\boldsymbol{\widehat{\ol b}}$ and
$\boldsymbol{\widehat{b}\left(x\right)}$}]
\label{thm:bx and bbar normality}  If the conditions of Lemma \ref{lem: bvw normality}
and Assumption \ref{assu:bx and bbar normality} are satisfied, then
\begin{align*}
\sqrt{n}\,\Omega_{2}^{-1/2}(\widehat{b}\left(x\right)-b\left(x\right)) & \dto N\left(0,I\right)\quad\text{and}\quad\sqrt{n}\,\O_{3}^{-1/2}(\widehat{\ol b}-\ol b)\dto N(0,I),
\end{align*}
where $\O_{2}$ and $\O_{3}$ are defined in (\ref{eq:def_Omega_2})
and (\ref{eq:Omega_3}), respectively.

Furthermore, 
\[
\widehat{\Omega}_{2}\Omega_{2}^{-1}\pto I\quad\text{and}\quad\widehat{\Omega}_{3}\Omega_{3}^{-1}\pto I,
\]
where $\widehat{\Omega}_{2}$ and $\widehat{\Omega}_{3}$ are defined
in (\ref{eq:def_Omega_2_hat}) and (\ref{eq:Omega_3^hat}), respectively.
\end{thm}
I present the main idea to the proof of Theorem \ref{thm:bx and bbar normality},
focusing on how the error propagates through the three-step analysis
and the roles of $\widehat{\Omega}_{2}$ and $\widehat{\Omega}_{3}$.
Define the functionals
\begin{align*}
a\left(b_{1},V\right) & \coloneqq\E\left[\rest{b_{1}\left(V,W\right)}X=x\right]=b\left(x\right),\\
\widehat{a}\left(b_{1},V\right) & :=\widehat{\E}\left[\rest{b_{1}\left(V,W\right)}X=x\right],\\
\widehat{a}\left(b_{1},\widehat{V}\right) & :=\widehat{\E}\left[\rest{b_{1}\left(\widehat{V},W\right)}X=x\right],\text{ and}\\
\widehat{a}\left(\widehat{b}_{1},\widehat{V}\right) & \coloneqq\widehat{\E}\left[\rest{\widehat{b}_{1}\left(\widehat{V},W\right)}X=x\right]=\widehat{b}\left(x\right).
\end{align*}
Then
\begin{equation}
\widehat{b}(x)-b(x)=\underbrace{\widehat{a}(b_{1},\widehat{V})-\widehat{a}(b_{1},V)}_{\text{(i) first-step error in }V}+\underbrace{\widehat{a}(\widehat{b}_{1},\widehat{V})-\widehat{a}(b_{1},\widehat{V})}_{\text{(ii) second-step error in }b_{1}}+\underbrace{\widehat{a}(b_{1},V)-a(b_{1},V)}_{\text{(iii) third-step sampling error}}.\label{eq:an-err-dec}
\end{equation}
This decomposition makes clear that the asymptotic covariance of $\sqrt{n}\,(\widehat{b}(x)-b(x))$,
denoted by $\Omega_{2}$ as in (\ref{eq:def_Omega_2}), must account
for errors of all three sources. In the proof of Theorem \ref{thm:bx and bbar normality},
I show that
\begin{equation}
\sqrt{n}\,\Omega_{2}^{-1/2}\big(\widehat{b}(x)-b(x)\big)=\frac{1}{\sqrt{n}}\sum_{i=1}^{n}\sum_{k=1}^{3}\psi_{ki}+o_{p}(1),\label{eq:if-func-1}
\end{equation}
where the influence functions $\psi_{1i},\ \psi_{2i},\text{ and }\psi_{3i}$
correspond exactly to the three error terms on the right-hand side
of (\ref{eq:an-err-dec}) up to proper normalization. Once (\ref{eq:if-func-1})
is established, I verify the Lindeberg-Feller condition on $\Psi_{in}:=n^{-1/2}\sum_{k=1}^{3}\psi_{ki}$,
which proves its convergence to the standard normal distribution and
that $\O_{2}$ is indeed the appropriate asymptotic covariance associated
with $\sqrt{n}\,(\widehat{b}(x)-b(x))$.

For feasible inference, a consistent estimator for $\Omega_{2}$ is
needed, and I show in the proof of Theorem \ref{thm:bx and bbar normality}
that $\widehat{\Omega}_{2}$ is consistent. Note that $\Omega_{2}$
may diverge because $\widehat{b}(x)$ converges more slowly than $n^{-1/2}$
as shown in Theorem \ref{thm:conv rate beta x beta bar}. Hence, all
convergence results are expressed in self-normalized way, for example,
\begin{equation}
\sqrt{n}\,\Omega_{2}^{-1/2}\big(\widehat{b}(x)-b(x)\big)\dto N(0,I)\quad\text{and}\quad\widehat{\Omega}_{2}\O_{2}^{-1}\pto I.
\end{equation}
The same argument applies to $\widehat{\O}_{3}$ with the basis function
$r^{M_{3}}\left(x\right)=1$ and the influence function capturing
serial correlation of estimating the APE across periods.

\section{Empirical Application\label{sec:Empirical-Illustration}}

In this section, I apply my procedure to estimate a heterogeneous
Cobb-Douglas production function for each of the five largest manufacturing
sectors in China. I obtain unconditional means of the output elasticities
(APE) and compare them with those derived using classic methods on
the same data set. Furthermore, I estimate conditional means of the
elasticities given the regressors (LAR). Results show that there are
significant across-firm variations in the output elasticities. I also
examine how heterogeneity in output elasticities relates to observed
characteristics and obtain intuitive results.

In Appendix \ref{sec:Additional-Empirical-Results}, I conduct several
robustness checks---including alternative price deflators, sample
trimming, alternative IV and $W_{i}$ constructions, city--year fixed
effects, and comparison with OLS and naive IV estimates---and the
results remain stable. In Appendix \ref{sec:Simulation}, I conduct
a production-function motivated simulation study to support the findings
in this application.

\subsection{Data and Methodology}

I use the China Annual Survey of Industrial Firms (CASIF), a longitudinal
micro-level dataset collected by the National Bureau of Statistics
of China that includes information on all state-owned industrial firms
and non-state-owned firms with annual sales above 5 million RMB (\textasciitilde US\$770K).
According to \citet*{brandt2017wto}, they account for 91\% of the
gross output, 71\% of employment, 97\% of exports, and 91\% of total
fixed assets in 2004, and thus are representative of industrial activities
in China. Many papers on topics such as firm behavior, international
trade, and growth theory have used the CASIF data (e.g., \citet*{hsieh2009misallocation}).

I focus on the five largest 2-digit sectors in terms of the number
of firms between 2004 and 2007. Note that the CASIF dataset spans
between 1998 and 2007. I choose year 2004 to 2007 to (i) ensure data
consistency due to the change in the Chinese Industry Classification
codes in 2003, (ii) avoid major structural breaks in the early 2000s
(e.g., China joined the WTO in 2001), and (iii) use the most recent
data. 

Following \citet*{brandt2014challenges}, appropriate price deflators
for inputs and outputs are applied separately. I preprocess the data
so that firms with strictly positive amounts of capital, employment,
value-added output, real wage expense, and real interest rate are
used for estimation. The final dataset consists of a balanced panel
of 11,317 firms over four years across five sectors. Summary statistics
are presented in Table \ref{tab:emp summary stat}.

\begin{table}
\caption{Summary Statistics\label{tab:emp summary stat}}

\smallskip{}

\centering{}\begin{threeparttable}%
\begin{tabular}{lccccc}
\toprule 
Variables & N & mean & sd & min & max\tabularnewline
\midrule
$y_{it}=\ln$$\left(\text{value-added output}\right)$ & 45,268 & 9.55 & 1.34 & 2.24 & 16.96\tabularnewline
$k_{it}=\ln$$\left(\text{capital}\right)$ & 45,268 & 9.17 & 1.56 & 0.98 & 16.84\tabularnewline
$l_{it}=\ln$$\left(\text{labor}\right)$ & 45,268 & 5.08 & 1.08 & 2.08 & 11.97\tabularnewline
$Z_{it,1}=\ln\ol r_{cs(i),t}^{\text{real},(-i)}$ & 45,268 & 0.73 & 0.45 & -6.02 & 6.94\tabularnewline
$Z_{it,2}=\ln\ol w_{cs(i),t}^{\text{real},(-i)}$ & 45,268 & 2.68 & 0.37 & 0.33 & 5.28\tabularnewline
Year & 4 & - & - & 2004 & 2007\tabularnewline
Firm ID & 11,317 & - & - & - & -\tabularnewline
Industry Code & 5 & - & - & - & -\tabularnewline
\bottomrule
\end{tabular}\begin{tablenotes}[flushleft]
\scriptsize
\item \textit{Notes:} 

(i) Output is measured as firm-level real value added in constant 1998 RMB (10,000 yuan). Capital is measured as firm-level real fixed capital stock (book value net of depreciation), deflated to constant 1998 RMB (10,000 yuan). Labor is measured as the number of employees (persons). The real interest rate is defined as real interest expenditures divided by total debt (percent). The real wage is average real salary per worker per year in constant 1998 RMB (10,000 yuan). 

(ii) Sample sizes by industry are: textile (3,072), chemicals (2,131), nonmetallic mineral products (1,758), general equipment (2,828), and transportation equipment (1,528). 

(iii)  $Z_{it,1}$ denotes the natural logarithm of debt-weighted average real interest rate of competitors in the same city--industry--year.  $Z_{it,2}$ denotes the natural logarithm of employment-weighted average real wage of competitors in the same city--industry--year.

\end{tablenotes}
\end{threeparttable}
\end{table}

The value-added heterogeneous production function is
\begin{align}
y_{it} & =k_{it}\b_{it,K}+l_{it}\b_{it,L}+\omega_{it}+\widetilde{\epsilon}_{it},\nonumber \\
\b_{it,K} & =\b_{K}\left(A_{i},\ve_{it},\upsilon_{it}\right),\ \b_{it,L}=\b_{L}\left(A_{i},\ve_{it},\upsilon_{it}\right),\ \omega_{it}=\omega\left(A_{i},\ve_{it},\upsilon_{it}\right),\nonumber \\
k_{it} & =g_{K}\left(Z_{it},A_{i},\eta_{it,K}\right),\ l_{it}=g_{L}\left(Z_{it},A_{i},\eta_{it,L}\right).\label{eq:emp prod fcn}
\end{align}
I highlight two features of model \eqref{eq:emp prod fcn}. First,
output elasticities $\b_{it}\coloneqq(\b_{it,K},\b_{it,L})'$ are
allowed to differ across firms and through time. Second, input choices
$X_{it}\coloneqq(k_{it},l_{it})'$ can be correlated with $\b_{it}$
through their dependence on the pairwise correlated random variables
$A_{i}$, $\eta_{it}$, and $\ve_{it}$. 

I use leave-one-out industry--city--year cell averages of real wages
and real interest rates faced by other firms in the same industry
and city, weighted by competitors\textquoteright{} employment (for
wages) and total debt (for interest rates) as IVs. I write them as
$Z_{it}\coloneqq(\ln\ol r_{cs(i),t}^{\text{real},(-i)},\,\ln\ol w_{cs(i),t}^{\text{real},(-i)})'$,
where $cs(i)$ represents city and sector of firm $i$. These instruments
are relevant because firms compete locally for labor and capital,
so variation in competitors\textquoteright{} cost conditions shifts
firm $i$\textquoteright s effective input prices and hence its optimal
input choices. At the same time, they are plausibly exogenous to firm
$i$\textquoteright s idiosyncratic productivity shocks and fixed
effects because firm $i$\textquoteright s own input prices are excluded
from the construction; identification therefore relies on competitors\textquoteright{}
cost variation rather than firm-level wage or interest-rate realizations.
In Appendix \ref{sec:Additional-Empirical-Results}, I examine the
exogeneity issue by using a more exogenous IV constructed at the provincial
level, and the results remain stable. 

For estimation, I take $W_{i}$ to be the mean through time of each
coordinate of $X_{it}$ by adapting the argument of \citet{mundlak1978pooling}
nonparametrically. I also examine the effect of further including
the mean through time of each coordinate of $Z_{it}$ and the results
reported in Table \ref{tab:CF-sensivitity} are similar. For all series
estimation steps, I use second-degree polynomial spline basis functions
with knots at the median, and a robustness check to this choice is
included in Appendix \ref{sec:Additional-Empirical-Results}. First,
I estimate each coordinate of $V_{it}\coloneqq(F_{\rest{k_{it}}Z_{it},W_{i}}(\rest{k_{it}}Z_{it},W_{i}),F_{\rest{l_{it}}Z_{it},W_{i}}(\rest{l_{it}}Z_{it},W_{i}))'$
by regressing $\ind(k_{it}\leq k)$ and $\ind(l_{it}\leq l)$ on the
basis functions of $(Z_{it},W_{i})$, respectively. Next, I estimate
$G_{it}\coloneqq\E[\rest{y_{it}}X_{it},V_{it},W_{i}]$ by regressing
$y_{it}$ on the basis functions of $(X_{it},\widehat{V}_{it},W_{i})$.
Estimation of $b_{1t}(V_{it},W_{i})\coloneqq\E[\rest{\b_{it}}V_{it},W_{i}]$
is then obtained by taking the partial derivative of $\widehat{G}_{it}(X_{it},\widehat{V}_{it},W_{i})$
with respect to $X_{it}$. Finally, I estimate the pooled APE $\ol b$
by simply averaging $\widehat{b}_{1t}(\widehat{V}_{it},W_{i})$ over
$i$ and $t$. For the LAR, as in (\ref{eq: estimator for beta bar}),
I regress $\widehat{b}_{1t}(\widehat{V}_{it},W_{i})$ on the basis
functions of $X_{it}$ for each $t$, then average over $t$ to obtain
$\widehat{\b}_{i,K}$ and $\widehat{\b}_{i,L}$ as a measure of heterogeneity
in output elasticities across firms. 

\subsection{Results }

\subsubsection*{Average Output Elasticities}

I compare my TERC estimates of average output elasticities with those
obtained from OP (\citet{olleypakes1996teltfp}), LP (\citet{levinsohnpetrin2003res}),
and ACF (\citet*{ackerberg2015identification}) applied to the same
dataset. Table \ref{tab:emp_betabar_results} contains the estimation
results.

\begin{table}
\noindent\caption{Estimates of Average Output Elasticities\label{tab:emp_betabar_results}}

\smallskip{}

\centering{}\begin{threeparttable}%
\begin{tabular}{lcccc}
\toprule 
Textile & OP & LP & ACF & TERC\tabularnewline
\midrule
Capital Elasticity & 0.359 & 0.252 & 0.300 & 0.415\tabularnewline
95\% CI & {[}0.300, 0.417{]} & {[}0.206, 0.297{]} & {[}0.227, 0.373{]} & {[}0.393, 0.441{]}\tabularnewline
Labor Elasticity & 0.470 & 0.175 & 0.567 & 0.452\tabularnewline
95\% CI & {[}0.442, 0.498{]} & {[}0.154, 0.196{]} & {[}0.440, 0.695{]} & {[}0.415, 0.485{]}\tabularnewline
\midrule 
Chemical & OP & LP & ACF & TERC\tabularnewline
\midrule 
Capital Elasticity & 0.294 & 0.288 & 0.344 & 0.463\tabularnewline
95\% CI & {[}0.223, 0.365{]} & {[}0.228, 0.348{]} & {[}0.155, 0.533{]} & {[}0.430, 0.497{]}\tabularnewline
Labor Elasticity & 0.296 & 0.113 & 0.378 & 0.311\tabularnewline
95\% CI & {[}0.253, 0.339{]} & {[}0.084, 0.143{]} & {[}-0.039, 0.795{]} & {[}0.263, 0.361{]}\tabularnewline
\midrule 
Nonmetallic Mineral & OP & LP & ACF & TERC\tabularnewline
\midrule 
Capital Elasticity & 0.697 & 0.311 & 0.236 & 0.371\tabularnewline
95\% CI & {[}0.499, 0.895{]} & {[}0.263, 0.358{]} & {[}0.038, 0.434{]} & {[}0.332, 0.410{]}\tabularnewline
Labor Elasticity & 0.353 & 0.071 & 0.601 & 0.311\tabularnewline
95\% CI & {[}0.311, 0.394{]} & {[}0.051, 0.091{]} & {[}-0.151, 1.354{]} & {[}0.255, 0.365{]}\tabularnewline
\midrule 
General Equipment & OP & LP & ACF & TERC\tabularnewline
\midrule 
Capital Elasticity & 0.416 & 0.246 & 0.176 & 0.433\tabularnewline
95\% CI & {[}0.267, 0.565{]} & {[}0.202, 0.291{]} & {[}0.016, 0.337{]} & {[}0.402, 0.469{]}\tabularnewline
Labor Elasticity & 0.444 & 0.071 & 0.927 & 0.521\tabularnewline
95\% CI & {[}0.406, 0.482{]} & {[}0.053, 0.089{]} & {[}0.635, 1.218{]} & {[}0.479, 0.564{]}\tabularnewline
\midrule 
Transportation Equipment & OP & LP & ACF & TERC\tabularnewline
\midrule
Capital Elasticity & 0.523 & 0.281 & 0.217 & 0.425\tabularnewline
95\% CI & {[}0.397, 0.649{]} & {[}0.216, 0.346{]} & {[}0.052, 0.381{]} & {[}0.393, 0.459{]}\tabularnewline
Labor Elasticity & 0.523 & 0.137 & 1.042 & 0.588\tabularnewline
95\% CI & {[}0.473, 0.573{]} & {[}0.100, 0.173{]} & {[}0.771, 1.313{]} & {[}0.536, 0.644{]}\tabularnewline
\bottomrule
\end{tabular}\begin{tablenotes}[flushleft]
\scriptsize
\item \textit{Notes:} 
For TERC, I implement the three-step series estimator as described above. To compute the CI, I use a subsampling method to draw $b=\lfloor4n_s^{3/4}\rfloor$  firms from each sector without replacement and repeat this 1,000 times. For the other methods, I use the Stata commands  (\texttt{prodest}, \citet{rovigatti2018theory}) for OP and LP and  (\texttt{acfest}, \citet{manjon2016production}) for ACF.
\end{tablenotes}
\end{threeparttable}
\end{table}

The $\widehat{\ol b}_{K}$ estimates from TERC fall within $\left[0.371,\,0.463\right]$
across the five sectors, broadly consistent with the results obtained
from applying OP, LP, and ACF to the same dataset. In contrast, the
$\widehat{\ol b}_{L}$ estimates exhibit greater discrepancies across
methods. The TERC estimates of $\widehat{\ol b}_{L}$ lie in $\left[0.311,0.588\right]$
across the five sectors. OP's labor elasticity estimates are close
to TERC's for the first three sectors but are smaller for the last
two. LP seems to produce smaller $\widehat{\ol b}_{L}$ estimates
across all sectors, whereas ACF generates larger $\widehat{\ol b}_{L}$
estimates for the general equipment and transportation equipment sectors.
It is worth noting that the ACF approach tends to produce wider confidence
intervals in certain sectors, which may partly reflect numerical instability
in the Stata implementation, an issue also discussed by \citet*{aureo2024production}.
The 95\% CIs for TERC estimates are reasonably tight across all sectors. 

I consider the estimation results from TERC reasonable based on two
pieces of empirical evidence. First, it is well documented in the
literature that output elasticities in Cobb-Douglas production function
estimation usually lie within $\left[0,1\right]$. Second, using Chinese
manufacturing data, \citet{hsieh2009misallocation} show that roughly
half of output accrues to capital. Under profit maximization with
a Cobb-Douglas production function, output elasticities correspond
to factor revenue shares. The magnitudes of the estimated capital
and labor elasticities can therefore be interpreted through this lens,
and my estimates are consistent with this result.

The broad similarity of average elasticities across methods is not
unexpected, as OP, LP, ACF, and the TERC approach are all designed
to address simultaneity between input choices and unobserved productivity.
While OP, LP, and ACF provide consistent estimates under a constant-coefficient
production function, TERC adds value by relaxing this restriction
and allowing output elasticities to vary across firms while still
addressing simultaneity via a control-function approach.
\begin{rem}[\textbf{Addressing Simultaneity Issue}]
\label{rem:OP_vs_TERC} OP, LP, and ACF address simultaneity via
a proxy-variable strategy: they invert a strictly monotone proxy input
demand---investment in OP and intermediate inputs in LP/ACF---to
recover a scalar productivity term $\o_{it}$, then impose a Markov-style
law of motion for $\o_{it}$ to form identifying moments. This requires
(i) strict monotonicity in $\o_{it}$ and (ii) that $\o_{it}$ is
the only econometric unobservable entering the proxy demand, which
rules out random coefficients that directly affect firms' input choices.
In contrast, my baseline TERC model allows input choices to correlate
with heterogeneous output elasticities through an arbitrary-dimensional
fixed effect $A_{i}$ and a time-varying shock $\eta_{it}$, with
productivity modeled flexibly as $\o_{it}=\o(A_{i},\ve_{it})$. Identification
proceeds via a control-function strategy using a sufficient statistic
$W_{i}$ for $A_{i}$ and a period-specific control $V_{it}$ for
$\eta_{it}$, so that conditional on $(V_{it},W_{i})$ the remaining
input variation is driven by exogenous instruments without taking
a Markov law for $\o_{it}$ as a primitive assumption.
\end{rem}

\subsubsection*{Cross-Firm Heterogeneity in Output Elasticities }

Next, I examine the distribution of conditional mean output elasticities
given the regressors. These conditional means summarize average elasticities
for firm subgroups defined by their capital and labor levels, and
thus provide a useful basis for designing more targeted, sector-specific
policies. They are also informative about the underlying distribution
of true elasticities. For instance, by the law of total variance,
the variance of these conditional means is a lower bound on the variance
of the true output elasticities. \citet{demirer2020production} argues
that the heterogeneity in output elasticities is largely explained
by across-firm variation. To study this issue, I average $\widehat{b}_{t,K}(X_{it})$
and $\widehat{b}_{t,L}(X_{it})$ for each firm through time and denote
them by $\widehat{\b}_{i,K}$ and $\widehat{\b}_{i,L}$, respectively.
These $\widehat{\b}_{i}$ estimates can be considered as a proxy for
their output elasticities. Then, I plot the histograms of $\widehat{\b}_{i,K}$
and $\widehat{\b}_{i,L}$ across firms for each sector.

In Figure \ref{fig:betakl_textile}, I present the histograms of $\widehat{\b}_{i,K}$
and $\widehat{\b}_{i,L}$ for the textile sector and indicate the
corresponding pooled APE labeled by ``Mean'' in the figure. The
left subplot of Figure \ref{fig:betakl_textile} corresponds to the
histogram of $\widehat{\b}_{i,K}$. All of its probability mass lies
between zero and one, with its mode at 0.4. Over 95\% of its probability
mass lies between 0.3 and 0.5.\textbf{ }The distribution of $\widehat{\b}_{i,K}$
is concentrated around the mean and symmetric, suggesting that these
firms tend to be homogeneous in capital efficiency. The right subplot
of Figure \ref{fig:betakl_textile} is the histogram of $\widehat{\b}_{i,L}$.
Again, the majority of its probability mass lies between zero and
one. The distribution of $\widehat{\b}_{i,L}$ is more dispersed than
that of $\widehat{\b}_{i,K}$ and is slightly left-skewed, with a
small number of textile firms exhibiting low labor efficiency.
\begin{center}
\begin{figure}
\centering
\begin{centering}
\centering\includegraphics[width=0.5\textwidth]{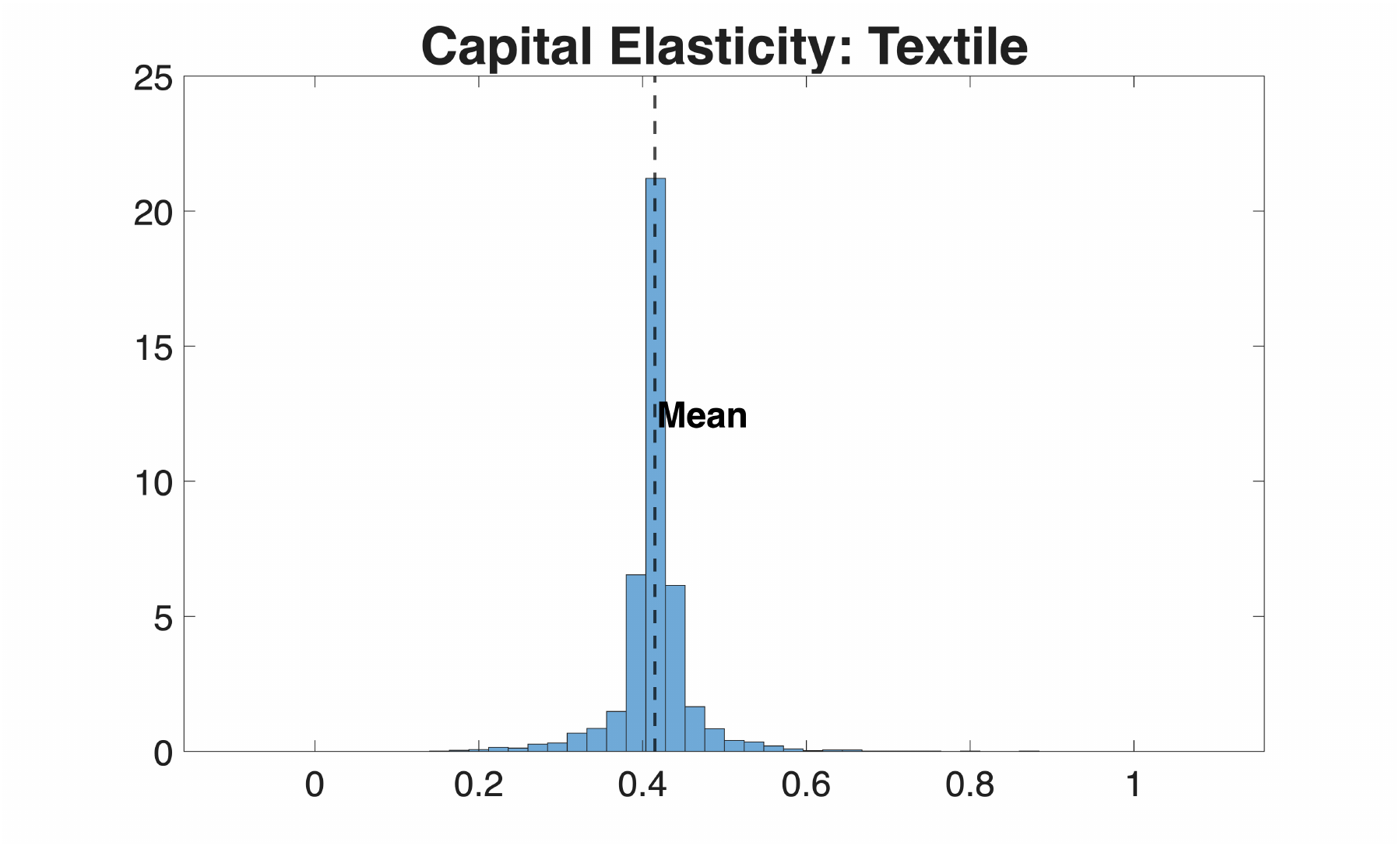}\includegraphics[width=0.5\textwidth]{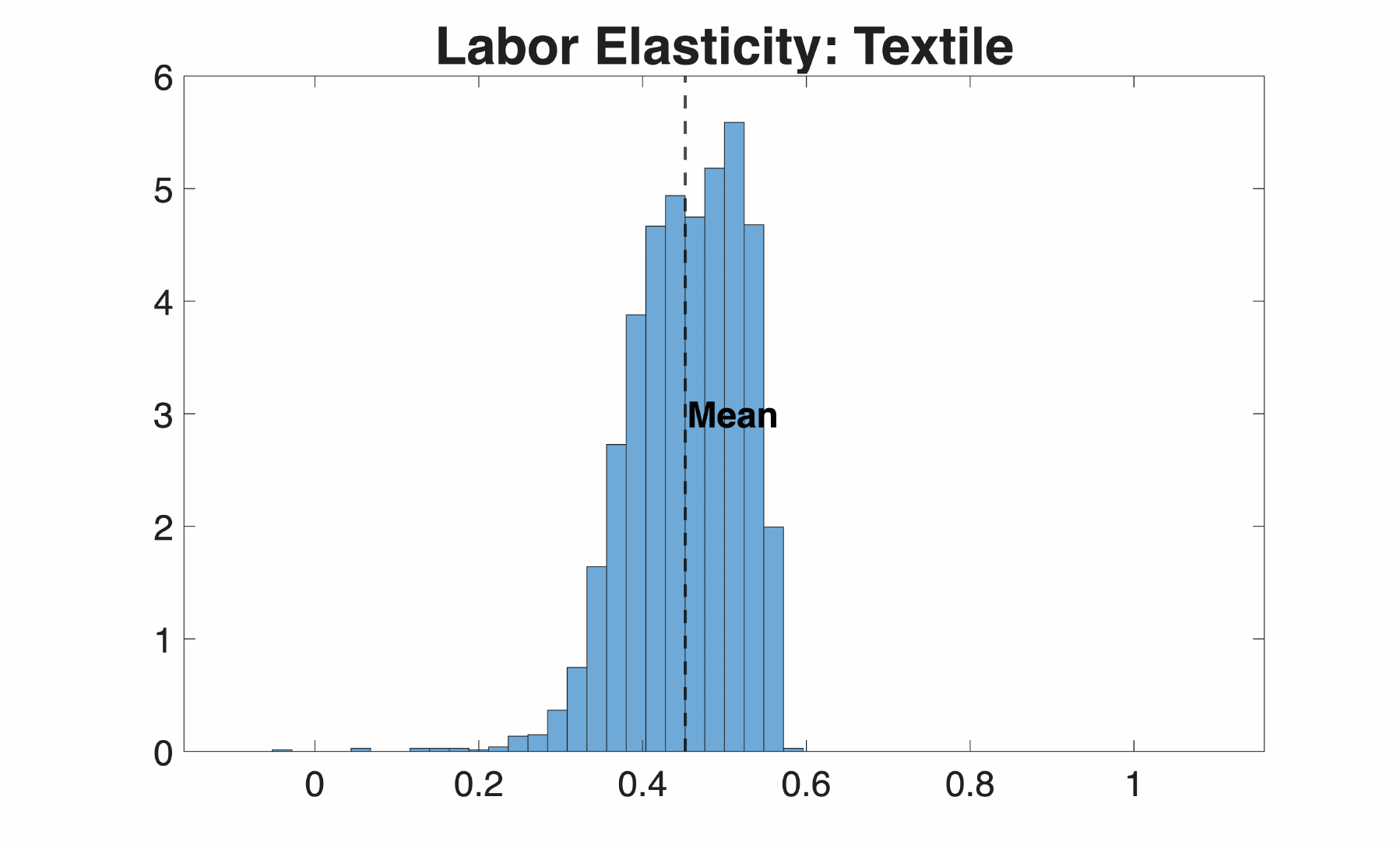}
\par\end{centering}
\caption{Distribution of Conditional Means of Output Elasticities: Textile\label{fig:betakl_textile}}
\end{figure}
\par\end{center}

Figure \ref{fig:betakl_4others} presents the histograms of $\widehat{\b}_{i,K}$
and $\widehat{\b}_{i,L}$ for the other sectors. I draw the following
two conclusions. First, for all sectors, the majority of $\widehat{\b}_{i,K}$
and $\widehat{\b}_{i,L}$ lie between zero and one, consistent with
the empirical evidence in \citet{hsieh2009misallocation}. Second,
there is substantial across-firm variations in both capital and labor
elasticities within each sector. The extent of heterogeneity, however,
differs across elasticities and sectors. For example, in the left
panel of Figure \ref{fig:betakl_4others}, capital elasticity exhibits
greater dispersion among chemical firms than among firms in the nonmetallic
minerals sector.
\begin{center}
\begin{figure}[H]
\centering

\includegraphics[width=0.5\textwidth]{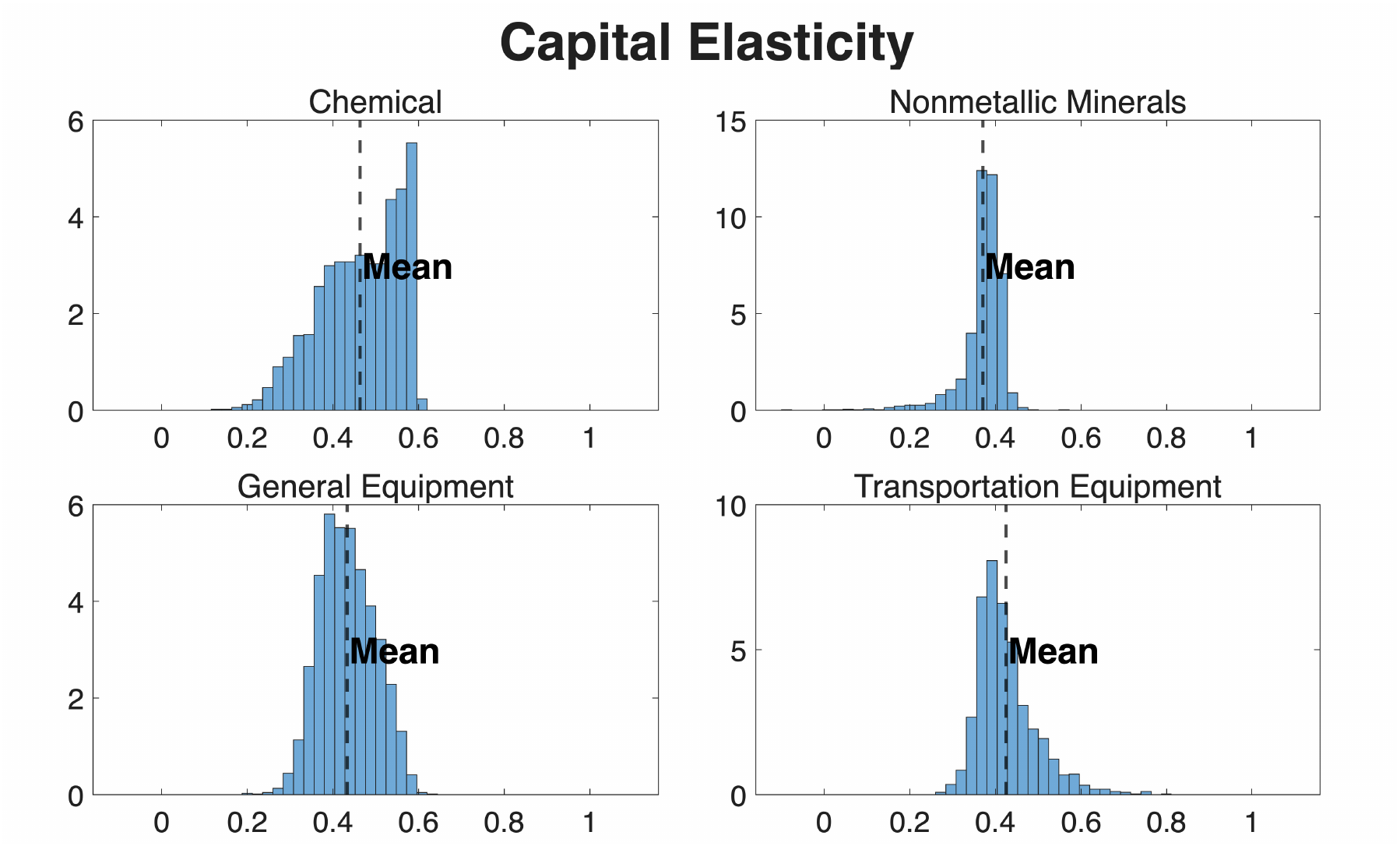}\includegraphics[width=0.5\textwidth]{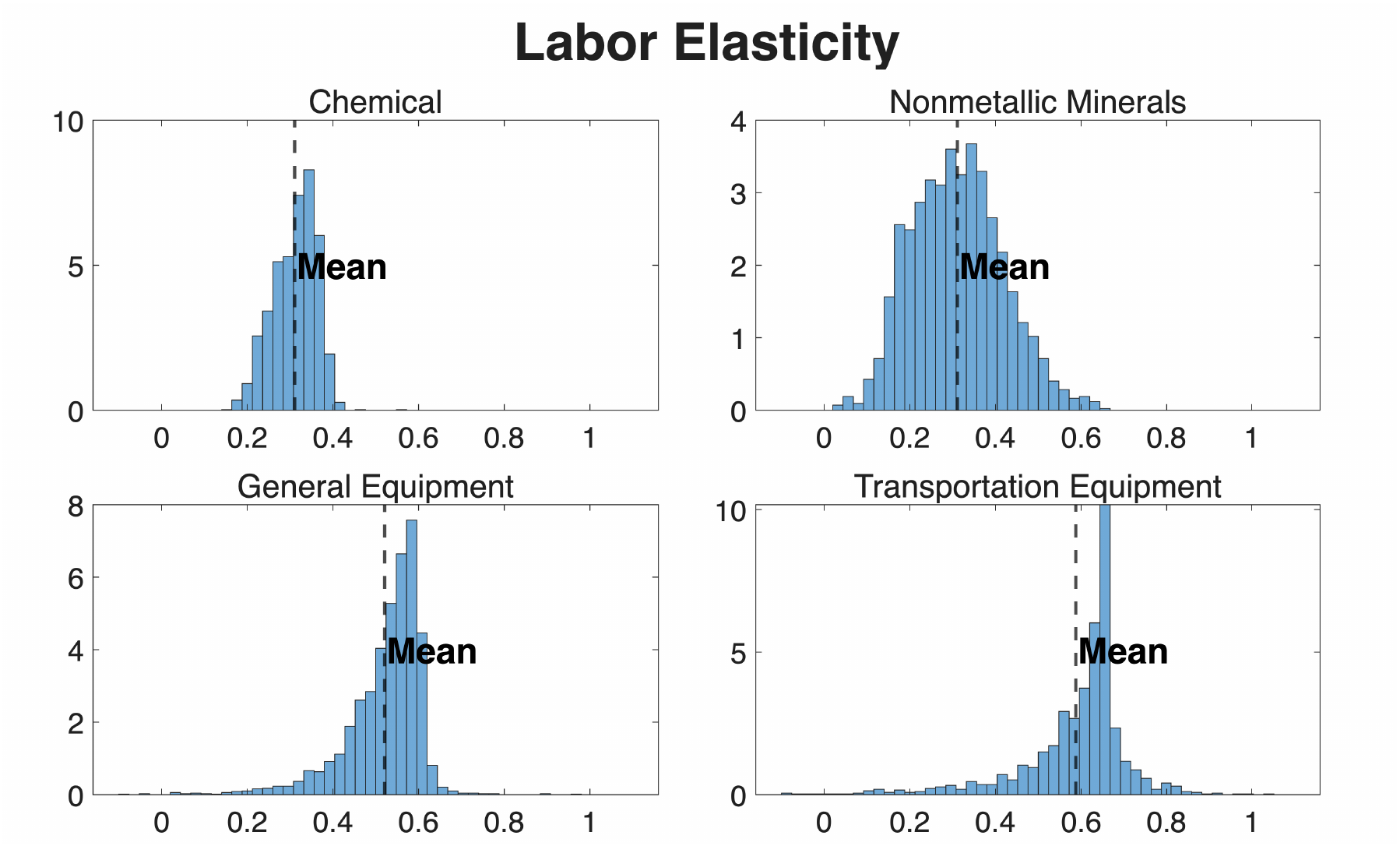}

\caption{Distribution of Conditional Means of Output Elasticities: Other Sectors\label{fig:betakl_4others}}
\end{figure}
\par\end{center}

\subsubsection*{Explaining the Heterogeneity}

I examine cross-firm heterogeneity in output elasticities by regressing
the estimated $\widehat{\b}_{i,K}$ and $\widehat{\b}_{i,L}$ on firm
characteristics. Specifically, using the CASIF dataset, I construct
five covariates: firm size, leverage, export status, industrial-park
location, and state ownership. All specifications include industry
and city fixed effects, and I report heteroskedasticity-robust standard
errors. 

\begin{table}
\caption{Explaining the Heterogeneity\label{tab:reg-beta-firm-obs}}

\smallskip{}

\centering{}\begin{threeparttable}%
\begin{tabular}{lcc}
\toprule 
 & $\widehat{\b}^{K}$ & $\widehat{\b}^{L}$\tabularnewline
\midrule
Firm Size & \begin{cellvarwidth}[t]
\centering
0.0328\\
(0.0006)
\end{cellvarwidth} & \begin{cellvarwidth}[t]
\centering
0.0002\\
(0.0012)
\end{cellvarwidth}\tabularnewline
Leverage & \begin{cellvarwidth}[t]
\centering
0.0004\\
(0.0024)
\end{cellvarwidth} & \begin{cellvarwidth}[t]
\centering
0.0261\\
(0.0044)
\end{cellvarwidth}\tabularnewline
Export Status & \begin{cellvarwidth}[t]
\centering
0.0098\\
(0.0010)
\end{cellvarwidth} & \begin{cellvarwidth}[t]
\centering
0.0023\\
(0.0018)
\end{cellvarwidth}\tabularnewline
Industrial Park & \begin{cellvarwidth}[t]
\centering
-0.0014\\
(0.0013)
\end{cellvarwidth} & \begin{cellvarwidth}[t]
\centering
-0.0043\\
(0.0026)
\end{cellvarwidth}\tabularnewline
State Ownership & \begin{cellvarwidth}[t]
\centering
0.0050\\
(0.0040)
\end{cellvarwidth} & \begin{cellvarwidth}[t]
\centering
-0.0534\\
(0.0076)
\end{cellvarwidth}\tabularnewline
\bottomrule
\end{tabular}\begin{tablenotes}[flushleft]
\scriptsize
\item \textit{Notes:} 

(i) Size is defined as the natural logarithm of total average assets. Leverage is measured as the debt-to-asset ratio. Exporting firms are defined as those with positive export delivery value. Industrial park firms are identified based on firm address information containing keywords such as ``development zone'' or ``industrial park.'' State-owned enterprises (SOEs) are defined as firms whose state capital accounts for more than 50 percent of total paid-in capital. 

(ii) The variables are measured as follows: levels for the elasticities; the natural logarithm of 10,000 yuan for firm size; a ratio for leverage; and binary indicators (0/1) for the last three variables in the table. 

(iii) The numbers in parentheses in the second row of each cell represent the standard errors.

\end{tablenotes}
\end{threeparttable}
\end{table}

The results summarized in Table \ref{tab:reg-beta-firm-obs} show
that larger firms have significantly higher capital elasticities,
while leverage is positively associated with labor elasticities. Exporters
exhibit modestly higher capital elasticities, and state-owned enterprises
have substantially lower labor elasticities. Conditional on these
controls, the industrial-park indicator is small. Industry and location
fixed effects are jointly significant, indicating that sectoral and
regional factors account for an important share of the heterogeneity.
The adjusted-$R^{2}$s are 0.58 and 0.57 for $\widehat{\b}^{K}$ and
$\widehat{\b}^{L}$, respectively, suggesting that over half of the
variation in estimated elasticities can be explained by observable
firm characteristics and fixed effects.

The results are generally intuitive and broadly consistent with existing
empirical evidence. Larger firms, for instance, often operate in more
capital-intensive segments or adopt technologies with stronger scale
economies, automation, and standardized production processes, so output
tends to be more responsive to capital at the margin (\citet{Syverson2011JEL}).
Leverage may also be associated with more labor-intensive operating
choices---for example, delaying or outsourcing capital investment,
relying on older or rented equipment, or placing greater emphasis
on variable inputs (\citet*{KalemliOzcanLaevenMoreno2022JEEA}). To
the extent that highly leveraged firms substitute away from capital
toward labor, the estimated production relationship will place greater
weight on labor. Finally, exporters tend to use capital more efficiently
(\citet{BernardJensen1999JIE}), whereas state-owned enterprises may
employ labor in excess of the efficient level due to political considerations
(\citet{Wen2025RESTAutocraticControl}), which is consistent with
their substantially lower labor elasticities relative to private firms.

\section{Conclusion\label{sec:Conclusion}}

This paper proposes a new TERC model in which regressors are correlated
with random coefficients through not only a fixed effect but also
a time-varying shock---an empirically relevant feature consistent
with optimizing behavior in many applications. I construct feasible
control variables for both the fixed effect and the time-varying shock,
and use the resulting residual variation in regressors to identify
the APE and LAR. I then develop three-step series estimators and establish
their convergence rates and asymptotic normality. In an application
to Chinese manufacturing, the estimates reveal substantial cross-firm
dispersion in output elasticities, part of which is explained by observable
firm characteristics.

I propose two directions for future research. First, beyond the low-order
moments studied here, policymakers may be interested in the full distribution
of random coefficients. One---admittedly demanding---route is to
identify moments of all orders by induction and recover the distribution
via moment determinacy (\citet{stoyanov2000krein}). Second, it remains
open whether elements of the approach can be extended to dynamic linear
or nonlinear panel models, such as those in \citet*{marx2024heterogeneous}
and \citet*{liu2024identification}. Doing so would likely require
additional structure governing how lagged outcomes (or state variables)
co-move with the random coefficients.

\newpage{}

\bibliographystyle{ecta}
\bibliography{CRC_PFE_LMU}

\newpage{}

\appendix
\counterwithin{table}{part}
\setcounter{table}{0}
\renewcommand{\thetable}{A.\arabic{table}}

\part*{Appendix}

Appendix \ref{sec:Proofs-addl-theory} provides proofs for all theorems
and lemmas in the main text, along with additional theoretical results.
Appendix \ref{sec:Additional-Empirical-Results} reports additional
empirical results. Appendix \ref{sec:Simulation} presents the simulation
results. 

\section{Proofs and Supplementary Theoretical Results\label{sec:Proofs-addl-theory}}

Appendix \ref{subsec:Proofs} contains proofs of all the lemmas and
theorems in the main text. Appendix \ref{subsec:Prop-1} introduces
Proposition \ref{prop:sufficient condition for assu 2}, which provides
sufficient conditions for Assumption \ref{assu:index exclusion} by
adapting the nonparametric exchangeability condition of \citet{altonji2005cross}.
Appendix \ref{subsec:As245-Discuss} further discusses Assumptions
\ref{assu:index exclusion}, \ref{assu:residual variation in X},
and \ref{assu:more_var_X}. 

\subsection{Proofs\label{subsec:Proofs}}
\begin{proof}[\textbf{Proof of Theorem \ref{thm:main id}}]
\label{proof sufficiency A} I first prove that $V_{it}$ is a control
for $\eta_{it}$ given $A_{i}$ and $W_{i}$. Then, I show $\E\left[\rest{\b_{it}}X_{it},V_{it},W_{i}\right]$
does not depend on $X_{it}$ via the law of iterated expectations.
Finally, I identify $\E\b_{it}$ and $\E\left[\rest{\b_{it}}X_{it}\right]$
by leveraging the residual variation in $X_{it}$ given $V_{it}$
and $W_{i}$. I present how the inclusion of exogenous shocks $\upsilon_{it}$
and $\widetilde{\epsilon}_{it}$ as well as exogenous coordinates
in the regressors affects the analysis at the end of this proof.

Without loss of generality, I assume that $d_{\eta}=d_{X}$ and that
each coordinate of $\eta_{it}$ enters the corresponding coordinate
of $X_{it}$.\footnote{This is without loss of generality because under Assumption \ref{assu:mono},
I can always redefine $\widetilde{\eta}$ to be the vector that collects
the coordinates of $\eta$ that enters each coordinate of $g$ function.} By Assumption \ref{assu:index exclusion}, I have $\rest{A_{i}\perp\left(X_{it},Z_{it}\right)}W_{i}$,
which implies $\rest{X_{it}\perp A_{i}}\left(Z_{it},W_{i}\right)$.
Thus, for each $l\in\left\{ 1,\ldots,d_{X}\right\} $ and any on-support
$\left(x_{l},z,a,w\right)$, I have
\begin{align}
 & \ F_{\rest{X_{it,l}}Z_{it},W_{i}}\left(\rest{x_{l}}z,w\right)\nonumber \\
= & \ F_{\rest{X_{it,l}}Z_{it},A_{i},W_{i}}\left(\rest{x_{l}}z,a,w\right)\nonumber \\
= & \ \P\left(\rest{g_{l}\left(z,a,\eta_{it,l}\right)\leq x_{l}}Z_{it}=z,A_{i}=a,W_{i}=w\right)\nonumber \\
= & \ \P\left(\rest{\eta_{it,l}\leq g_{l}^{-1}\left(x_{l},z,a\right)}A_{i}=a,W_{i}=w\right)\nonumber \\
= & \ F_{\rest{\eta_{it,l}}A_{i},W_{i}}\left(\rest{g_{l}^{-1}\left(x_{l},z,a\right)}a,w\right),\label{eq:cond_U}
\end{align}
where the first equality holds by $\rest{X_{it}\perp A_{i}}\left(Z_{it},W_{i}\right)$,
the second uses (\ref{eq:main eqn 2 X}), the third holds by Assumptions
\ref{assu:mono} and \ref{assu:control for eta}\ref{enu:assu 3 Z indep eta ve},
and the last holds by definition. By \eqref{eq:main eqn 2 X}, the
random variable $\eta_{it,l}=g_{l}^{-1}\left(X_{it,l},Z_{it},A_{i}\right)$
for each $l$, so that plugging in gives
\begin{equation}
V_{it,l}\coloneqq F_{\rest{X_{it,l}}Z_{it},W_{i}}\left(\rest{X_{it,l}}Z_{it},W_{i}\right)=F_{\rest{\eta_{it,l}}A_{i},W_{i}}\left(\rest{\eta_{it,l}}A_{i},W_{i}\right),\label{eq:Vitequalsccdf}
\end{equation}
which establishes a one-to-one mapping to $\eta_{it,l}$ given $A_{i}$
and $W_{i}$ under Assumption \ref{assu:control for eta}\ref{enu:assu 3 control eta CDF mono}.
Let $V_{it}\coloneqq(V_{it,1},\dots,V_{it,d_{X}})'$. By \eqref{eq:Vitequalsccdf},
$V_{it}$ uniquely determines the vector of $\eta_{it}$ given $A_{i}$
and $W_{i}$.

Next, I have 
\begin{align}
 & \ \E\left[\rest{\b_{it}}X_{it},A_{i},V_{it},W_{i}\right]\nonumber \\
= & \ \E\left[\rest{\b\left(A_{i},\ve_{it}\right)}g\left(Z_{it},A_{i},\eta_{it}\right),A_{i},V_{it},W_{i}\right]\nonumber \\
= & \ \E\left[\rest{\b\left(A_{i},\ve_{it}\right)}A_{i},V_{it},W_{i}\right]\nonumber \\
\eqqcolon & \ b_{2}\left(A_{i},V_{it},W_{i}\right),\label{eq:EbgivenXAVW}
\end{align}
where the second equality holds because conditioning on $\left(A_{i},V_{it},W_{i}\right)$
is equivalent to fixing $\left(A_{i},\eta_{it},W_{i}\right)$ by \eqref{eq:Vitequalsccdf},
and thus the residual variation in $X_{it}$ is driven solely by $Z_{it}$
which is independent of $\ve_{it}$ given $\left(A_{i},\eta_{it},W_{i}\right)$
by Assumption \ref{assu:control for eta}\ref{enu:assu 3 Z indep eta ve}.
I exclude $X_{it}$ from the conditioning set of $\E\left[\rest{\b_{it}}X_{it},A_{i},V_{it},W_{i}\right]$
and write it out as $b_{2}\left(A_{i},V_{it},W_{i}\right)$. 

Then, by \eqref{eq:EbgivenXAVW}, I have
\begin{align}
 & \ \E\left[\rest{\b_{it}}X_{it},V_{it},W_{i}\right]\nonumber \\
= & \ \E\left[\rest{b_{2}\left(A_{i},V_{it},W_{i}\right)}X_{it},V_{it},W_{i}\right]\nonumber \\
= & \ \E\left[\rest{\E\left[\rest{b_{2}\left(A_{i},V_{it},W_{i}\right)}X_{it},Z_{it},V_{it},W_{i}\right]}X_{it},V_{it},W_{i}\right]\nonumber \\
= & \ \E\left[\rest{\E\left[\rest{b_{2}\left(A_{i},V_{it},W_{i}\right)}X_{it},Z_{it},W_{i}\right]}X_{it},V_{it},W_{i}\right]\nonumber \\
= & \ \E\left[\rest{\int b_{2}\left(a,V_{it},W_{i}\right)f_{\rest{A_{i}}X_{it},Z_{it},W_{i}}\left(\rest aX_{it},Z_{it},W_{i}\right)\mu\left(da\right)}X_{it},V_{it},W_{i}\right]\nonumber \\
= & \ \E\left[\rest{\int b_{2}\left(a,V_{it},W_{i}\right)f_{\rest{A_{i}}W_{i}}\left(\rest aW_{i}\right)\mu\left(da\right)}X_{it},V_{it},W_{i}\right]\nonumber \\
\eqqcolon & \ b_{1}\left(V_{it},W_{i}\right),\label{eq:EbetagivenXVW}
\end{align}
where the first and second equalities hold by the law of iterated
expectations, the third holds because $V_{it}$ is a measurable function
of $X_{it},\ Z_{it},$ and $W_{i}$, all of which are also conditioned
on, and the fifth equality holds by $\rest{\left(X_{it},Z_{it}\right)\perp A_{i}}W_{i}$
by Assumption \ref{assu:index exclusion}. 

Finally, given \eqref{eq:EbetagivenXVW}, I have 
\begin{equation}
\E\left[\rest{Y_{it}}X_{it},V_{it},W_{i}\right]=X_{it}'b_{1}\left(V_{it},W_{i}\right).\label{eq:maineqid}
\end{equation}
When Assumption \ref{assu:residual variation in X} is satisfied,
I pre-multiply both sides of \eqref{eq:maineqid} by $X_{it}$ and
take conditional expectation of both sides conditioning on $V_{it}$
and $W_{i}$:
\[
\E\left[\rest{\E\left[\rest{X_{it}Y_{it}}X_{it},V_{it},W_{i}\right]}V_{it},W_{i}\right]=\E\left[\rest{X_{it}X_{it}'}V_{it},W_{i}\right]b_{1}\left(V_{it},W_{i}\right),
\]
which identifies $b_{1}\left(V_{it},W_{i}\right)$
\[
b_{1}\left(V_{it},W_{i}\right)=\left(\E\left[\rest{X_{it}X_{it}'}V_{it},W_{i}\right]\right)^{-1}\E\left[\rest{X_{it}Y_{it}}V_{it},W_{i}\right].
\]
When Assumption \ref{assu:more_var_X} is used, I take partial derivative
of both sides of \eqref{eq:maineqid} with respect to $X_{it}$ and
obtain
\[
b_{1}\left(V_{it},W_{i}\right)=\partial\E\left[\rest{Y_{it}}X_{it},V_{it},W_{i}\right]/\partial X_{it}.
\]

Given $b_{1}\left(V_{it},W_{i}\right)$, I use the law of iterated
expectations to identify $\ol b$ and $b(X_{it})$ by 
\begin{equation}
\ol b=T^{-1}\sum_{t=1}^{T}\E\left[b_{1}\left(V_{it},W_{i}\right)\right]\text{ and }b\left(X_{it}\right)=\E\left[\rest{b_{1}\left(V_{it},W_{i}\right)}X_{it}\right],\label{eq:idlarape}
\end{equation}
respectively.

When ex-post shock $\widetilde{\epsilon}_{it}$ is included additively
in \eqref{eq:Yeq} and $\E\widetilde{\epsilon}_{it}=0$, equations
\eqref{eq:EbetagivenXVW} and \eqref{eq:maineqid} still hold, so
the identification result \eqref{eq:idlarape} holds without any changes.
When ex-post shock $\upsilon_{it}$ is also included in $\b_{it}\coloneqq\b\left(A_{i},\ve_{it},\upsilon_{it}\right)$,
since $b(X_{it})$ is already a time-varying function of $X_{it}$
due to the distribution of $(\ve_{it},\eta_{it})$ being possibly
time-varying, adding $\upsilon_{it}$ which is independent of everything
else does not affect the identification proof.

Finally, when exogenous regressors are included in \eqref{eq:Yeq},
I let $X_{it}=\left(U_{it}',Z_{it,1}'\right)'$ and $Z_{it}=\left(Z_{it,1}',Z_{it,2}'\right)'$,
and rewrite model \eqref{eq:Yeq}--\eqref{eq:main eqn 2 X} to be
\begin{align*}
Y_{it} & =X_{it}'\b\left(A_{i},\ve_{it}\right),\\
U_{it} & =g\left(Z_{it},A_{i},\eta_{it}\right).
\end{align*}
Then, I replace $X_{it}$ by $U_{it}$ in Assumptions \ref{assu:mono}--\ref{assu:control for eta}
as well as in the definition of $V_{it}$, and the analysis goes through
as before. Note that I keep $X_{it}$ in Assumption \ref{assu:residual variation in X}
or \ref{assu:more_var_X} because I need to perturb the whole $X_{it}$
vector instead of just endogenous coordinates $U_{it}$ in \eqref{eq:maineqid}
to identify $b_{1}\left(V_{it},W_{i}\right)$. As discussed in Appendix
\ref{subsec:As245-Discuss}, the residual variation requirement of
Assumption \ref{assu:residual variation in X} or \ref{assu:more_var_X}
is easier to be satisfied when exogenous $Z_{it,1}$ is included in
$X_{it}$ as the support of $Z_{it,2}$ is unaffected by $V_{it}$
and $W_{i}$.
\end{proof}
\begin{proof}[\textbf{Proof of Theorem \ref{thm:conv rate beta x beta bar}}]
\label{proof conv rates} I denote $\sum_{i=1}^{n}$ by $\sum_{i}$
and omit all $t$-subscripts. As the result for a finite-dimensional
vector-valued $\b$ can be established by proving for each of its
coordinates and combining the results using the triangle inequality,
I assume $\b$ is a scalar in this proof. I focus on $\widehat{b}\left(x\right)$,
since the result for $\widehat{\ol b}$ follows immediately by setting
$r^{M_{3}}\equiv1$. Let $p_{i}^{m_{2}}\coloneqq p^{m_{2}}\left(V_{i},W_{i}\right)$,
$r_{i}\coloneqq r^{M_{3}}\left(X_{i}\right)$, and $\ol p_{i}=\ol p^{M_{2}}\left(S_{i}\right)$.
Following \citet{imbens2009identification}, I can normalize $\E p_{i}^{m_{2}}p_{i}^{m_{2}}{}'=I_{m_{2}}$
by Assumption \ref{assu: rates-V and G} and $R\coloneqq\E r_{i}r_{i}'=I_{M_{3}}$
by Assumption \ref{assu: rates beta x beta bar}, which imply $\ol P\coloneqq\E\ol p_{i}\ol p_{i}'=\E\left[I_{d_{X}}\otimes\left(p_{i}^{m_{2}}p_{i}^{m_{2}}{}'\right)\right]=I_{M_{2}}$.
Furthermore, I have $\lambda_{\text{min}}\left(\widehat{R}\right)\geq C>0$
with probability approaching one by \citet{newey1997convergence}.
Let $\widetilde{B}\coloneqq\left(b_{1}\left(\widehat{v}_{1},w_{1}\right),\ldots,b_{1}\left(\widehat{v}_{n},w_{n}\right)\right)'$.

By \eqref{eq: estimator for beta bar},
\begin{align}
 & \ \norm{n\widehat{R}^{1/2}\left(\widehat{\rho}^{M_{3}}-\rho^{M_{3}}\right)}^{2}/4\nonumber \\
\leq & \ \left(\widehat{B}-\widetilde{B}\right)'r\widehat{R}^{-1}r'\left(\widehat{B}-\widetilde{B}\right)+\left(\widetilde{B}-B\right)'r\widehat{R}^{-1}r'\left(\widetilde{B}-B\right)\nonumber \\
 & \ +\left(B-B^{X}\right)'r\widehat{R}^{-1}r'\left(B-B^{X}\right)+\left(B^{X}-R\rho^{M_{3}}\right)'r\widehat{R}^{-1}r'\left(B^{X}-r\rho^{M_{3}}\right),\label{eq:expan norm}
\end{align}
where $B\coloneqq\left(b_{1}\left(v_{1},w_{1}\right),\ldots,b_{1}\left(v_{n},w_{n}\right)\right)'$
and $B^{X}\coloneqq\left(b\left(x_{1}\right),\ldots,b\left(x_{n}\right)\right)'$.
Hence, $\xi=B-B^{X}$. I analyze the RHS of \eqref{eq:expan norm}
term by term.

By Lemma S.5 of \citet{imbens2009identification}, $\norm{n^{-1}\sum_{i}\widehat{\overline{p}}_{i}\widehat{\overline{p}}_{i}'-I}=o_{p}\left(1\right)$.
Then,
\begin{align}
 & \ n^{-2}\left(\widehat{B}-\widetilde{B}\right)'r\widehat{R}^{-1}r'\left(\widehat{B}-\widetilde{B}\right)\leq Cn^{-1}\left(\widehat{B}-\widetilde{B}\right)'\left(\widehat{B}-\widetilde{B}\right)\nonumber \\
= & \ Cn^{-1}\sum_{i}\left(\widehat{\overline{p}}_{i}'\left(\widehat{\alpha}^{M_{2}}-\alpha^{M_{2}}\right)+\left(\widehat{\overline{p}}_{i}'\alpha^{M_{2}}-b_{1}\left(\widehat{v}_{i},w_{i}\right)\right)\right)^{2}\nonumber \\
\leq & \ C\norm{\widehat{\alpha}^{M_{2}}-\alpha^{M_{2}}}^{2}+C\sup_{s\in\mathcal{S}}\norm{\ol p^{M_{2}}\left(s\right)'\alpha^{M_{2}}-b_{1}\left(v,w\right)}^{2}=O_{p}\left(\Delta_{2n}^{2}\right),\label{eq:beta vw hat}
\end{align}
where the first inequality holds because $n^{-1}r\widehat{R}^{-1}r'$
is idempotent, the last inequality holds by the Cauchy--Schwarz inequality
(CS) and $\norm{n^{-1}\sum_{i}\widehat{\overline{p}}_{i}\widehat{\overline{p}}_{i}'-I}=o_{p}\left(1\right)$,
and the last equality uses Lemma \ref{lem: rates of V and G}.

Next,
\begin{align}
n^{-2}\left(\widetilde{B}-B\right)'r\widehat{R}^{-1}r'\left(\widetilde{B}-B\right)\leq & \ Cn^{-1}\sum_{i}\left(b_{1}\left(\widehat{v}_{i},w_{i}\right)-b_{1}\left(v_{i},w_{i}\right)\right)^{2}\nonumber \\
\leq & \ Cn^{-1}\sum_{i}\left(\widehat{v}_{i}-v_{i}\right)^{2}=O_{p}\left(\D_{1n}^{2}\right),\label{eq:b tilda - b vw}
\end{align}
where the last inequality holds by the mean value theorem and Assumption
\ref{assu: rates-V and G} and the equality holds by Lemma \ref{lem: rates of V and G}
and Markov's inequality.

Finally, for the last two terms on the right-hand side of \eqref{eq:expan norm},
\begin{align*}
 & \ n^{-2}\E\left[\rest{\left(B-B^{X}\right)'r\widehat{R}^{-1}r'\left(B-B^{X}\right)}{\bf X}\right]\\
= & \ n^{-2}\text{tr}\left\{ \E\left[\rest{\xi'r\widehat{R}^{-1}r'\xi}{\bf X}\right]\right\} =n^{-2}\text{tr}\left\{ \E\left[\rest{\xi\xi'}{\bf X}\right]r\widehat{R}^{-1}r'\right\} \\
\leq & \ n^{-2}\text{tr}\left\{ CIr\widehat{R}^{-1}r'\right\} =Cn^{-1}\text{tr}\left\{ \widehat{R}^{-1}\widehat{R}\right\} =Cn^{-1}M_{3},
\end{align*}
and 
\begin{align}
n^{-2}\left(B^{X}-R\rho^{M_{3}}\right)'r\widehat{R}^{-1}r'\left(B^{X}-R\rho^{M_{3}}\right)\leq & \ n^{-1}\norm{B^{X}-R\rho^{M_{3}}}^{2}=O_{p}\left(M_{3}^{-2d_{3}/j_{3}}\right).\label{eq:-3}
\end{align}

By $\lambda_{\text{min}}\left(\widehat{R}\right)\geq C>0$ and conditional
Markov's (CM) inequality (i.e., $\E\left[\rest{\abs{Y_{n}}}Z_{n}\right]=O_{p}\left(r_{n}\right)$
implies $Y_{n}=O_{p}\left(r_{n}\right)$),
\begin{equation}
\norm{\widehat{\rho}^{M_{3}}-\rho^{M_{3}}}^{2}=O_{p}\left(\D_{2n}^{2}+n^{-1}M_{3}+M_{3}^{-2d_{3}/j_{3}}\right)\eqqcolon O_{p}\left(\D_{3n}^{2}\right),\label{eq:eta M conv rate}
\end{equation}
which yields
\begin{align*}
\int\norm{\widehat{b}\left(x\right)-b\left(x\right)}^{2}dF\left(x\right)\leq & \int\left(r^{M_{3}}\left(x\right)'\left(\widehat{\rho}^{M_{3}}-\rho^{M_{3}}\right)+\left(r^{M_{3}}\left(x\right)'\rho^{M_{3}}-b\left(x\right)\right)\right)^{2}dF\left(x\right)\\
\leq & \ 2\norm{\widehat{\rho}^{M_{3}}-\rho^{M_{3}}}^{2}+2\sup_{x\in\mathcal{X}}\abs{b\left(x\right)-r^{M_{3}}\left(x\right)'\rho^{M_{3}}}^{2}=O_{p}\left(\D_{3n}^{2}\right),
\end{align*}
and
\begin{align*}
\sup_{x\in\mathcal{X}}\norm{\widehat{b}\left(x\right)-b\left(x\right)} & \leq\sup_{x\in\mathcal{X}}\norm{r^{M_{3}}\left(x\right)}\norm{\widehat{\rho}^{M_{3}}-\rho^{M_{3}}}+\sup_{x\in\mathcal{X}}\abs{b\left(x\right)-r^{M_{3}}\left(x\right)'\rho^{M_{3}}}\\
 & =O_{p}\left(\zeta\left(M_{3}\right)\D_{3n}\right).\tag*{\qedhere}
\end{align*}
\end{proof}
\begin{proof}[\textbf{Proof of Lemma \ref{lem: bvw normality}}]
\label{proof normality beta vw} Define 
\begin{align}
\O_{1} & \coloneqq\ol p^{M_{2}}\left(v,w\right)'P^{-1}\left(\Sigma+\Sigma_{1}\right)P^{-1}\ol p^{M_{2}}\left(v,w\right),\ \Sigma\coloneqq\E p_{i}p_{i}'u_{i}^{2},\label{eq:def_Omega_1}\\
\Sigma_{1} & \coloneqq\E\ol{\mu}_{i}^{I}\ol{\mu}_{i}^{I'},\ p_{i}\coloneqq p^{M_{2}}\left(S_{i}\right),\ q_{i}\coloneqq q^{M_{1}}\left(X_{i},Z_{i},W_{i}\right),\nonumber \\
\ol{\mu}_{i}^{I} & \coloneqq\E\left[\rest{G_{V}\left(S_{j}\right)\tau_{V}\left(V_{j}\right)p_{j}q_{j}'Q^{-1}q_{i}v_{ji}}\mathcal{I}_{i}\right],\ u_{i}\coloneqq Y_{i}-G\left(S_{i}\right),\nonumber \\
G_{V}\left(S_{j}\right) & \coloneqq\rest{\partial G\left(s\right)/\partial v}_{s=S_{j}},\text{ and }v_{ji}\coloneqq\ind\left\{ x_{i}\leq x_{j}\right\} -F\left(\rest{x_{j}}z_{i},w_{i}\right).\nonumber 
\end{align}
and 
\begin{align}
\widehat{\Omega}_{1} & \coloneqq\ol p^{M_{2}}\left(v,w\right)'\widehat{P}^{-1}\left(\widehat{\Sigma}+\widehat{\Sigma}_{1}\right)\widehat{P}^{-1}\ol p^{M_{2}}\left(v,w\right),\label{eq:def_Omega_1_hat}\\
\widehat{\Sigma} & \coloneqq n^{-1}\sum_{i=1}^{n}\widehat{p}_{i}\widehat{p}_{i}'\left(y_{i}-\widehat{G}\left(\widehat{s}_{i}\right)\right)^{2},\ \widehat{\Sigma}_{1}\coloneqq n^{-1}\sum_{i=1}^{n}\widehat{\ol{\mu}}_{i}^{I}\widehat{\ol{\mu}}_{i}^{I}{}',\nonumber \\
\widehat{\ol{\mu}}_{i}^{I} & \coloneqq n^{-1}\sum_{j=1}^{n}\widehat{G}_{V}\left(\widehat{s}_{j}\right)\widehat{p}_{j}q_{j}'\widehat{Q}^{-1}q_{i}\widehat{v}_{ji},\text{ and }\widehat{v}_{ji}\coloneqq\left(\ind\left\{ x_{i}\leq x_{j}\right\} -\widehat{F}\left(\rest{x_{j}}z_{i},w_{i}\right)\right).\nonumber 
\end{align}

IN02 have proved asymptotic normality for known and scalar-valued
functionals of $G\left(s\right)$. I apply their results to $c'\widehat{b}_{1}\left(v,w\right)$
for any constant vector $c'c=1$ and obtain
\begin{align}
c'\sqrt{n}\Omega_{1}^{-1/2}\left(\widehat{b}_{1}\left(v,w\right)-b_{1}\left(v,w\right)\right) & \dto N\left(0,1\right)\ \text{and}\nonumber \\
\left(c'\Omega_{1}c\right)^{-1}\left[c'\left(\widehat{\Omega}_{1}-\Omega_{1}\right)c\right] & \pto0.\label{eq:bvw step 1}
\end{align}
By \eqref{eq:bvw step 1} and Assumption \ref{assu:bvw normality}\ref{enu:bvw normality  6 jiii of andrews 1991},
\begin{equation}
c'\left(c_{1n}\widehat{\Omega}_{1}-c_{1n}\Omega_{1}\right)c\pto0,\label{eq:bvw step 2 pf}
\end{equation}
which implies
\begin{equation}
c_{1n}\widehat{\Omega}_{1}\pto\ol{\Omega}_{1}.\label{eq:bvw step 3}
\end{equation}
Then,
\begin{align}
 & \ \sqrt{n}\widehat{\Omega}_{1}^{-1/2}\left(\widehat{b}_{1}\left(v,w\right)-b_{1}\left(v,w\right)\right)\nonumber \\
= & \ \left(c_{1n}\widehat{\Omega}_{1}\right)^{-1/2}\left(c_{1n}\Omega_{1}\right)^{1/2}\sqrt{n}\Omega_{1}^{-1/2}\left(\widehat{b}_{1}\left(v,w\right)-b_{1}\left(v,w\right)\right)\nonumber \\
\dto & \ \ol{\Omega}_{1}^{-1/2}\ol{\Omega}_{1}^{1/2}N\left(0,I\right)=_{d}N\left(0,I\right),\label{eq:bvw normality}
\end{align}
where the convergence holds by \eqref{eq:bvw step 1}, the Cram\'er--Wold
device, \eqref{eq:bvw step 3}, Assumption \ref{assu:bvw normality},
and the continuous mapping theorem.
\end{proof}
\begin{proof}[\textbf{Proof of Theorem \ref{thm:bx and bbar normality}}]
\label{proof normality beta x} Similarly to the proof of Lemma \ref{lem: bvw normality},
by the Cram\'er--Wold device it suffices to consider the case when
$b\left(x\right)$ is a scalar. First, I derive the influence functions
for $\widehat{b}\left(x\right)$ that correctly account for the estimation
errors from each step and prove its asymptotic normality. Then, I
show consistency for the estimator of the variance of $\widehat{b}\left(x\right)$.
I write $r^{M_{3}}\left(x\right)$ as $r\left(x\right)$ when there
is no confusion. The proof relies on the results from IN02, and I
point out differences.

Define 
\begin{align}
\Omega_{2} & \coloneqq\Omega_{21}+\Omega_{22},\text{ where}\nonumber \\
\Omega_{21} & \coloneqq\E\left(A_{1}P^{-1}p_{i}u_{i}\right)\left(A_{1}P^{-1}p_{i}u_{i}\right)',\nonumber \\
\O_{22} & \coloneqq\E\left(A_{1}P^{-1}\ol{\mu}_{i}^{I}-r\left(x\right)'\left(\ol{\mu}_{i}^{II}+r_{i}\xi_{i}\right)\right)\left(A_{1}P^{-1}\ol{\mu}_{i}^{I}-r\left(x\right)'\left(\ol{\mu}_{i}^{II}+r_{i}\xi_{i}\right)\right)',\nonumber \\
A_{1} & \coloneqq r\left(x\right)'\E r_{i}\ol p_{i}',\nonumber \\
\ol{\mu}_{i}^{II} & \coloneqq\E\left[\rest{r_{j}b_{1,V}\left(V_{j},W_{j}\right)q_{j}'Q^{-1}q_{i}v_{ji}}\mathcal{I}_{i}\right],\text{ and}\nonumber \\
b_{1,V}\left(V_{j},W_{j}\right) & \coloneqq\rest{\partial b_{1}\left(v,w\right)/\partial v}_{v=V_{j},w=W_{j}}.\label{eq:def_Omega_2}
\end{align}

Let $F\coloneqq\Omega_{2}^{-1/2}$, which is well-defined because
$\Omega_{2}=\Omega_{21}+\Omega_{22}$ and
\begin{align}
\Omega_{21} & =A_{1}P^{-1}\left(\E p_{i}p_{i}'u_{i}^{2}\right)P^{-1}A_{1}'\nonumber \\
 & =A_{1}P^{-1}\left(\E\left[p_{i}p_{i}'\E\left(\rest{u_{i}^{2}}X_{i},Z_{i},W_{i}\right)\right]\right)P^{-1}A_{1}'\nonumber \\
 & \geq CA_{1}P^{-1}A_{1}'=Cr\left(x\right)'\left(\E r_{i}\ol p_{i}'\right)P^{-1}\left(\E\ol p_{i}r_{i}'\right)r\left(x\right)>0,\label{eq: F well defined}
\end{align}
where the first inequality holds by Assumption \ref{assu:bvw normality}\ref{enu:var y bdd below}
and the last inequality holds by Assumption \ref{assu:bx and bbar normality}\ref{enu:as on Erp}.

Define functionals: 
\begin{align*}
\widehat{a}\left(\widehat{b}_{1},\widehat{V}\right) & \coloneqq\widehat{\E}\left[\rest{\widehat{b}_{1}\left(\widehat{V},W\right)}X=x\right]=\widehat{b}\left(x\right),\\
\widehat{a}\left(b_{1},\widehat{V}\right) & \coloneqq\widehat{\E}\left[\rest{b_{1}\left(\widehat{V},W\right)}X=x\right],\\
\widehat{a}\left(b_{1},V\right) & \coloneqq\widehat{\E}\left[\rest{b_{1}\left(V,W\right)}X=x\right],\text{ and}\\
a\left(b_{1},V\right) & =\E\left[\rest{b_{1}\left(V,W\right)}X=x\right]=b\left(x\right).
\end{align*}
I expand 
\begin{align}
 & \ \sqrt{n}F\left(\widehat{a}\left(\widehat{b}_{1},\widehat{V}\right)-a\left(b_{1},V\right)\right)\nonumber \\
= & \ \sqrt{n}F\left(\widehat{a}\left(\widehat{b}_{1},\widehat{V}\right)-\widehat{a}\left(b_{1},\widehat{V}\right)+\widehat{a}\left(b_{1},\widehat{V}\right)-\widehat{a}\left(b_{1},V\right)+\widehat{a}\left(b_{1},V\right)-a\left(b_{1},V\right)\right)\nonumber \\
= & \ n^{-1/2}\sum_{i}\left(\psi_{1i}+\psi_{2i}+\psi_{3i}\right)+o_{p}\left(1\right),\label{eq:expand a hat}
\end{align}
and show that 
\begin{align}
\psi_{1i} & =H_{1}\left(p_{i}u_{i}-\text{\ensuremath{\ol{\mu}_{i}^{I}}}\right),\ \psi_{2i}=H_{2}\ol{\mu}_{i}^{II},\text{ and }\psi_{3i}=H_{2}r_{i}\xi_{i},\label{eq:psi 1i - 3i-1}
\end{align}
where $H_{1}\coloneqq FA_{1}P^{-1}$, $A_{1}=r\left(x\right)'R^{-1}\E r_{i}\ol p_{i}'=r\left(x\right)'\E r_{i}\ol p_{i}'$,
$r_{i}\coloneqq r\left(x_{i}\right)$, $P\coloneqq\E p_{i}p_{i}'$,
$p_{i}\coloneqq p^{M_{2}}\left(X_{i},V_{i},W_{i}\right)$, $u_{i}\coloneqq y_{i}-G\left(s_{i}\right)$,
$\text{\ensuremath{\ol{\mu}_{i}^{I}}}\coloneqq\E\left[\rest{G_{V}\left(S_{j}\right)\tau_{V}\left(V_{j}\right)p_{j}q_{j}'Q^{-1}q_{i}v_{ji}}\mathcal{I}_{i}\right],$
$G_{V}\left(S_{j}\right)\coloneqq\rest{\partial G\left(s\right)/\partial v}_{s=S_{j}},$
$v_{ji}\coloneqq\ind\left\{ x_{i}\leq x_{j}\right\} -F\left(\rest{x_{j}}z_{i},w_{i}\right)$,
$H_{2}\coloneqq FA_{2}R^{-1}=FA_{2}$, $A_{2}\coloneqq r\left(x\right)'$,
$\ol{\mu}_{i}^{II}\coloneqq\E\left[\rest{r_{j}b_{1,V}\left(V_{j},W_{j}\right)q_{j}'Q^{-1}q_{i}v_{ji}}\mathcal{I}_{i}\right]$,
$b_{1,V}\left(V_{j},W_{j}\right)=\rest{\partial b_{1}\left(v,w\right)/\partial v}_{v=V_{j}},$
and $\xi_{i}=b_{1}\left(v_{i},w_{i}\right)-b\left(x_{i}\right)$.

Let $\widehat{H}_{1}\coloneqq F\widehat{A}_{1}\widehat{P}^{-1}$,
$\widehat{A}_{1}\coloneqq r\left(x\right)'\widehat{R}^{-1}n^{-1}\sum_{i=1}^{n}r_{i}\widehat{\ol p}_{i}'$,
$\widehat{R}\coloneqq n^{-1}\sum_{i=1}^{n}r_{i}r_{i}'$, $\widehat{\ol p}_{i}\coloneqq\ol p\left(\widehat{s}_{i}\right)$,
$\widehat{H}_{2}\coloneqq FA_{2}\widehat{R}^{-1}$, $G\coloneqq\left(G\left(s_{1}\right),\ldots,G\left(s_{n}\right)\right)'$,
and $\widetilde{G}\coloneqq\left(G\left(\widehat{s}_{1}\right),\ldots,G\left(\widehat{s}_{n}\right)\right)'$.
First, for $\psi_{1i}$,
\begin{align}
 & \ \sqrt{n}F\left(\widehat{a}\left(\widehat{b}_{1},\widehat{V}\right)-\widehat{a}\left(b_{1},\widehat{V}\right)\right)\nonumber \\
= & \ n^{-1/2}Fr\left(x\right)'\widehat{R}^{-1}r'\left(\widehat{B}-\widetilde{B}\right)\nonumber \\
= & \ n^{-1/2}Fr\left(x\right)'\widehat{R}^{-1}r'\left(n^{-1}\widehat{\ol p}\widehat{P}^{-1}\widehat{p}'Y-\widetilde{B}\right)\nonumber \\
= & \ n^{-1/2}Fr\left(x\right)'\widehat{R}^{-1}r'\left[n^{-1}\widehat{\ol p}\widehat{P}^{-1}\widehat{p}'\left(Y-G+G-\widetilde{G}+\widetilde{G}-\widehat{p}\ol{\alpha}^{M_{2}}\right)+\left(\widehat{\ol p}\ol{\alpha}^{M_{2}}-\widetilde{B}\right)\right]\nonumber \\
= & \ n^{-1/2}\sum_{i}\widehat{H}_{1}\widehat{p}_{i}\left[u_{i}-\left(G\left(\widehat{s}_{i}\right)-G\left(s_{i}\right)\right)\right]+n^{-1/2}\widehat{H}_{1}\widehat{p}'\left(\widetilde{G}-\widehat{p}\ol{\alpha}^{M_{2}}\right)\nonumber \\
 & \ +n^{-1/2}\widehat{H}_{2}r'\left(\widehat{\ol p}\ol{\alpha}^{M_{2}}-\widetilde{B}\right)\nonumber \\
\eqqcolon & \ D_{11}+D_{12}+D_{13}.\label{eq: psi 1i expansion}
\end{align}
I show $D_{11}=n^{-1/2}\sum_{i}\psi_{1i}+o_{p}\left(1\right)$, $D_{12}=o_{p}\left(1\right)$,
and $D_{13}=o_{p}\left(1\right)$.

The proof of
\begin{equation}
D_{11}=n^{-1/2}\sum_{i}\psi_{1i}+o_{p}\left(1\right)\label{eq:D11}
\end{equation}
is analogous to that of Lemma B7 and B8 of IN02, except that I need
to establish $\norm{\widehat{H}_{1}-H_{1}}=o_{p}\left(1\right)$ and
$\norm{\widehat{H}_{2}-H_{2}}=o_{p}\left(1\right)$. To prove these
two claims, notice that 
\begin{equation}
\norm{H_{1}}=O\left(1\right)\text{ and }\norm{H_{2}}=O\left(1\right),\label{eq:H1 H2 H1 hat H2 hat}
\end{equation}
because $\norm{H_{1}}^{2}\leq CA_{1}A_{1}'/\Omega_{2}\leq C$ and
$\norm{H_{2}}^{2}=A_{2}A_{2}'/\Omega_{2}\leq CA_{1}A_{1}'/\Omega_{2}\leq C$.
In addition, $\norm{\widehat{P}-P}=o_{p}\left(1\right)$, $\norm{\widehat{R}-I}=o_{p}\left(1\right)$,
and $\norm{n^{-1}\sum_{i}r_{i}\widehat{\ol p}_{i}'-\E r_{i}\ol p_{i}'}=o_{p}\left(1\right)$
by Lemma A3 of IN02 and \citet{newey1997convergence}. Furthermore,
$\lambda_{\min}\left(P\right)\geq C>0$ and $\norm P=O\left(1\right)$
by Assumption \ref{assu: rates-V and G}. By Slutsky's theorem, $\norm{\widehat{R}^{-1}-I}=o_{p}\left(1\right)$
and $\norm{\widehat{P}^{-1}-P^{-1}}=o_{p}\left(1\right)$. By the
CS inequality and Lemma A3 of IN02,
\begin{align}
\norm{n^{-1}\sum_{i}r_{i}\left(\ol p_{i}-\widehat{\ol p}_{i}\right)'}^{2} & \leq n^{-1}\sum_{i}\norm{r_{i}}^{2}\times n^{-1}\sum_{i}\norm{\widehat{\ol p}_{i}-\ol p_{i}}^{2}\nonumber \\
 & =O_{p}\left(M_{3}\zeta_{1}\left(m_{2}\right)^{2}\D_{1n}^{2}\right)=o_{p}\left(1\right).\label{eq:pi bar ol}
\end{align}
Therefore, by the triangle inequality with probability approaching
one, 
\begin{align}
 & \ \norm{\widehat{H}_{1}-H_{1}}^{2}\nonumber \\
= & \ \norm{F\widehat{A}_{1}\widehat{P}^{-1}-FA_{1}P^{-1}}^{2}\nonumber \\
\leq & \ 2\norm{F\left(\widehat{A}_{1}-A_{1}\right)\widehat{P}^{-1}}^{2}+2\norm{FA_{1}\left(\widehat{P}^{-1}-P^{-1}\right)}^{2}\nonumber \\
= & \ 2\norm{F\left(r\left(x\right)'\left(I+o_{p}\left(1\right)\right)\left(\E r_{i}\ol p_{i}'+o_{p}\left(1\right)\right)-r\left(x\right)'\E r_{i}\ol p_{i}'\right)\widehat{P}^{-1}}^{2}\nonumber \\
 & \ +2\norm{FA_{1}P^{-1}\left(P-\widehat{P}\right)\widehat{P}^{-1}}^{2}\nonumber \\
\leq & \ \norm{H_{2}}^{2}o_{p}\left(1\right)+\norm{H_{1}}^{2}o_{p}\left(1\right)=o_{p}\left(1\right),\label{eq:H1 hat conv to H1}
\end{align}
and similarly $\norm{\widehat{H}_{2}-H_{2}}=o_{p}\left(1\right)$,
which establish \eqref{eq:D11}.

Next, recall that by Assumption \ref{assu:bvw normality}
\begin{align*}
n^{-1}\left(\widetilde{G}-\widehat{p}\ol{\alpha}^{M_{2}}\right)'\left(\widetilde{G}-\widehat{p}\ol{\alpha}^{M_{2}}\right)= & \ O_{p}\left(M_{2}^{-2d_{4}/j_{4}}\right)\text{ and}\\
n^{-1}\left(\widehat{\ol p}\ol{\alpha}^{M_{2}}-\widetilde{B}\right)'\left(\widehat{\ol p}\ol{\alpha}^{M_{2}}-\widetilde{B}\right)= & \ O_{p}\left(M_{2}^{-2d_{4}/j_{4}}\right).
\end{align*}
Therefore, for $D_{12}$, I have
\begin{align}
\abs{n^{-1/2}\widehat{H}_{1}\widehat{p}'\left(\widetilde{G}-\widehat{p}\ol{\alpha}^{M_{2}}\right)}^{2} & \leq n\left[\widehat{H}_{1}\widehat{P}\widehat{H}_{1}'\right]\left[n^{-1}\left(\widetilde{G}-\widehat{p}\ol{\alpha}^{M_{2}}\right)'\left(\widetilde{G}-\widehat{p}\ol{\alpha}^{M_{2}}\right)\right]\nonumber \\
 & \leq\norm{\widehat{H}_{1}}^{2}O_{p}\left(nM_{2}^{-2d_{4}/j_{4}}\right)=o_{p}\left(1\right).\label{eq:D12}
\end{align}

For $D_{13}$, similarly to \eqref{eq:D12}, I have
\begin{align}
\abs{n^{-1/2}\widehat{H}_{2}r'\left(\widehat{\ol p}\ol{\alpha}^{M_{2}}-\widetilde{B}\right)}^{2} & \leq n\left[\widehat{H}_{2}\widehat{R}\widehat{H}_{2}\right]\left[n^{-1}\left(\widehat{\ol p}\ol{\alpha}^{M_{2}}-\widetilde{B}\right)'\left(\widehat{\ol p}\ol{\alpha}^{M_{2}}-\widetilde{B}\right)\right]\nonumber \\
 & \leq\norm{\widehat{H}_{2}}^{2}O_{p}\left(nM_{2}^{-2d_{4}/j_{4}}\right)=o_{p}\left(1\right).\label{eq:D13}
\end{align}

Combining the results for $D_{11}$, $D_{12,}$ and $D_{13}$, I have
\begin{equation}
\psi_{1i}=H_{1}\left(p_{i}u_{i}-\text{\ensuremath{\ol{\mu}_{i}^{I}}}\right).\label{eq:psi 1i result}
\end{equation}

To prove $\psi_{2i}=H_{2}\ol{\mu}_{i}^{II}$, by Taylor expansion
it is true that
\begin{align}
 & \ \sqrt{n}F\left(\widehat{a}\left(b_{1},\widehat{V}\right)-\widehat{a}\left(b_{1},V\right)\right)\nonumber \\
= & \ n^{-1/2}Fr\left(x\right)'\widehat{R}^{-1}r'\left(\widetilde{B}-B\right)\nonumber \\
= & \ \widehat{H}_{2}n^{-1/2}\sum_{i}r_{i}\left(b_{1}\left(\widehat{v}_{i},w_{i}\right)-b_{1}\left(v_{i},w_{i}\right)\right)\nonumber \\
= & \ \widehat{H}_{2}n^{-1/2}\sum_{i}r_{i}b_{1,V}\left(v_{i},w_{i}\right)\left(\widehat{v}_{i}-v_{i}\right)+\widehat{H}_{2}n^{-1/2}\sum_{i}r_{i}b_{1,VV}\left(\widetilde{v}_{i},w_{i}\right)\left(\widehat{v}_{i}-v_{i}\right)^{2}/2\nonumber \\
\eqqcolon & \ D_{21}+D_{22},\label{eq: D21 D22}
\end{align}
where $b_{1,VV}\left(v_{i},w_{i}\right)\coloneqq\rest{\partial^{2}b_{1}\left(v,w\right)/\partial v^{2}}_{v=v_{i},w=w_{i}}$
and $\widetilde{v}_{i}$ lies between $v_{i}$ and $\widehat{v}_{i}$.
I prove $D_{21}=n^{-1/2}\sum_{i}H_{2}\ol{\mu}_{i}^{II}+o_{p}\left(1\right)$
and $D_{22}=o_{p}\left(1\right)$.

For $D_{21}$, 
\begin{align}
D_{21}= & \ \widehat{H}_{2}n^{-1/2}\sum_{i}r_{i}b_{1,V}\left(v_{i},w_{i}\right)\left(\widehat{v}_{i}-v_{i}\right)\nonumber \\
= & \ H_{2}n^{-1/2}\sum_{i}r_{i}b_{1,V}\left(v_{i},w_{i}\right)\D_{i}^{I}+\left(\widehat{H}_{2}-H_{2}\right)n^{-1/2}\sum_{i}r_{i}b_{1,V}\left(v_{i},w_{i}\right)\left(\widehat{v}_{i}-v_{i}\right)\nonumber \\
 & +H_{2}n^{-1/2}\sum_{i}r_{i}b_{1,V}\left(v_{i},w_{i}\right)\left(\D_{i}^{II}+\D_{i}^{III}\right)\nonumber \\
\eqqcolon & \ D_{211}+D_{212}+D_{213},\label{eq:D21 expand}
\end{align}
where
\begin{align}
 & \delta_{ij}=F_{\rest{X_{i}}Z_{j},W_{j}}\left(\rest{x_{i}}z_{j},w_{j}\right)-q_{j}'\gamma^{M_{1}}\left(x_{i}\right),\ \Delta_{i}^{I}=q_{i}'\widehat{Q}^{-1}n^{-1}\sum_{j}q_{j}v_{ij},\nonumber \\
 & \Delta_{i}^{II}=q_{i}'\widehat{Q}^{-1}n^{-1}\sum_{j}q_{j}\delta_{ij},\text{ and }\Delta_{i}^{III}=-\delta_{ii}.\label{eq: D21 components}
\end{align}
Following an argument identical to the proof of Lemma B7 in IN02,
\begin{equation}
D_{211}=n^{-1/2}\sum_{i}H_{2}\ol{\mu}_{i}^{II}+o_{p}\left(1\right).\label{eq:D211}
\end{equation}

For $D_{212}$, 
\begin{align}
\abs{D_{212}}^{2} & \leq Cn\left[\left(\widehat{H}_{2}-H_{2}\right)\widehat{R}\left(\widehat{H}_{2}-H_{2}\right)'\right]\left[n^{-1}\sum_{i}\left(\widehat{v}_{i}-v_{i}\right)^{2}\right]\nonumber \\
 & =O_{p}\left\{ n\left(n^{-1}\zeta\left(M_{3}\right)^{2}M_{3}\right)\D_{1n}^{2}\right\} =o_{p}\left(1\right).\label{eq:D212}
\end{align}

For $D_{213}$, 
\begin{equation}
\abs{D_{213}}^{2}\leq Cn\left[H_{2}\widehat{R}H_{2}'\right]\left[n^{-1}\sum_{i}\left(\left(\D_{i}^{II}\right)^{2}+\left(\D_{i}^{III}\right)^{2}\right)\right]=O_{p}\left(nM_{1}^{1-2d_{1}/j_{1}}\right)=o_{p}\left(1\right),\label{eq:D213}
\end{equation}
where the first equality is established in the proof of Theorem 4
in IN02.

Next, for $D_{22}$, 
\begin{align}
\abs{D_{22}} & \leq C\sqrt{n}\norm{\widehat{H}_{2}}\sup_{x\in\mathcal{X}}\norm{r\left(x\right)}\abs{n^{-1}\sum_{i}\left(\widehat{v}_{i}-v_{i}\right)^{2}}\nonumber \\
 & =O_{p}\left(\sqrt{n}\zeta\left(M_{3}\right)\D_{1n}^{2}\right)=o_{p}\left(1\right).\label{eq:D22}
\end{align}

Combining the results for $D_{21}$ and $D_{22}$,
\begin{equation}
\sqrt{n}F\left(\widehat{a}\left(b_{1},\widehat{V}\right)-\widehat{a}\left(b_{1},V\right)\right)=n^{-1/2}\sum_{i}H_{2}\ol{\mu}_{i}^{II}+o_{p}\left(1\right).\label{eq:D21}
\end{equation}

To show $\psi_{3i}=H_{2}r_{i}\xi_{i}$, I expand 
\begin{align}
 & \ \sqrt{n}F\left(\widehat{a}\left(b_{1},V\right)-a\left(b_{1},V\right)\right)\nonumber \\
= & \ n^{-1/2}\sum_{i}\widehat{H}_{2}r_{i}b_{1}\left(v_{i},w_{i}\right)-\sqrt{n}Fb\left(x\right)\nonumber \\
= & \ n^{-1/2}\sum_{i}H_{2}r_{i}\left(b_{1}\left(v_{i},w_{i}\right)-b\left(x_{i}\right)\right)+n^{-1/2}\sum_{i}\left(\widehat{H}_{2}-H_{2}\right)r_{i}\left(b_{1}\left(v_{i},w_{i}\right)-b\left(x_{i}\right)\right)\nonumber \\
 & +n^{-1/2}\sum_{i}\widehat{H}_{2}r_{i}\left(b\left(x_{i}\right)-r_{i}'\rho^{M_{3}}\right)-\sqrt{n}F\left(b\left(x\right)-r\left(x\right)'\rho^{M_{3}}\right)\nonumber \\
\eqqcolon & \ D_{31}+D_{32}+D_{33}+D_{34},
\end{align}
where I use the definition of $\widehat{H}_{2}$ and $\widehat{R}$
for the decomposition. Recall that $D_{31}=n^{-1/2}\sum_{i}H_{2}r_{i}\xi_{i}$
by the definition of $\xi_{i}$. Thus, I only need to show $D_{32}$,
$D_{33}$, and $D_{34}$ are all $o_{p}\left(1\right)$.

For $D_{32}$, 
\begin{align}
\E\left[\rest{\abs{D_{32}}^{2}}{\bf X}\right] & =n^{-1}\left(\widehat{H}_{2}-H_{2}\right)r'\E\left[\rest{\xi\xi'}{\bf X}\right]r\left(\widehat{H}_{2}-H_{2}\right)'\nonumber \\
 & \leq C\left(\widehat{H}_{2}-H_{2}\right)\widehat{R}\left(\widehat{H}_{2}-H_{2}\right)'\nonumber \\
 & \leq C\norm{\widehat{H}_{2}-H_{2}}^{2}\left(1+\norm{\widehat{R}-I}\right)\nonumber \\
 & =O_{p}\left\{ \norm{\widehat{H}_{2}-H_{2}}^{2}\right\} \nonumber \\
 & =O_{p}\left(n^{-1}\zeta\left(M_{3}\right)^{2}M_{3}\right)=o_{p}\left(1\right),
\end{align}
where the first inequality holds by Assumption \ref{assu: rates beta x beta bar}\ref{enu:rates bx xi bound 2nd order}
and the fact that $\widehat{H}_{2}$ and $r$ are functions of $X$
only, the second equality holds by $\norm{\widehat{R}-I}=o_{p}\left(1\right)$,
and the third equality follows similarly as in equation (A.1) and
(A.6) of \citet{newey1997convergence}. Therefore, $D_{32}=o_{p}\left(1\right)$
by the CM inequality.

For $D_{33}$, by the CS inequality, 
\begin{align}
\abs{D_{33}}^{2} & \leq n\left(\widehat{H}_{2}\widehat{R}\widehat{H}_{2}'\right)n^{-1}\sum_{i}\left(b\left(x_{i}\right)-r_{i}'\rho^{M_{3}}\right)^{2}\nonumber \\
 & =O_{p}\left(nM_{3}^{-2d_{3}/j_{3}}\right)=o_{p}\left(1\right),
\end{align}
where the first equality holds by Assumption \ref{assu: rates beta x beta bar}\ref{enu:approx beta x series}.

Finally, for $D_{34}$, 
\begin{equation}
\abs{D_{34}}^{2}=nF^{2}\left(b\left(x\right)-r\left(x\right)'\rho^{M_{3}}\right)^{2}=O_{p}\left(nM_{3}^{-2d_{3}/j_{3}}\right)=o_{p}\left(1\right).
\end{equation}

Combining the results for $D_{31},$ $D_{32},$ $D_{33}$, and $D_{34}$,
I have
\begin{equation}
\sqrt{n}F\left(\widehat{a}\left(b_{1},V\right)-a\left(b_{1},V\right)\right)=n^{-1/2}\sum_{i}H_{2}r_{i}\xi_{i}+o_{p}\left(1\right).\label{eq:a hat - a third step}
\end{equation}

In sum, I have proved
\begin{align}
\sqrt{n}F\left(\widehat{a}\left(\widehat{b}_{1},\widehat{V}\right)-a\left(b_{1},V\right)\right) & =n^{-1/2}\sum_{i}\left(\psi_{1i}+\psi_{2i}+\psi_{3i}\right)+o_{p}\left(1\right),\label{eq: influence functions}
\end{align}
where
\begin{equation}
\psi_{1i}=H_{1}\left(p_{i}u_{i}-\text{\ensuremath{\ol{\mu}_{i}^{I}}}\right),\ \psi_{2i}=H_{2}\ol{\mu}_{i}^{II},\text{ and }\psi_{3i}=H_{2}r_{i}\xi_{i}.\label{eq:componenet of psis}
\end{equation}
Furthermore, observe that
\begin{equation}
H_{1}p_{i}u_{i}\perp\left(H_{1}\ol{\mu}_{i}^{I},H_{2}\ol{\mu}_{i}^{II},H_{2}r_{i}\xi_{i}\right)\label{eq:independence between psis}
\end{equation}
because $\E\left[\rest{u_{i}}X_{i},V_{i},W_{i}\right]=0$.

Let $\Psi_{in}=n^{-1/2}\left(\psi_{1i}+\psi_{2i}+\psi_{3i}\right)$.
It is clear that $\E\Psi_{in}=0$ and $\V\left(\Psi_{in}\right)=n^{-1}$.
For any $\epsilon>0$, under Assumptions \ref{assu:bvw normality}
and \ref{assu:bx and bbar normality}, 
\begin{align}
 & \ n\E\left[\ind\left\{ \abs{\Psi_{in}}>\epsilon\right\} \Psi_{in}^{2}\right]\nonumber \\
\leq & \ n\epsilon^{2}\E\left[\ind\left\{ \abs{\Psi_{in}}>\epsilon\right\} \left(\Psi_{in}/\epsilon\right)^{4}\right]\leq n\epsilon^{-2}\E\Psi_{in}^{4}\nonumber \\
\leq & \ Cn^{-1}\E\left[\left(H_{1}p_{i}u_{i}\right)^{4}+\left(H_{1}\ol{\mu}_{i}^{I}\right)^{4}+\left(H_{2}\ol{\mu}_{i}^{II}\right)^{4}+\left(H_{2}r_{i}\xi_{i}\right)^{4}\right]\nonumber \\
\leq & \ Cn^{-1}\left(\zeta\left(M_{2}\right)^{2}M_{2}+\zeta\left(M_{2}\right)^{4}\zeta\left(M_{1}\right)^{4}M_{1}+\zeta\left(M_{3}\right)^{4}\zeta\left(M_{1}\right)^{4}M_{1}+\zeta\left(M_{3}\right)^{2}M_{3}\right)\gto0,
\end{align}
where the last inequality holds by Lemma B5 of IN02. 

Then, by the Lindeberg--Feller CLT, 
\begin{equation}
\sqrt{n}\Omega_{2}^{-1/2}\left(\widehat{b}\left(x\right)-b\left(x\right)\right)\dto N\left(0,1\right).\label{eq:LFCLT with Omega 2}
\end{equation}

To construct a feasible confidence interval, one needs a consistent
estimator of the covariance matrix $\O_{2}$. Define 
\begin{align}
\widehat{\Omega}_{2} & \coloneqq\widehat{\Omega}_{21}+\widehat{\Omega}_{22},\text{ where}\nonumber \\
\widehat{\Omega}_{21} & \coloneqq\widehat{A}_{1}\widehat{P}^{-1}\left(n^{-1}\sum_{i}\widehat{p}_{i}\widehat{p}_{i}'\widehat{u}_{i}^{2}\right)\widehat{P}^{-1}\widehat{A}_{1}',\nonumber \\
\widehat{\O}_{22} & \coloneqq n^{-1}\sum_{i}\left(\widehat{A}_{1}\widehat{P}^{-1}\widehat{\ol{\mu}}_{i}^{I}-r\left(x\right)'\widehat{R}^{-1}\left(\widehat{\ol{\mu}}_{i}^{II}+r_{i}\widehat{\xi}_{i}\right)\right)\left(\widehat{A}_{1}\widehat{P}^{-1}\widehat{\ol{\mu}}_{i}^{I}-r\left(x\right)'\widehat{R}^{-1}\left(\widehat{\ol{\mu}}_{i}^{II}+r_{i}\widehat{\xi}_{i}\right)\right)',\nonumber \\
\widehat{A}_{1} & \coloneqq r\left(x\right)'\widehat{R}^{-1}\left(n^{-1}\sum_{i}r_{i}\ol p\left(\widehat{s}_{i}\right)'\right),\nonumber \\
\widehat{\ol{\mu}}_{i}^{II} & \coloneqq n^{-1}\sum_{j}r_{j}\widehat{b}_{1,V}\left(\widehat{v}_{j},w_{j}\right)q_{j}'\widehat{Q}^{-1}q_{i}\widehat{v}_{ji},\text{ and }\nonumber \\
\widehat{\xi}_{i} & \coloneqq\widehat{b}_{1}\left(\widehat{v}_{i},w_{i}\right)-\widehat{b}\left(x_{i}\right).\label{eq:def_Omega_2_hat}
\end{align}
I show $\widehat{\Omega}_{2}/\Omega_{2}-1\pto0$. 

Recall that 
\begin{equation}
\Omega_{2}=\E\left(A_{1}P^{-1}p_{i}u_{i}\right)^{2}+\E\left(A_{1}P^{-1}\ol{\mu}_{i}^{I}-A_{2}\left(\ol{\mu}_{i}^{II}+r_{i}\xi_{i}\right)\right)^{2}=\Omega_{21}+\Omega_{22}\label{eq:Omega 2 def}
\end{equation}
and
\begin{equation}
\widehat{\Omega}_{2}=n^{-1}\sum_{i}\left(\widehat{A}_{1}\widehat{P}^{-1}\widehat{p}_{i}\widehat{u}_{i}\right)^{2}+n^{-1}\sum_{i}\left(\widehat{A}_{1}\widehat{P}^{-1}\widehat{\ol{\mu}}_{i}^{I}-\widehat{A}_{2}\widehat{R}^{-1}\left(\widehat{\ol{\mu}}_{i}^{II}+r_{i}\widehat{\xi}_{i}\right)\right)^{2}\eqqcolon\widehat{\Omega}_{21}+\widehat{\Omega}_{22}.\label{eq:Omega 2 hat}
\end{equation}

The proof of $\widehat{\Omega}_{21}/\Omega_{2}-\Omega_{21}/\Omega_{2}\pto0$
is almost identical to the proof of Lemma B10 of IN02, except that
$\widehat{A}_{1}$ instead of $A_{1}$ appears in the definition of
$\widehat{H}_{1}$. Nonetheless, I have shown $\norm{\widehat{H}_{1}-H_{1}}=o_{p}\left(1\right)$.
Thus, the proof of Lemma B10 of IN02 directly applies.

For $\widehat{\Omega}_{22}$, I need to show
\begin{align}
n^{-1}\sum_{i}\left(\widehat{H}_{1}\widehat{\ol{\mu}}_{i}^{I}-H_{1}\ol{\mu}_{i}^{I}\right)^{2} & =o_{p}\left(1\right),\nonumber \\
n^{-1}\sum_{i}\left(\widehat{H}_{2}\widehat{\ol{\mu}}_{i}^{II}-H_{2}\ol{\mu}_{i}^{II}\right)^{2} & =o_{p}\left(1\right),\text{ and}\nonumber \\
n^{-1}\sum_{i}\left(\widehat{H}_{2}r_{i}\widehat{\xi}_{i}-H_{2}r_{i}\xi_{i}\right)^{2} & =o_{p}\left(1\right).\label{eq:Omega 22 consistency}
\end{align}
The first two convergence results have been established in Lemma B9
of IN02. For the last one, 
\begin{align}
 & \ \widehat{H}_{2}r_{i}\widehat{\xi}_{i}-H_{2}r_{i}\xi_{i}\nonumber \\
= & \ \widehat{H}_{2}r_{i}\left(\widehat{\xi}_{i}-\xi_{i}\right)+\left(\widehat{H}_{2}-H_{2}\right)r_{i}\xi_{i}\nonumber \\
= & \ \widehat{H}_{2}r_{i}\left(\widehat{b}_{1}\left(\widehat{v}_{i},w_{i}\right)-\widehat{b}\left(x_{i}\right)-b_{1}\left(v_{i},w_{i}\right)+b\left(x_{i}\right)\right)+\left(\widehat{H}_{2}-H_{2}\right)r_{i}\xi_{i}\nonumber \\
= & \ \widehat{H}_{2}r_{i}\left(\widehat{b}_{1}\left(\widehat{v}_{i},w_{i}\right)-b_{1}\left(\widehat{v}_{i},w_{i}\right)\right)+\widehat{H}_{2}r_{i}\left(b_{1}\left(\widehat{v}_{i},w_{i}\right)-b_{1}\left(v_{i},w_{i}\right)\right)\nonumber \\
 & +\widehat{H}_{2}r_{i}\left(b\left(x_{i}\right)-\widehat{b}\left(x_{i}\right)\right)+\left(\widehat{H}_{2}-H_{2}\right)r_{i}\xi_{i}\nonumber \\
\eqqcolon & \ D_{41i}+D_{42i}+D_{43i}+D_{44i}.
\end{align}

For $D_{41}$, 
\begin{align}
n^{-1}\sum_{i}D_{41i}^{2} & \leq\norm{\widehat{H}_{2}}^{2}\sup_{x\in\mathcal{X}}\norm{r\left(x\right)}^{2}n^{-1}\sum_{i}\left(\widehat{b}_{1}\left(\widehat{v}_{i},w_{i}\right)-b_{1}\left(\widehat{v}_{i},w_{i}\right)\right)^{2}\nonumber \\
 & \leq C\zeta\left(M_{3}\right)^{2}n^{-1}\sum_{i}\left[\left(\widehat{\ol p}_{i}'\left(\widehat{\alpha}^{M_{2}}-\alpha^{M_{2}}\right)\right)^{2}+\left(\widehat{\ol p}_{i}'\alpha^{M_{2}}-b_{1}\left(\widehat{v}_{i},w_{i}\right)\right)^{2}\right]\nonumber \\
 & =O_{p}\left(\zeta\left(M_{3}\right)^{2}\D_{2n}^{2}\right)=o_{p}\left(1\right),\label{eq:D41}
\end{align}
where the second inequality holds by $\norm{\widehat{H}_{2}}=O_{p}\left(1\right)$
and Assumption \ref{assu:bx and bbar normality}\ref{enu:as on Erp}
and the first equality holds by \eqref{eq:beta vw hat}.

For $D_{42}$, 
\begin{align}
n^{-1}\sum_{i}D_{42i}^{2} & \leq\norm{\widehat{H}_{2}}^{2}\sup_{x\in\mathcal{X}}\norm{r\left(x\right)}^{2}n^{-1}\sum_{i}\left(b_{1}\left(\widehat{v}_{i},w_{i}\right)-b_{1}\left(v_{i},w_{i}\right)\right)^{2}\nonumber \\
 & \leq C\zeta\left(M_{3}\right)^{2}n^{-1}\sum_{i}\left(\widehat{v}_{i}-v_{i}\right)^{2}=O_{p}\left(\zeta\left(M_{3}\right)^{2}\D_{1n}^{2}\right)=o_{p}\left(1\right),\label{eq:D42}
\end{align}
where the first equality holds by Lemma \ref{lem: rates of V and G}.

The proof of $n^{-1}\sum_{i}D_{43i}^{2}=o_{p}\left(1\right)$ is completely
analogous to \eqref{eq:D41} and is omitted.

For $D_{44}$, 
\begin{align}
\E\left[\rest{n^{-1}\sum_{i}D_{44i}^{2}}{\bf X}\right] & =\left(\widehat{H}_{2}-H_{2}\right)n^{-1}\sum_{i}r_{i}r_{i}'\E\left(\rest{\xi_{i}^{2}}X_{i}\right)\left(\widehat{H}_{2}-H_{2}\right)'\nonumber \\
 & \leq C\left(\widehat{H}_{2}-H_{2}\right)\widehat{R}\left(\widehat{H}_{2}-H_{2}\right)'\nonumber \\
 & \leq C\norm{\widehat{H}_{2}-H_{2}}^{2}=o_{p}\left(1\right),\label{eq:D44}
\end{align}
where the first equality holds by the fact that both $\widehat{H}_{2}$
and $r_{i}$ are functions of ${\bf X}$, the first inequality holds
by Assumption \ref{assu: rates beta x beta bar}\ref{enu:rates bx xi bound 2nd order},
and the last inequality uses $\norm{\widehat{R}-I}=o_{p}\left(1\right)$.
Then, by the CM inequality, 
\begin{equation}
n^{-1}\sum_{i}D_{44i}^{2}=o_{p}\left(1\right).\label{eq:D_44 result}
\end{equation}

Combining the results for $D_{41}$, $D_{42}$, $D_{43}$, and $D_{44}$,
I have
\begin{equation}
n^{-1}\sum_{i}\left(\widehat{H}_{2}r_{i}\widehat{\xi}_{i}-H_{2}r_{i}\xi_{i}\right)^{2}=o_{p}\left(1\right).\label{eq:D4}
\end{equation}
Therefore, by \eqref{eq:Omega 22 consistency},
\begin{align}
 & \ n^{-1}\sum_{i}\left(\left(\widehat{H}_{1}\widehat{\ol{\mu}}_{i}^{I}-\widehat{H}_{2}\widehat{\ol{\mu}}_{i}^{II}-\widehat{H}_{2}r_{i}\widehat{\xi}_{i}\right)-\left(H_{1}\ol{\mu}_{i}^{I}-H_{2}\ol{\mu}_{i}^{II}-H_{2}r_{i}\xi_{i}\right)\right)^{2}\nonumber \\
\leq & \ 3n^{-1}\sum_{i}\left(\widehat{H}_{1}\widehat{\ol{\mu}}_{i}^{I}-H_{1}\ol{\mu}_{i}^{I}\right)^{2}+3n^{-1}\sum_{i}\left(\widehat{H}_{2}\widehat{\ol{\mu}}_{i}^{II}-H_{2}\ol{\mu}_{i}^{II}\right)^{2}\nonumber \\
 & +3n^{-1}\sum_{i}\left(\widehat{H}_{2}r_{i}\widehat{\xi}_{i}-H_{2}r_{i}\xi_{i}\right)^{2}=o_{p}\left(1\right).
\end{align}
Since $\E\left(H_{1}\ol{\mu}_{i}^{I}-H_{2}\ol{\mu}_{i}^{II}-H_{2}r_{i}\xi_{i}\right)^{2}=\Omega_{22}/\Omega_{2}\leq1$,
by Markov's inequality and Lemma B6 of IN02, 
\begin{equation}
\abs{\widehat{\Omega}_{22}/\Omega_{2}-n^{-1}\sum_{i}\left(H_{1}\ol{\mu}_{i}^{I}-H_{2}\ol{\mu}_{i}^{II}-H_{2}r_{i}\xi_{i}\right)^{2}}=o_{p}\left(1\right).\label{eq:intermediate step}
\end{equation}
By the law of large numbers, 
\begin{equation}
\abs{n^{-1}\sum_{i}\left(H_{1}\ol{\mu}_{i}^{I}-H_{2}\ol{\mu}_{i}^{II}-H_{2}r_{i}\xi_{i}\right)^{2}-\Omega_{22}/\Omega_{2}}=o_{p}\left(1\right).\label{eq: int step 2}
\end{equation}
Therefore, by the triangle inequality, 
\begin{equation}
\widehat{\Omega}_{22}/\Omega_{2}-\Omega_{22}/\Omega_{2}=o_{p}\left(1\right).\label{eq:Omega 22 hat result}
\end{equation}

Combining the results for $\widehat{\Omega}_{21}$ and $\widehat{\Omega}_{22}$,
I obtain
\begin{equation}
\widehat{\Omega}_{2}/\Omega_{2}-1\pto0.\label{eq:consistency-Omega_2}
\end{equation}

Next, for the APE estimator $\widehat{\ol b}$, define $b_{t}^{\text{APE}}=\E\b_{it}$
and thus $\ol b=T^{-1}\sum_{t=1}^{T}b_{t}^{\text{APE}}$. I first
estimate $b_{t}^{\text{APE}}$ by setting $r^{M_{3}}(x)=1$ in (\ref{eq: estimator for beta bar})
for $\widehat{b}_{t}(x)$. Hence, the asymptotic normality proof for
$\widehat{b}_{t}(x)$ directly applies to $\widehat{b}_{t}^{\text{APE}}$.
In particular, by adapting to (\ref{eq: influence functions}) with
$r^{M_{3}}(x)=1$, the influence function for $\widehat{b}_{t}^{\text{APE}}$
in period $t$ is
\begin{equation}
\psi_{it}=A_{3t}P_{t}^{-1}(p_{it}u_{it}+\ol{\mu}_{it}^{I})-(\ol{\mu}_{it}^{III}+\xi_{it}),\label{eq:psi_b^APE}
\end{equation}
where $A_{3t}=\E\ol p_{it}'$ and $\ol{\mu}_{it}^{III}=\E\left[\rest{b_{1t,V}\left(V_{jt},W_{j}\right)q_{jt}'Q_{t}^{-1}q_{it}v_{ji,t}}\mathcal{I}_{i,t}\right]$. 

Let 
\begin{equation}
\O_{3}:=\V\left(\dfrac{1}{T}\sum_{t=1}^{T}\psi_{it}\right)=\dfrac{1}{T^{2}}\sum_{s=1}^{T}\sum_{t=1}^{T}\text{Cov}(\psi_{it},\psi_{is}).\label{eq:Omega_3}
\end{equation}
Following the same argument as for $\widehat{b}(x)$, I have 
\begin{equation}
\sqrt{n}\,\O_{3}^{-1/2}(\widehat{\ol b}-\ol b)=\sqrt{n}\,\O_{3}^{-1/2}\left[\dfrac{1}{T}\sum_{t=1}^{T}(\widehat{b}_{t}^{\text{APE}}-b_{t}^{\text{APE}})\right]=\dfrac{1}{\sqrt{n}}\sum_{i=1}^{n}\left(\O_{3}^{-1/2}\,\dfrac{1}{T}\sum_{t=1}^{T}\psi_{it}\right)+o_{p}(1)\label{eq:b^bar-inf}
\end{equation}
and the asymptotic normality follows by the Lindeberg-Feller CLT:
\begin{equation}
\sqrt{n}\,\O_{3}^{-1/2}(\widehat{\ol b}-\ol b)\dto N(0,I).\label{eq:conv-b^bar}
\end{equation}
Finally, by (\ref{eq:Omega_3}), an estimator $\widehat{\O}_{3}$
for $\O_{3}$ is written as 
\begin{align}
\widehat{\O}_{3} & =\dfrac{1}{T^{2}}\sum_{s=1}^{T}\sum_{t=1}^{T}\left(\dfrac{1}{n}\sum_{i=1}^{n}\widehat{\psi}_{it}\widehat{\psi}_{is}'\right),\label{eq:Omega_3^hat}
\end{align}
where $\widehat{\psi}_{it}=\widehat{A}_{3t}\widehat{P}_{t}^{-1}(\widehat{p}_{it}\widehat{u}_{it}+\widehat{\ol{\mu}}_{it}^{I})-(\widehat{\ol{\mu}}_{it}^{III}+\widehat{\xi}_{it})$,
$\widehat{A}_{3t}=n^{-1}\sum_{i}\widehat{\ol p}_{it}'$, and $\widehat{\ol{\mu}}_{it}^{III}=n^{-1}\sum_{j}\widehat{b}_{1t,V}\left(\widehat{v}_{jt},w_{j}\right)q_{jt}'\widehat{Q}^{-1}q_{it}\widehat{v}_{ji,t}.$
The consistency is established following the same argument as in deriving
(\ref{eq:consistency-Omega_2}):
\begin{equation}
\widehat{\O}_{3}/\O_{3}-1\pto0.\label{eq:consistency-Omega^3}
\end{equation}
\end{proof}
\begin{rem}[\textbf{Degree of Basis Functions}]
\label{rem:asym-tuning-para}The three degrees $M_{1},M_{2},M_{3}$
correspond to the series dimensions used to estimate $V_{t}(x,z,w)$,
$G_{t}(x,v,w)$, and $b_{t}(x)$, respectively. For consistency I
require $M_{i}\to\infty$ while $M_{i}/n\to0$, so that approximation
bias and sampling variance both vanish. Under Assumption \ref{assu: rates-V and G},
the first-step estimator satisfies
\[
\text{MSE}(\widehat{V}_{t})=O\left(\underbrace{M_{1}^{1-2d_{1}/j_{1}}}_{\text{bias}}+\underbrace{M_{1}/n}_{\text{variance}}\right),
\]
where the extra factor $M_{1}$ in the bias term comes from the uniform-in-$x$
analysis with $\mathbf{1}\{X_{it}\le x\}$ as the dependent variable
(as in IN02). For steps 2--3 the analysis is analogous, with two
additional points: (i) since the model implies $G_{t}$ is linear
in $x$, I use the reduced basis $p^{M_{2}}(x,v,w)=x\otimes p^{m_{2}}(v,w)$,
which reduces dimension ($M_{2}=m_{2}d_{X}$); and (ii) first-step
errors propagate, so the error for $\widehat{b}_{t}(x)$ aggregates
across steps
\[
\text{MSE}(\widehat{b}_{t}(x))=\left(M_{1}^{1-2d_{1}/j_{1}}+M_{1}/n\right)+\left(m_{2}^{-2d_{2}/j_{2}}+m_{2}/n\right)+\left(M_{3}^{-2d_{3}/j_{3}}+M_{3}/n\right).
\]
Sup-norm results additionally impose standard bounds on basis functions
and derivatives (Assumption \ref{assu: rates beta x beta bar}\ref{enu:sup-norm-r-bd}),
satisfied by common choices such as splines and power series. Finally,
Assumption \ref{assu:bvw normality} imposes ``not too fast'' growth
on $M_{i}$ for asymptotic normality; for instance, for $\widehat{\ol b}$
(which does not involve $M_{3}$), with splines it suffices that $M_{1}=o(n^{1/7})$
and $M_{2}=o(n^{1/7})$, which is feasible when $d_{1}/j_{1}\ge4$
and $d_{2}/j_{2}\ge4$.
\end{rem}

\subsection{Nonparametric Justification of the Index Exclusion Assumption\label{subsec:Prop-1}}

Next, I introduce a proposition that supplies sufficient conditions
for Assumption\textbf{ }\ref{assu:index exclusion} by adapting the
exchangeability condition of \citet{altonji2005cross}.
\begin{prop}[\textbf{Sufficient Conditions for Assumption \ref{assu:index exclusion}}]
\label{prop:sufficient condition for assu 2}Suppose the following
conditions hold:
\begin{enumerate}[label=\textup{(\alph*)}]
\item \label{enu:prop1}$f_{\rest{\eta_{i}}A_{i}}=f_{\rest{\widetilde{\eta}_{i}}A_{i}}$
for any permutation $\widetilde{\eta}_{i}$ of the $\eta_{i}$ vector
in time,
\item \label{enu:prop2}$Z_{it}\perp\left(A_{i},\eta_{it}\right)$ and the
support of $\left(X_{it},Z_{it}\right)$ is compact, and,
\item \label{enu:prop3}$f_{\rest{A_{i}}{\bf X}_{i},{\bf Z}_{i}}$ is continuous
in $\left({\bf X}_{i},{\bf Z}_{i}\right)$.
\end{enumerate}
Then, Assumption \ref{assu:index exclusion} is satisfied.
\end{prop}
\begin{proof}[\textbf{Proof of Proposition \ref{prop:sufficient condition for assu 2}}]
To begin with, I use the first condition $f_{\rest{\eta_{i}}A_{i}}=f_{\rest{\widetilde{\eta}_{i}}A_{i}}$
to establish the exchangeability condition
\begin{equation}
f_{\rest{A_{i}}{\bf X}_{i},{\bf Z}_{i}}=f_{\rest{A_{i}}\widetilde{{\bf X}}_{i},\widetilde{{\bf Z}}_{i}},\label{eq:exch A given X, Z}
\end{equation}
where $\left(\widetilde{{\bf X}}_{i},\widetilde{{\bf Z}}_{i}\right)$
is any permutation of $\left({\bf X}_{i},{\bf Z}_{i}\right)$ in time.
It is worth emphasizing that \eqref{eq:exch A given X, Z}, which
will be proved in the following, is different from Assumption 2.3
of \citet{altonji2005cross}. In particular, \citet{altonji2005cross}
assume an exchangeability condition involving both $A$ and $\eta_{t}$,
i.e., $f_{\rest{A_{i},\eta_{it}}{\bf X}_{i}}=f_{\rest{A_{i},\eta_{it}}{\bf \widetilde{X}}_{i}},$
which effectively rules out time-varying endogeneity through the random
coefficients because it requires the density of $\eta_{it}$ given
$X_{it}=x_{it}$ to be the same as that given $X_{it}=x_{is}$ for
any $s\neq t$. Instead, I prove \eqref{eq:exch A given X, Z} under
a more primitive assumption of $f_{\rest{\eta_{i}}A_{i}}=f_{\rest{\widetilde{\eta}_{i}}A_{i}}$
that is compatible with time-varying endogeneity through the random
coefficients.

Without loss of generality, I prove \eqref{eq:exch A given X, Z}
for $T=2$ since any ordering of $\left\{ 1,\ldots,T\right\} $ for
finite $T$ can be achieved via a finite number of pairwise permutations.
Then, I prove that one can construct $W_{i}$ such that Assumption
\ref{assu:index exclusion} holds. For simplicity of notation, I assume
$X_{it}$ and $Z_{it}$ are both scalars and suppress $i$ subscripts
in all variables. Thus, in this proof all subscripts of the variables
denote the time period.

By condition \ref{enu:prop1} of Proposition \ref{prop:sufficient condition for assu 2},

\begin{equation}
f_{A,\eta_{1},\eta_{2}}\left(a,h_{1},h_{2}\right)=f_{A,\eta_{1},\eta_{2}}\left(a,h_{2},h_{1}\right).\label{eq:imply}
\end{equation}
Let $g^{-1}\left(X,Z,A\right)$ denote the inverse function of $g\left(Z,A,\eta\right)$
with respect to $\eta$, which exists by Assumption \ref{assu:mono}.
Define $h_{1}=g^{-1}\left(x_{1},z_{1},a\right)$ and $h_{2}=g^{-1}\left(x_{2},z_{2},a\right)$.
Calculate the determinants of the Jacobians as
\begin{align*}
D_{1}\coloneqq & \abs{\begin{array}{ccc}
\dfrac{\partial A}{\partial X_{1}} & \dfrac{\partial A}{\partial X_{2}} & \dfrac{\partial A}{\partial A}\\
\dfrac{\partial g^{-1}\left(X_{1},Z_{1},A\right)}{\partial X_{1}} & \dfrac{\partial g^{-1}\left(X_{1},Z_{1},A\right)}{\partial X_{2}} & \dfrac{\partial g^{-1}\left(X_{1},Z_{1},A\right)}{\partial A}\\
\dfrac{\partial g^{-1}\left(X_{2},Z_{2},A\right)}{\partial X_{1}} & \dfrac{\partial g^{-1}\left(X_{2},Z_{2},A\right)}{\partial X_{2}} & \dfrac{\partial g^{-1}\left(X_{2},Z_{2},A\right)}{\partial A}
\end{array}}_{\begin{array}{c}
\left(X_{1},X_{2},Z_{1},Z_{2},A\right)\\
=\left(x_{1},x_{2},z_{1},z_{2},a\right)
\end{array}}\\
= & \abs{\begin{array}{ccc}
0 & 0 & 1\\
\dfrac{\partial g^{-1}\left(X_{1},Z_{1},A\right)}{\partial X_{1}} & 0 & \dfrac{\partial g^{-1}\left(X_{1},Z_{1},A\right)}{\partial A}\\
0 & \dfrac{\partial g^{-1}\left(X_{2},Z_{2},A\right)}{\partial X_{2}} & \dfrac{\partial g^{-1}\left(X_{2},Z_{2},A\right)}{\partial A}
\end{array}}_{\begin{array}{c}
\left(X_{1},X_{2},Z_{1},Z_{2},A\right)\\
=\left(x_{1},x_{2},z_{1},z_{2},a\right)
\end{array}}\\
= & \rest{\partial g^{-1}\left(X,Z,A\right)/\partial X}_{\left(X,Z,A\right)=\left(x_{1},z_{1},a\right)}\times\rest{\partial g^{-1}\left(X,Z,A\right)/\partial X}_{\left(X,Z,A\right)=\left(x_{2},z_{2},a\right)},
\end{align*}
and similarly
\begin{align*}
D_{2}\coloneqq & \abs{\begin{array}{ccc}
\dfrac{\partial g\left(Z_{1},A,\eta_{1}\right)}{\partial A} & \dfrac{\partial g\left(Z_{1},A,\eta_{1}\right)}{\partial\eta_{1}} & \dfrac{\partial g\left(Z_{1},A,\eta_{1}\right)}{\partial\eta_{2}}\\
\dfrac{\partial g\left(Z_{2},A,\eta_{2}\right)}{\partial A} & \dfrac{\partial g\left(Z_{2},A,\eta_{2}\right)}{\partial\eta_{1}} & \dfrac{\partial g\left(Z_{2},A,\eta_{2}\right)}{\partial\eta_{2}}\\
\dfrac{\partial A}{\partial A} & \dfrac{\partial A}{\partial\eta_{1}} & \dfrac{\partial A}{\partial\eta_{2}}
\end{array}}_{\begin{array}{c}
\left(Z_{1},Z_{2},A,\eta_{1},\eta_{2}\right)\\
=\left(z_{2},z_{1},a,h_{2},h_{1}\right)
\end{array}}\\
= & \rest{\partial g\left(Z,A,\eta\right)/\partial\eta}_{\left(Z,A,\eta\right)=\left(z_{2},a,h_{2}\right)}\times\rest{\partial g\left(Z,A,\eta\right)/\partial\eta}_{\left(Z,A,\eta\right)=\left(z_{1},a,h_{1}\right)}.
\end{align*}
Then, 
\begin{align}
 & \ f_{\rest{X_{1},X_{2},A}Z_{1},Z_{2}}\left(\rest{x_{1},x_{2},a}z_{1},z_{2}\right)\nonumber \\
= & \ f_{\rest{A,\eta_{1},\eta_{2}}Z_{1},Z_{2}}\left(\rest{a,g^{-1}\left(x_{1},z_{1},a\right),g^{-1}\left(x_{2},z_{2},a\right)}z_{1},z_{2}\right)\abs{D_{1}}\nonumber \\
= & \ f_{\rest{A,\eta_{1},\eta_{2}}Z_{1},Z_{2}}\left(\rest{a,g^{-1}\left(x_{2},z_{2},a\right),g^{-1}\left(x_{1},z_{1},a\right)}z_{2},z_{1}\right)\abs{D_{1}}\nonumber \\
= & \ f_{\rest{X_{1},X_{2},A}Z_{1},Z_{2}}\left(\rest{x_{2},x_{1},a}z_{2},z_{1}\right)\abs{D_{2}D_{1}}\nonumber \\
= & \ f_{\rest{X_{1},X_{2},A}Z_{1},Z_{2}}\left(\rest{x_{2},x_{1},a}z_{2},z_{1}\right),\label{eq:last cov}
\end{align}
where the first equality holds by the change of variables for $\eta_{1}$
and $\eta_{2}$, the second equality uses \eqref{eq:imply} and $Z\perp\left(A,\eta\right)$,
the third equality holds by $X_{1}=g\left(z_{2},a,g^{-1}\left(x_{2},z_{2},a\right)\right)=x_{2}$
and $X_{2}=g\left(z_{1},a,g^{-1}\left(x_{1},z_{1},a\right)\right)=x_{1}$,
and the last equality uses the fact that the product of derivatives
of inverse functions equals one.

Given \eqref{eq:last cov}, I integrate $A$ out in \eqref{eq:last cov}
to obtain
\begin{align}
f_{\rest{X_{1},X_{2}}Z_{1},Z_{2}}\left(\rest{x_{1},x_{2}}z_{1},z_{2}\right) & =f_{\rest{X_{1},X_{2}}Z_{1},Z_{2}}\left(\rest{x_{2},x_{1}}z_{2},z_{1}\right),\label{eq:marginal}
\end{align}
which implies 
\begin{align}
 & \ f_{\rest AX_{1},X_{2},Z_{1},Z_{2}}\left(\rest ax_{1},x_{2},z_{1},z_{2}\right)\nonumber \\
= & \ f_{\rest{X_{1},X_{2},A}Z_{1},Z_{2}}\left(\rest{x_{1},x_{2},a}z_{1},z_{2}\right)/f_{\rest{X_{1},X_{2}}Z_{1},Z_{2}}\left(\rest{x_{1},x_{2}}z_{1},z_{2}\right)\nonumber \\
= & \ f_{\rest{X_{1},X_{2},A}Z_{1},Z_{2}}\left(\rest{x_{2},x_{1},a}z_{2},z_{1}\right)/f_{\rest{X_{1},X_{2}}Z_{1},Z_{2}}\left(\rest{x_{2},x_{1}}z_{2},z_{1}\right)\nonumber \\
= & \ f_{\rest AX_{1},X_{2},Z_{1},Z_{2}}\left(\rest ax_{2},x_{1},z_{2},z_{1}\right),\label{eq:reason}
\end{align}
where the second equality uses \eqref{eq:last cov} and \eqref{eq:marginal}.

The rest of the proof follows the same argument as in Section 2.2
of \citet{altonji2005cross}. Specifically, I show that for any on-support
$a$, the conditional density $f_{\rest AX_{1},X_{2},Z_{1},Z_{2}}\left(\rest ax_{1},x_{2},z_{1},z_{2}\right)$
can be approximated arbitrarily closely by a function of the form
$f_{\rest AW}\left(\rest aw\right)$. To construct $W$, I follow
\citet{weyl1939classical} and let $\varphi_{1}\left(u\right)=\sum_{t=1}^{T}u_{t}$,
$\varphi_{2}\left(u,v\right)=\sum_{t\neq s}^{T}u_{t}v_{s}$, ...,
$\varphi_{T}\left(u,v,\ldots,w\right)=\sum_{t\neq s\neq...\neq k}^{T}u_{t}v_{s}\ldots w_{k}$,
where $u,v,\ldots,w$ are generic $T\times1$ vectors. Then, I substitute
each column vector of ${\bf D}_{i}=\left(\X_{i},{\bf Z}_{i}\right)$
for each of the arguments $u,v,\ldots,w$ (repetitions included) to
construct $W_{i}$. See Chapter II.3 of \citet{weyl1939classical}
for details of the construction and its proof.

By condition \ref{enu:prop2} of Proposition \ref{prop:sufficient condition for assu 2},
the supports of $X$ and $Z$ are compact. By condition \ref{enu:prop3}
of Proposition \ref{prop:sufficient condition for assu 2}, $f_{\rest AX_{1},X_{2},Z_{1},Z_{2}}$
is continuous in $\left(X_{1},X_{2},Z_{1},Z_{2}\right)$. Therefore,
by the Stone-Weierstrass theorem, there exists a function $f_{\rest AX_{1},X_{2},Z_{1},Z_{2}}^{w}$
that is a polynomial in $\left(X_{1},X_{2},Z_{1},Z_{2}\right)$ over
a compact set with the property that for any fixed $\d$ that is arbitrarily
close to 0, 
\begin{equation}
\max_{x_{1},x_{2}\in\mathcal{X},z_{1},z_{2}\in\mathcal{Z}}\abs{f_{\rest AX_{1},X_{2},Z_{1},Z_{2}}\left(\rest ax_{1},x_{2},z_{1},z_{2}\right)-f_{\rest AX_{1},X_{2},Z_{1},Z_{2}}^{w}\left(\rest ax_{1},x_{2},z_{1},z_{2}\right)}\leq\d.\label{eq:poly approx}
\end{equation}

Let 
\begin{align*}
 & \ol f_{\rest AX_{1},X_{2},Z_{1},Z_{2}}\left(\rest ax_{1},x_{2},z_{1},z_{2}\right)\\
\coloneqq & \left[f_{\rest AX_{1},X_{2},Z_{1},Z_{2}}\left(\rest ax_{1},x_{2},z_{1},z_{2}\right)+f_{\rest AX_{1},X_{2},Z_{1},Z_{2}}\left(\rest ax_{2},x_{1},z_{2},z_{1}\right)\right]/2!
\end{align*}
denote the simple average of $f_{\rest AX_{1},X_{2},Z_{1},Z_{2}}\left(\rest ax_{1},x_{2},z_{1},z_{2}\right)$
over all $T!$ (here $T=2$) unique permutations of $\left(x_{t},z_{t}\right)$,
and similarly for $\ol f_{\rest AX_{1},X_{2},Z_{1},Z_{2}}^{w}\left(\rest ax_{1},x_{2},z_{1},z_{2}\right)$.
By \eqref{eq:reason}, 
\[
\ol f_{\rest AX_{1},X_{2},Z_{1},Z_{2}}\left(\rest ax_{1},x_{2},z_{1},z_{2}\right)=f_{\rest AX_{1},X_{2},Z_{1},Z_{2}}\left(\rest ax_{1},x_{2},z_{1},z_{2}\right).
\]
By \eqref{eq:reason}, \eqref{eq:poly approx}, and the triangle inequality,
\begin{align}
 & \ \abs{f_{\rest AX_{1},X_{2},Z_{1},Z_{2}}\left(\rest ax_{1},x_{2},z_{1},z_{2}\right)-\ol f_{\rest AX_{1},X_{2},Z_{1},Z_{2}}^{w}\left(\rest ax_{1},x_{2},z_{1},z_{2}\right)}\nonumber \\
= & \ \abs{\ol f_{\rest AX_{1},X_{2},Z_{1},Z_{2}}\left(\rest ax_{1},x_{2},z_{1},z_{2}\right)-\ol f_{\rest AX_{1},X_{2},Z_{1},Z_{2}}^{w}\left(\rest ax_{1},x_{2},z_{1},z_{2}\right)}\nonumber \\
\leq & \ T!\times\left(\d/T!\right)=\d.\label{eq:approximation}
\end{align}
Since $f^{w}$ can be chosen to make $\d$ arbitrarily small, equation
\eqref{eq:approximation} implies that $f_{\rest AX_{1},X_{2},Z_{1},Z_{2}}\left(\rest ax_{1},x_{2},z_{1},z_{2}\right)$
can be approximated arbitrarily closely by a polynomial $\ol f^{w}$
that is symmetric in $\left(x_{t},z_{t}\right)$ pairs for $t=1,2$.
Furthermore, by the fundamental theorem of symmetric functions, for
any $a$ on its support, $\ol f^{w}$ can be written as a polynomial
function of the elementary symmetric functions $W$ of $\left(\left(x_{1},z_{1}\right),\left(x_{2},z_{2}\right)\right)$.
I denote this function by $f_{\rest AW}\left(\rest aw\right)$ and
obtain that $f_{\rest AX_{1},X_{2},Z_{1},Z_{2}}\left(\rest ax_{1},x_{2},z_{1},z_{2}\right)$
can be approximated arbitrarily closely by $f_{\rest AW}\left(\rest aW\right)$.
Let $\d\gto0$ in \eqref{eq:approximation}. Note that to let $\d\to0$
the degree of polynomials in the approximating function $f^{w}$ needs
to increase. However, it does not affect the dimension of $W$ since
by the fundamental theorem of symmetric functions, the elements of
$W$ have a fixed degree of $T$. The key is that there are two different
degrees of polynomials: one is the degree of the approximating functions
$f^{w}$, and the other is the degree of the arguments of $W$.\footnote{For example, one may let the degree $P_{1}$ and $P_{2}$ of a polynomial
function $h\left(x,y\right)=\sum_{i=1}^{P_{1}}\left(x+y\right)^{i}+\sum_{i=1}^{P_{2}}\left(xy\right)^{i}$
to increase while keeping the degree of its symmetric arguments (in
this example, $W=(x+y,xy)$) fixed. See also footnotes 9 and 10 of
\citet{altonji2005cross} for a detailed discussion.}

Then, for any $t\in\left\{ 1,...,T\right\} $ and on-support $\left(x_{t},z_{t},a,w\right)$
\[
f_{\rest AX_{t},Z_{t},W}\left(\rest ax_{t},z_{t},w\right)=f_{\rest AW}\left(\rest aw\right).\tag*{\qedhere}
\]
\end{proof}

\subsection{Further Discussion of Identification Assumptions\label{subsec:As245-Discuss}}

When the exchangeability conditions are used to justify Assumption
\ref{assu:index exclusion}, I present a symmetry-based argument in
the discussion of Assumption \ref{assu:residual variation in X} that
justifies Assumptions \ref{assu:index exclusion} and \ref{assu:residual variation in X}
simultaneously. I provide an example here to clarify this point. Suppose
$T=2$ and $(X_{it},Z_{it})\in\R^{2}$. By Proposition \ref{prop:sufficient condition for assu 2},
when $W_{i}$ includes averages through time of the polynomials of
$\left(X_{it},Z_{it}\right)$ up to the second degree, Assumption
\ref{assu:index exclusion} holds. Suppose $(X_{i1},X_{i2},Z_{i1},Z_{i2})=\left(x_{1},x_{2},z_{1},z_{2}\right)$
satisfies $\{V_{it}=v,W_{i}=w\}$ for $t=1,2$ where $x_{1},x_{2},z_{1},z_{2}\in\R^{2}$
and $x_{1}$ and $x_{2}$ are linearly independent. Then, by definition
of $V_{it}$ and the symmetry of $W_{i}$ in $\left(X_{it},Z_{it}\right)$
through $t$, $(X_{i1},X_{i2},Z_{i1},Z_{i2})=\left(x_{2},x_{1},z_{2},z_{1}\right)$
must also satisfy $V_{it}=v$ and $W_{i}=w$ for $t=1,2$. Therefore,
given $V_{it}=v$ and $W_{i}=w$ there are two linearly independent
vectors of $x_{1}$ and $x_{2}$ that $X_{it}$ can take, satisfying
Assumption \ref{assu:residual variation in X}.

However, Assumption \ref{assu:more_var_X} requires more residual
variation in $X_{it}$ given $(V_{it},W_{i})$ and can be restrictive
if nonparametric justifications are used for Assumption \ref{assu:index exclusion}.
I summarize the proposal by \citet{altonji2005cross} to achieve such
residual variation. Specifically, \citet{altonji2005cross} suggest
to 

(i) leverage unconditional variation in the exogenous regressors that
do not enter $W_{i}$, and I provide more details at the end of this
section; 

(ii) restrict the conditional distribution of $A_{i}$ given $W_{i}$
to depend on not all but only a subset of coordinates of $W_{i}$,
which is testable by comparing the fit of $\E\left[\rest{Y_{it}}X_{it},V_{it},W_{i}\right]$
with $W_{i}$ replaced by a proper subset of it; 

(iii) assume $f_{\rest{A_{i}}W_{i}}$ to depend on $W_{i}$ only through
a linear combination of the elements of $W_{i}$, i.e., $f_{\rest{A_{i}}W_{i}}=f_{\rest{A_{i}}\sum_{k=1}^{d_{W}}c_{k}W_{i,k}}$
where $W_{i}=(W_{i,1},\ldots,W_{i,d_{W}})'$ and the $c_{k}$s are
unknown parameters, which implies that 
\[
\E\left[\rest{Y_{it}}X_{it},V_{it},W_{i}\right]=\E\left[\rest{Y_{it}}X_{it},V_{it},\sum_{k=1}^{d_{W}}c_{k}W_{i,k}\right]
\]
and the $c_{k}$s can be estimated using \citet{ichimura1991semiparametric}; 

(iv) impose a priori restrictions on model \eqref{eq:Yeq}--\eqref{eq:main eqn 2 X}
such that $W_{i}$ enters $\E\left[\rest{Y_{it}}X_{it},V_{it},W_{i}\right]$
in a parametric way (e.g., 
\[
\E\left[\rest{\b_{it}}V_{it},W_{i}\right]=\sum_{k=1}^{d_{W}}c_{k}W_{i,k}+\sum_{l=1}^{d_{V}}d_{l}V_{it,l}
\]
 when $d_{X}=1$); and, 

(v) directly impose functional restrictions on $\E\left[\rest{Y_{it}}X_{it},V_{it},W_{i}\right]$.

Finally, to see how the \textit{unconditional} variations in the exogenous
regressors help to satisfy the required residual variation in endogenous
regressors, suppose $X_{it,1}$ is endogenous while $X_{it,-1}$ (all
other coordinates of $X_{it}$ except $X_{it,1}$) are exogenous.
Suppose the conditional support of $X_{it,1}$ given $(V_{it},W_{i})$
only contains a non-zero singleton, while the unconditional support
of $X_{it,-1}$ contains a small ball of positive radius. Denote $b_{1}\left(V_{it},W_{i}\right)=\E\left[\rest{\b_{it}}V_{it},W_{i}\right]$.
Then, I can identify the first-order moments of $\b_{it}$ by taking
partial derivative with respect to $X_{it,-1}$ on both sides of 
\begin{equation}
\E\left[\rest{Y_{it}}X_{it},V_{it},W_{i}\right]=X_{it,1}b_{1,1}\left(V_{it},W_{i}\right)+X_{it,-1}'b_{1,-1}\left(V_{it},W_{i}\right)\label{eq:exo-x-deri}
\end{equation}
to identify 
\begin{equation}
b_{1,-1}\left(V_{it},W_{i}\right)=\p\E\left[\rest{Y_{it}}X_{it},V_{it},W_{i}\right]/\p X_{it,-1}.\label{eq:b1_-1_id}
\end{equation}
Finally, identification of $b_{1,1}\left(V_{it},W_{i}\right)$ can
be obtained as
\[
b_{1,1}\left(V_{it},W_{i}\right)=\left(\E\left[\rest{Y_{it}}X_{it},V_{it},W_{i}\right]-X_{it,-1}'b_{1,-1}\left(V_{it},W_{i}\right)\right)/X_{it,1}.
\]
Note that there are two conditions on $X_{it,-1}$ needed for the
argument to work. First, $X_{it,-1}$ needs to be exogenous, so that
it does not enter the construction of $V_{it}$ and $W_{i}$. Because
of this, when taking partial derivative with respect to $X_{it,-1}$
in (\ref{eq:exo-x-deri}), it directly leads to (\ref{eq:b1_-1_id}).
Second, $X_{it,-1}$ needs to be continuous vector to enable the derivative-based
argument. When $X_{it,-1}$ is discrete, one may use arithmetic differencing
technique to identify $b_{1,-1}\left(V_{it},W_{i}\right)$.

\section{Additional Empirical Results\label{sec:Additional-Empirical-Results}}

I conduct several robustness checks of the empirical findings in Section
\ref{sec:Empirical-Illustration}.

\textbf{Measurement error}. To address measurement error that is typical
in firm-level data, I go beyond the baseline cleaning (removing outliers,
negative, and extreme values) and re-estimate the model using alternative
deflators and excluding firms in the top and bottom 1\% of total output.
Table \ref{tab:robust-meas-err} reports the results; the estimates
largely remain in line with the baseline.

\textbf{Instrument validity}. I probe instrument validity by varying
the geographic aggregation (provincial rather than finer units), using
unweighted competitor averages, and adding city fixed effects. As
shown in Table \ref{tab:robust-IV}, the main conclusion is unchanged.
The specifications with city fixed effects yield noisier estimates,
likely because many firms have few within-city competitors, making
the within-city demeaning numerically less stable.

\textbf{Control-function specification}. I assess sensitivity to the
control-function specification along two dimensions. First, I consider
an alternative construction of $W$ that includes averages of both
the instruments and the regressors. Second, I re-estimate the series
control function using alternative bases: second-degree polynomials
and third-degree cubic splines with a knot at the median. Table \ref{tab:CF-sensivitity}
shows that the estimates are essentially unchanged. One systematic
difference is that cubic splines tend to deliver larger point estimates
and wider confidence intervals across industries---consistent with
the simulation evidence in Table \ref{tab:sim vary order}---suggesting
a curse-of-dimensionality cost when the basis becomes too rich.

\textbf{Comparison with classic methods.} I further compare the TERC
estimates with standard OLS and a naive IV approach in Table \ref{tab:OLS_IV_Comp}.
Overall, the TERC method yields higher capital elasticities and lower
labor elasticities than OLS in three sectors; in the remaining two
sectors, the estimates are broadly similar. These patterns echo the
findings of \citet{olleypakes1996teltfp}, albeit in a different context
and using a different dataset. By contrast, the naive IV estimates
are economically implausible, underscoring the importance of accounting
for potentially nonlinear, nonseparable heterogeneity through time-varying
random coefficients. 

Taken together, these exercises indicate that the results are robust
to a wide range of alternative specifications and sample restrictions. 

\begin{table}
\caption{Robustness to Measurement Error\label{tab:robust-meas-err}}

\smallskip{}

\centering{}\begin{threeparttable}{\small{}%
\begin{tabular}{lccc}
\toprule 
Textile & TERC & Alt Deflator & Non-Extreme Output\tabularnewline
\midrule
Capital Elasticity & 0.415 & 0.403 & 0.408\tabularnewline
95\% CI & {[}0.393, 0.441{]} & {[}0.379, 0.428{]} & {[}0.385, 0.436{]}\tabularnewline
Labor Elasticity & 0.452 & 0.472 & 0.441\tabularnewline
95\% CI & {[}0.415, 0.485{]} & {[}0.436, 0.505{]} & {[}0.405, 0.475{]}\tabularnewline
\midrule 
Chemical & TERC & Alt Deflator & Non-Extreme Output\tabularnewline
\midrule 
Capital Elasticity & 0.463 & 0.420 & 0.463\tabularnewline
95\% CI & {[}0.430, 0.497{]} & {[}0.387, 0.453{]} & {[}0.429, 0.492{]}\tabularnewline
Labor Elasticity & 0.311 & 0.439 & 0.293\tabularnewline
95\% CI & {[}0.263, 0.361{]} & {[}0.387, 0.489{]} & {[}0.245, 0.339{]}\tabularnewline
\midrule 
Nonmetallic Mineral & TERC & Alt Deflator & Non-Extreme Output\tabularnewline
\midrule 
Capital Elasticity & 0.371 & 0.377 & 0.361\tabularnewline
95\% CI & {[}0.332, 0.410{]} & {[}0.341, 0.417{]} & {[}0.324, 0.398{]}\tabularnewline
Labor Elasticity & 0.311 & 0.314 & 0.291\tabularnewline
95\% CI & {[}0.255, 0.365{]} & {[}0.258, 0.365{]} & {[}0.238, 0.341{]}\tabularnewline
\midrule 
General Equipment & TERC & Alt Deflator & Non-Extreme Output\tabularnewline
\midrule 
Capital Elasticity & 0.433 & 0.434 & 0.422\tabularnewline
95\% CI & {[}0.402, 0.469{]} & {[}0.401, 0.474{]} & {[}0.393, 0.457{]}\tabularnewline
Labor Elasticity & 0.521 & 0.526 & 0.522\tabularnewline
95\% CI & {[}0.479, 0.564{]} & {[}0.477, 0.569{]} & {[}0.479, 0.566{]}\tabularnewline
\midrule 
Transportation Equipment & TERC & Alt Deflator & Non-Extreme Output\tabularnewline
\midrule
Capital Elasticity & 0.425 & 0.453 & 0.429\tabularnewline
95\% CI & {[}0.393, 0.459{]} & {[}0.415, 0.493{]} & {[}0.396, 0.463{]}\tabularnewline
Labor Elasticity & 0.588 & 0.566 & 0.571\tabularnewline
95\% CI & {[}0.536, 0.644{]} & {[}0.508, 0.632{]} & {[}0.515, 0.631{]}\tabularnewline
\bottomrule
\end{tabular}}\begin{tablenotes}[flushleft]
\scriptsize
\item \textit{Notes:} ``Alt Deflator'' refers to the price deflators constructed by \cite{brandt2014challenges}. ``Non-Extreme Output'' stands for the subsample that excludes the top and bottom 1\% firms in terms of total output.
\end{tablenotes}
\end{threeparttable}
\end{table}

\begin{table}
\caption{IV Validity Check\label{tab:robust-IV}}

\smallskip{}

\centering{}\begin{threeparttable}{\small{}%
\begin{tabular}{lcccc}
\toprule 
Textile & TERC & Provincial IV & Unweighted IV & Location FE\tabularnewline
\midrule
Capital Elasticity & 0.415 & 0.393 & 0.414 & 0.410\tabularnewline
95\% CI & {[}0.393, 0.441{]} & {[}0.368, 0.417{]} & {[}0.390, 0.442{]} & {[}0.386, 0.436{]}\tabularnewline
Labor Elasticity & 0.452 & 0.458 & 0.460 & 0.427\tabularnewline
95\% CI & {[}0.415, 0.485{]} & {[}0.423, 0.494{]} & {[}0.425, 0.494{]} & {[}0.392, 0.462{]}\tabularnewline
\midrule 
Chemical & TERC & Provincial IV & Unweighted IV & Location FE\tabularnewline
\midrule 
Capital Elasticity & 0.463 & 0.444 & 0.459 & 0.449\tabularnewline
95\% CI & {[}0.430, 0.497{]} & {[}0.411, 0.477{]} & {[}0.425, 0.492{]} & {[}0.409, 0.490{]}\tabularnewline
Labor Elasticity & 0.311 & 0.321 & 0.303 & 0.360\tabularnewline
95\% CI & {[}0.263, 0.361{]} & {[}0.274, 0.371{]} & {[}0.257, 0.352{]} & {[}0.300, 0.411{]}\tabularnewline
\midrule 
Nonmetallic Mineral & TERC & Provincial IV & Unweighted IV & Location FE\tabularnewline
\midrule 
Capital Elasticity & 0.371 & 0.309 & 0.346 & 0.410\tabularnewline
95\% CI & {[}0.332, 0.410{]} & {[}0.258, 0.353{]} & {[}0.303, 0.387{]} & {[}0.370, 0.455{]}\tabularnewline
Labor Elasticity & 0.311 & 0.289 & 0.321 & 0.381\tabularnewline
95\% CI & {[}0.255, 0.365{]} & {[}0.227, 0.345{]} & {[}0.263, 0.377{]} & {[}0.321, 0.429{]}\tabularnewline
\midrule 
General Equipment & TERC & Provincial IV & Unweighted IV & Location FE\tabularnewline
\midrule 
Capital Elasticity & 0.433 & 0.433 & 0.434 & 0.414\tabularnewline
95\% CI & {[}0.402, 0.469{]} & {[}0.400, 0.467{]} & {[}0.404, 0.471{]} & {[}0.378, 0.452{]}\tabularnewline
Labor Elasticity & 0.521 & 0.535 & 0.520 & 0.572\tabularnewline
95\% CI & {[}0.479, 0.564{]} & {[}0.492, 0.581{]} & {[}0.479, 0.563{]} & {[}0.523, 0.618{]}\tabularnewline
\midrule 
Transportation Equipment & TERC & Provincial IV & Unweighted IV & Location FE\tabularnewline
\midrule
Capital Elasticity & 0.425 & 0.408 & 0.424 & 0.382\tabularnewline
95\% CI & {[}0.393, 0.459{]} & {[}0.372, 0.440{]} & {[}0.393, 0.460{]} & {[}0.346, 0.418{]}\tabularnewline
Labor Elasticity & 0.588 & 0.589 & 0.587 & 0.612\tabularnewline
95\% CI & {[}0.536, 0.644{]} & {[}0.538, 0.649{]} & {[}0.538, 0.645{]} & {[}0.558, 0.673{]}\tabularnewline
\bottomrule
\end{tabular}}\begin{tablenotes}[flushleft]
\scriptsize
\item \textit{Notes:} ``Provincial IV'' denotes the leave-one-out weighted input prices of the competitors in the same province (c.f., baseline IV constructed at city level), with the weights given by the firm's total debt (for interest rate) and number of employees (for wages). ``Unweighted IV'' stands for the leave-one-out unweighted input prices of the competitors in the same city. ``Location FE'' includes a city fixed effect in the three-step series regression.
\end{tablenotes}
\end{threeparttable}
\end{table}

\begin{table}
\caption{Sensitivity to Control-Function Specification\label{tab:CF-sensivitity}}

\smallskip{}

\centering{}\begin{threeparttable}{\small{}%
\begin{tabular}{lcccc}
\toprule 
Textile & TERC & Alt $W$ & Alt Basis I & Alt Basis II\tabularnewline
\midrule
Capital Elasticity & 0.415 & 0.416 & 0.385 & 0.445\tabularnewline
95\% CI & {[}0.393, 0.441{]} & {[}0.394, 0.443{]} & {[}0.364, 0.407{]} & {[}0.423, 0.478{]}\tabularnewline
Labor Elasticity & 0.452 & 0.448 & 0.438 & 0.486\tabularnewline
95\% CI & {[}0.415, 0.485{]} & {[}0.413, 0.482{]} & {[}0.406, 0.469{]} & {[}0.451, 0.528{]}\tabularnewline
\midrule 
Chemical & TERC & Alt $W$ & Alt Basis I & Alt Basis II\tabularnewline
\midrule 
Capital Elasticity & 0.463 & 0.462 & 0.434 & 0.490\tabularnewline
95\% CI & {[}0.430, 0.497{]} & {[}0.429, 0.495{]} & {[}0.402, 0.464{]} & {[}0.450, 0.534{]}\tabularnewline
Labor Elasticity & 0.311 & 0.312 & 0.250 & 0.349\tabularnewline
95\% CI & {[}0.263, 0.361{]} & {[}0.265, 0.362{]} & {[}0.206, 0.291{]} & {[}0.297, 0.406{]}\tabularnewline
\midrule 
Nonmetallic Mineral & TERC & Alt $W$ & Alt Basis I & Alt Basis II\tabularnewline
\midrule 
Capital Elasticity & 0.371 & 0.357 & 0.334 & 0.370\tabularnewline
95\% CI & {[}0.332, 0.410{]} & {[}0.317, 0.396{]} & {[}0.298, 0.369{]} & {[}0.320, 0.415{]}\tabularnewline
Labor Elasticity & 0.311 & 0.303 & 0.331 & 0.303\tabularnewline
95\% CI & {[}0.255, 0.365{]} & {[}0.245, 0.356{]} & {[}0.280, 0.386{]} & {[}0.239, 0.358{]}\tabularnewline
\midrule 
General Equipment & TERC & Alt $W$ & Alt Basis I & Alt Basis II\tabularnewline
\midrule 
Capital Elasticity & 0.433 & 0.428 & 0.407 & 0.442\tabularnewline
95\% CI & {[}0.402, 0.469{]} & {[}0.398, 0.464{]} & {[}0.377, 0.441{]} & {[}0.408, 0.484{]}\tabularnewline
Labor Elasticity & 0.521 & 0.513 & 0.496 & 0.530\tabularnewline
95\% CI & {[}0.479, 0.564{]} & {[}0.471, 0.556{]} & {[}0.455, 0.540{]} & {[}0.486, 0.579{]}\tabularnewline
\midrule 
Transportation Equipment & TERC & Alt $W$ & Alt Basis I & Alt Basis II\tabularnewline
\midrule
Capital Elasticity & 0.425 & 0.424 & 0.411 & 0.416\tabularnewline
95\% CI & {[}0.393, 0.459{]} & {[}0.393, 0.458{]} & {[}0.383, 0.445{]} & {[}0.376, 0.451{]}\tabularnewline
Labor Elasticity & 0.588 & 0.586 & 0.525 & 0.616\tabularnewline
95\% CI & {[}0.536, 0.644{]} & {[}0.539, 0.644{]} & {[}0.471, 0.574{]} & {[}0.563, 0.678{]}\tabularnewline
\bottomrule
\end{tabular}}\begin{tablenotes}[flushleft]
\scriptsize
\item \textit{Notes:} ``Alt $W$'' means including the time average of both the IVs and regressors into the $W$ vector. ``Alt Basis I'' represents the second-degree polynomial basis function. ``Alt Basis II'' denotes the third-degree spline basis function. 
\end{tablenotes}
\end{threeparttable}
\end{table}

\begin{table}
\caption{Comparison with OLS and Naive IV Estimates\label{tab:OLS_IV_Comp}}

\smallskip{}

\centering{}\begin{threeparttable}{\small{}%
\begin{tabular}{lccc}
\toprule 
Textile & TERC & OLS & Naive IV\tabularnewline
\midrule
Capital Elasticity & 0.415 & 0.379 & 1.309\tabularnewline
95\% CI & {[}0.393, 0.441{]} & {[}0.358, 0.400{]} & {[}0.889, 1.729{]}\tabularnewline
Labor Elasticity & 0.452 & 0.509 & 0.266\tabularnewline
95\% CI & {[}0.415, 0.485{]} & {[}0.481, 0.537{]} & {[}-0.281, 0.813{]}\tabularnewline
\midrule 
Chemical & TERC & OLS & Naive IV\tabularnewline
\midrule 
Capital Elasticity & 0.463 & 0.452 & 0.555\tabularnewline
95\% CI & {[}0.430, 0.497{]} & {[}0.419, 0.484{]} & {[}-0.094, 1.204{]}\tabularnewline
Labor Elasticity & 0.311 & 0.339 & -0.757\tabularnewline
95\% CI & {[}0.263, 0.361{]} & {[}0.291, 0.386{]} & {[}-1.302, -0.212{]}\tabularnewline
\midrule 
Nonmetallic Mineral & TERC & OLS & Naive IV\tabularnewline
\midrule 
Capital Elasticity & 0.371 & 0.382 & -0.670\tabularnewline
95\% CI & {[}0.332, 0.410{]} & {[}0.354, 0.411{]} & {[}-4.605, 3.265{]}\tabularnewline
Labor Elasticity & 0.311 & 0.384 & -3.512\tabularnewline
95\% CI & {[}0.255, 0.365{]} & {[}0.340, 0.427{]} & {[}-9.599, 2.575{]}\tabularnewline
\midrule 
General Equipment & TERC & OLS & Naive IV\tabularnewline
\midrule 
Capital Elasticity & 0.433 & 0.389 & 0.951\tabularnewline
95\% CI & {[}0.402, 0.469{]} & {[}0.363, 0.416{]} & {[}0.765, 1.138{]}\tabularnewline
Labor Elasticity & 0.521 & 0.501 & -0.619\tabularnewline
95\% CI & {[}0.479, 0.564{]} & {[}0.464, 0.538{]} & {[}-1.220, -0.019{]}\tabularnewline
\midrule 
Transportation Equipment & TERC & OLS & Naive IV\tabularnewline
\midrule
Capital Elasticity & 0.425 & 0.442 & 0.723\tabularnewline
95\% CI & {[}0.393, 0.459{]} & {[}0.408, 0.475{]} & {[}0.523, 0.924{]}\tabularnewline
Labor Elasticity & 0.588 & 0.560 & 0.508\tabularnewline
95\% CI & {[}0.536, 0.644{]} & {[}0.509, 0.612{]} & {[}-0.110, 1.127{]}\tabularnewline
\bottomrule
\end{tabular}}\begin{tablenotes}[flushleft]
\scriptsize
\item \textit{Notes:} "OLS" denotes the pooled OLS method with constant coefficients to estimate the output elasticities. "Naive IV" represents two-stage least squares method where the IVs are the same as in TERC estimates. CIs are obtained using standard Stata command.
\end{tablenotes}
\end{threeparttable}
\end{table}

\section{Simulation\label{sec:Simulation}}

In this section, I examine the finite-sample performance of the series
estimators via a simulation study. A discussion of the data generating
process (DGP) motivated by production function applications is first
provided. Then, I present the baseline results of estimating the APE
$\ol b\coloneqq T^{-1}\sum_{t=1}^{T}\E\b_{it}$. Next, I show how
the method performs for estimating the LAR $b_{t}\left(x\right)\coloneqq\E\left[\rest{\b_{it}}X_{it}=x\right]$.
Finally, as robustness checks, I (i) vary the number of firms and
periods, (ii) change the degree of basis functions used for series
estimation, and (iii) include ex-post shocks $\widetilde{\epsilon}_{it}$
and $\upsilon_{it}$.

\subsection{DGP}

The baseline revenue-based DGP is
\[
Y_{it}=k_{it}\b_{it,K}+l_{it}\b_{it,L}+\omega_{it}+\widetilde{\epsilon}_{it},
\]
where $\omega_{it},\ \b_{it,K},\ \text{and }\b_{it,L}$ are all functions
of $\left(A_{i},\ve_{it},\upsilon_{it}\right)$, $k_{it}$ and $l_{it}$
are the natural logs of optimal capital and labor calculated from
the solution to firm $i$'s profit maximization problem, and $Y_{it}$
is the natural log of value-added output measured in dollars. Let
$\mathcal{U}\left[a,b\right]$ denote the uniform distribution over
$\left[a,b\right]$. I draw $A_{i}\sim_{i.i.d.}\mathcal{U}\left[1,2\right]$
and $\ve_{it}\sim_{i.i.d.}\mathcal{U}\left[1,2\right]$. I let true
$\omega_{it}=\ln\left(A_{i}+\ve_{it}/2+1\right)$, $\b_{it,K}=\left(A_{i}+\ve_{it}\right)/10$,
and $\b_{it,L}=\left(A_{i}+\ve_{it}\right)/10$, and write $\b_{it}=\left(\omega_{it},\b_{it,K},\b_{it,L}\right)'$.
I compute the true $\ol{\omega}\coloneqq T^{-1}\sum_{t=1}^{T}\E\omega_{it}=1.1736,$
$\ol b_{K}\coloneqq T^{-1}\sum_{t=1}^{T}\E\b_{it,K}=.3$, and $\ol b_{L}\coloneqq T^{-1}\sum_{t=1}^{T}\E\b_{it,L}=.3$.
The range of the random coefficients are set such that the second-order
condition for the profit maximization problem is satisfied. For the
baseline results, I let ex-post shocks $\widetilde{\epsilon}_{it}=\upsilon_{it}=0$
and investigate their impacts as robustness checks later. Suppose
capital and labor choices are made separately (e.g., investment and
hiring decisions made by different departments) based on ex-ante signals
about $\ve_{it}$ with noise $\lambda_{it,K}$ and $\lambda_{it,L}$:
$\eta_{it,K}=\ve_{it}+\lambda_{it,K}$ and $\eta_{it,L}=\ve_{it}+\lambda_{it,L}$,
where $\lambda_{it,K}\text{ and }\lambda_{it,L}\sim_{i.i.d.}\mathcal{U}\left[-.05,.05\right]$.
Since $Y_{it}$ is measured in dollars which is the case with most
real production datasets, the price of output $P_{it}$ is assumed
to be 1. I draw each element of the IVs $Z_{it}=\left(r_{it},w_{it}\right)'$
from $\mathcal{U}\left[0,\ln3\right]$ independent of each other and
all other variables (in particular, $Z_{it}\perp(\ve_{it},\eta_{it})'$),
and solve the firm's profit maximization problem to obtain 
\begin{align*}
k_{it} & =\dfrac{r_{it}-\left(A_{i}+\eta_{it,K}\right)\left(r_{it}-w_{it}\right)/10-\ln\left(\left(A_{i}+\eta_{it,K}\right)/10\right)-\ln\left(A_{i}+\eta_{it,K}/2+1\right)}{\left(A_{i}+\eta_{it,K}\right)/5-1},\\
l_{it} & =\dfrac{w_{it}-\left(A_{i}+\eta_{it,L}\right)\left(w_{it}-r_{it}\right)/10-\ln\left(\left(A_{i}+\eta_{it,L}\right)/10\right)-\ln\left(A_{i}+\eta_{it,L}/2+1\right)}{\left(A_{i}+\eta_{it,L}\right)/5-1}.
\end{align*}

Let $X_{it}=\left(k_{it},l_{it}\right)'$ denote the two endogenous
choice variables. It is clear that $X_{it}$ is correlated with $\b_{it}$
in each period via $\left(A_{i},\eta_{it,K},\eta_{it,L}\right)$ in
a nonseparable way. I use $n$, $T$, and $M$ to denote the total
number of firms, periods, and simulations, respectively. The simulated
dataset consists of $\left\{ X_{it}^{(m)},Y_{it}^{(m)},Z_{it}^{(m)}\right\} $
for $i=1,\ldots,n$, $t=1,\ldots,T$, and $m=1,\ldots,M$, which are
used to construct $\widehat{\ol b}^{(m)}$ and $\widehat{b}_{t}^{(m)}\left(x\right)$
to estimate $\ol b$ and $b_{t}\left(x\right)$, respectively, via
the estimation procedure outlined in Subsection \ref{subsec:Estimation}.
I evaluate the performance of the estimators by their biases, root-mean
squared errors (rMSE), and mean normed deviations (MND), with the
explicit mathematical definitions provided in the tables below.

\subsection{Results}

For the baseline results, I set $n=1,000$ and $T=2$, and use a second-degree
polynomial spline basis with its knot at the median for all series
estimation steps. I run $M=1,000$ simulations. For each $i$, I construct
$W_{i}$ as the mean over time of each coordinate of $X_{it}$. Following
the theory, I set $V_{it}\coloneqq\left(F_{\rest{k_{it}}Z_{it},W_{i}}\left(\rest{k_{it}}Z_{it},W_{i}\right),F_{\rest{l_{it}}Z_{it},W_{i}}\left(\rest{l_{it}}Z_{it},W_{i}\right)\right)'$.
The performance of $\widehat{\ol{\omega}}$, $\widehat{\ol b}_{K},$
and $\widehat{\ol b}_{L}$ is summarized in Table \ref{tab:firstperf}.
For notational simplicity, I use $\sum_{m}$ for $\sum_{m=1}^{M}$
and $\sum_{i,t,m}$ for $\sum_{i=1}^{n}\sum_{t=1}^{T}\sum_{m=1}^{M}$.

\begin{table}
\caption{Performance of $\widehat{\protect\ol b}$ \label{tab:firstperf}}

\smallskip{}

\centering{}%
\begin{tabular}{ccccc}
\toprule 
 & Formula & $\widehat{\ol{\omega}}$ & $\widehat{\ol b}_{K}$ & $\widehat{\ol b}_{L}$\tabularnewline
\midrule
Bias & $M^{-1}\sum_{m}\left(\widehat{\ol b}_{d}^{\left(m\right)}-\ol b_{d}\right)/\abs{\ol b_{d}}$ & 2.81\% & 3.22\% & 3.24\%\tabularnewline
rMSE & $\sqrt{M^{-1}\sum_{m}\left(\widehat{\ol b}_{d}^{\left(m\right)}-\ol b_{d}\right)^{2}}/\abs{\ol b_{d}}$ & 2.89\% & 3.65\% & 3.62\%\tabularnewline
\multirow{1}{*}{MND} & $M^{-1}\sum_{m}\abs{\widehat{\ol b}_{d}^{\left(m\right)}-\ol b_{d}}/\abs{\ol b_{d}}$ & 2.81\% & 3.26\% & 3.26\%\tabularnewline
\bottomrule
\end{tabular}
\end{table}

The first row of Table \ref{tab:firstperf} reports the normalized
bias of each coordinate of $\widehat{\ol b}$. The bias is reasonably
small across all three coordinates, with a magnitude between 2.81\%
and 3.24\% of the length of the corresponding coordinate of $\ol b$.
The second row shows the normalized rMSE of each coordinate of $\widehat{\ol b}$.
My method achieves low normalized rMSEs between 2.89\% and 3.65\%
of the size of the corresponding coordinate $\ol b_{d}$ of $\ol b$.
The last row presents the normalized MNDs of each coordinate of $\widehat{\ol b}$.
Again, the method performs well with an MND between 2.81\% and 3.26\%
of the size of the corresponding coordinate $\ol b_{d}$ of $\ol b$.

$\ $

Next, I investigate the performance of $\widehat{b}_{t}\left(x\right)$,
which is obtained following the estimation procedure outlined in Subsection
\ref{subsec:Estimation}. For the true $b_{t}\left(x\right)$, due
to the complex dependence structure of $X_{it}$ on $\b_{it}$, there
is no analytical solution available. Therefore, for each $t$ I pool
the observations across all $M$ simulations ($n\times M$ observations
for each $t$) and approximate $b_{t}\left(x\right)$ by regressing
the true $\b_{it}$ on the same spline basis functions of $X_{it}$.

Table \ref{tab:beta_x_perf} summarizes the results. Note that by
definition the bias of $\widehat{b}_{t}\left(x\right)$ is the same
as that of $\widehat{\ol b}$, so I omit it here. The normalized rMSE
of $\widehat{b}_{t}\left(x\right)$ is bigger than that of $\widehat{\ol b}$,
with a magnitude between 3.36\% and 5.59\% of the size of the corresponding
$\ol b_{d}$. The normalized MND follows a similar pattern. The performance
of $\widehat{b}_{t}\left(x\right)$ for estimating $b_{t}\left(x\right)$
is not as good as that of $\widehat{\ol b}$ for $\ol b$, which is
expected because (i) $b_{t}\left(x\right)$ is a function rather than
a finite-dimensional vector and (ii) there is approximation error
in calculating the true $b_{t}\left(x\right)$ via simulations.

\begin{table}
\caption{Performance of $\widehat{b}_{t}\left(x\right)$\label{tab:beta_x_perf}}

\smallskip{}

\centering{}{\small{}%
\begin{tabular}{ccccc}
\toprule 
 & Formula & $\widehat{\omega}_{t}\left(x\right)$ & $\widehat{b}_{t,K}\left(x\right)$ & $\widehat{b}_{t,L}\left(x\right)$\tabularnewline
\midrule
rMSE & $\sqrt{\left(NTM\right)^{-1}\sum_{i,t,m}\left(\widehat{b}_{t,d}^{(m)}\left(x_{it}^{(m)}\right)-b_{t,d}\left(x_{it}^{(m)}\right)\right)^{2}}/\abs{\ol b_{d}}$ & 3.36\% & 5.57\% & 5.59\%\tabularnewline
\multirow{1}{*}{MND} & $\left(NTM\right)^{-1}\sum_{i,t,m}\abs{\widehat{b}_{t,d}^{(m)}\left(x_{it}^{(m)}\right)-b_{t,d}\left(x_{it}^{(m)}\right)}/\abs{\ol b_{d}}$ & 2.87\% & 4.36\% & 4.37\%\tabularnewline
\bottomrule
\end{tabular}}{\small\par}
\end{table}

$\ $

To show how robust my method is in estimating $\ol b$ and $b_{t}\left(x\right)$,
I conduct another set of exercises. I evaluate the performance of
my TERC estimator using rMSE defined as 
\[
\sqrt{M^{-1}\sum_{m}\norm{\widehat{\ol b}^{(m)}-\ol b}^{2}}/\norm{\ol b}
\]
for the vector $\ol b$, and 
\[
\sqrt{\left(NTM\right)^{-1}\sum_{i,t,m}\norm{\widehat{b}_{t}^{(m)}\left(x_{it}^{(m)}\right)-b_{t}\left(x_{it}^{(m)}\right)}^{2}}/\norm{\ol b}
\]
for the vector $b_{t}\left(x\right)$. First, I vary $n$ and $T$,
and present the results in Table \ref{tab:sim vary N T}. As expected,
a larger $n$ is good for overall performance. However, the magnitude
in the improvement of performance is mild, possibly because with more
agents I need more data to control for the increasing degree of heterogeneity.
On the other hand, I find that the method performs reasonably well
even with a small sample size of $n=500$. Having a larger $T$ improves
the performance, which is again expected as I can exploit more information
from repeated observations of the same individual to better control
for the fixed effect $A_{i}$.

\begin{table}[H]
\caption{Performance under Varying $n$ and $T$ \label{tab:sim vary N T}}

\smallskip{}

\centering{}%
\begin{tabular}{ccc}
\toprule 
\multirow{1}{*}{$n$ ($T=2$)} & \multicolumn{1}{c}{rMSE: $\ol b$} & \multicolumn{1}{c}{rMSE: $b_{t}\left(x\right)$}\tabularnewline
\midrule
$500$ & 3.25\% & 4.23\%\tabularnewline
$1,000$ & 2.99\% & 3.69\%\tabularnewline
\multirow{1}{*}{$2,000$} & 2.87\% & 3.43\%\tabularnewline
\bottomrule
\end{tabular}%
\begin{tabular}{ccc}
\toprule 
\multirow{1}{*}{$T$ ($n=1,000$)} & \multicolumn{1}{c}{rMSE: $\ol b$} & \multicolumn{1}{c}{rMSE: $b_{t}\left(x\right)$}\tabularnewline
\midrule
2 & 2.99\% & 3.69\%\tabularnewline
3 & 2.74\% & 3.38\%\tabularnewline
\multirow{1}{*}{4} & 2.66\% & 3.27\%\tabularnewline
\bottomrule
\end{tabular}
\end{table}

Second, I vary the degree of the spline bases used to construct the
series estimators for all steps. The knot is placed at the median
for all specifications. Table \ref{tab:sim vary order} contains the
results. I find that increasing the degree of basis functions from
one to two improves estimation accuracy significantly. When I increase
the degree of basis functions from two to three, the performance deteriorates.
When the degree becomes larger, there are more regressors in each
step of regression, leading to a possible over-fitting problem. Based
on this result, I use the second-degree splines in the empirical application.
Note that one may also use the AIC criterion to select the degree
of basis functions.

\begin{table}[H]
\caption{Performance under Varying Degree of Basis Functions \label{tab:sim vary order}}

\smallskip{}

\centering{}%
\begin{tabular}{ccc}
\toprule 
\multirow{1}{*}{Degree of Basis Functions} & rMSE: $\ol b$ & rMSE: $b_{t}\left(x\right)$\tabularnewline
\midrule
\multirow{1}{*}{1} & 5.27\% & 6.41\%\tabularnewline
2 & 2.99\% & 3.69\%\tabularnewline
3 & 4.82\% & 6.40\%\tabularnewline
\bottomrule
\end{tabular}
\end{table}

Lastly, I examine how including the ex-post shocks $\widetilde{\epsilon}_{it}$
and $\upsilon_{it}$ into the model affects the performance of my
estimators. I draw $\widetilde{\epsilon}_{it}\sim\mathcal{U}\left[-.25,.25\right]$,
the ex-post shock to the main equation \eqref{eq:Yeq}, independently
of all the other variables. $\widetilde{\epsilon}_{it}$ can also
be interpreted as an ex-post shock to $\omega_{it}$. For $\upsilon_{it}$,
since ex-post shock to $\omega_{it}$ has been considered by $\widetilde{\epsilon}_{it}$,
I draw $\upsilon_{it}^{s}\sim\mathcal{U}\left[-.1,.1\right]$ independently
of all the other variables and use $\b_{it,s}^{\text{new}}=\b_{it,s}+\upsilon_{it,s}$
in \eqref{eq:Yeq} for $s\in\left\{ K,L\right\} $. Results are presented
in Table \ref{tab:sim meas error}. As expected, adding ex-post errors
$\widetilde{\epsilon}_{it}$ or $\upsilon_{it}$ negatively affects
the performance of the estimators. When $\widetilde{\epsilon}_{it}$
is included, the rMSE of $\widehat{\ol b}$ for estimating $\ol b$
increases from 2.99\% to 3.64\% and the rMSE of $\widehat{b}_{t}\left(x\right)$
for estimating $b_{t}\left(x\right)$ rises from 3.64\% to 5.43\%.
The effect of including $\upsilon_{it}$ on the performance is similar.

\begin{table}[H]
\caption{Performance with and without Ex Post Shocks \label{tab:sim meas error}}

\smallskip{}

\centering{}%
\begin{tabular}{cccccc}
\toprule 
\multirow{1}{*}{Add $\epsilon_{it}$ to $Y_{it}$?} & \multicolumn{1}{c}{rMSE: $\ol b$} & rMSE: $b_{t}\left(x\right)$ & Add $\upsilon_{it}$ to $\b_{it}$? & rMSE: $\ol b$ & rMSE: $b_{t}\left(x\right)$\tabularnewline
\midrule
No & 2.99\% & 3.69\% & No & 2.99\% & 3.69\%\tabularnewline
Yes & 3.64\% & 5.43\% & Yes & 3.57\% & 5.19\%\tabularnewline
\bottomrule
\end{tabular}
\end{table}

\end{document}